\documentclass[authoryear]{elsarticle}

\usepackage{latexsym}
\usepackage{booktabs}
\usepackage{geometry}
\usepackage{amsthm}
\usepackage{stmaryrd}
\usepackage{longtable}
\usepackage{subfig}
\usepackage{verbatim}
\usepackage{array}

\newcolumntype{C}[1]{>{\centering\arraybackslash}p{#1}}

\usepackage{soul}

\usepackage{color}
\usepackage[export]{adjustbox}
\usepackage{listings}
\usepackage{epstopdf} 
\usepackage{todonotes}
\usepackage{mathtools}
\usepackage[ruled,linesnumbered]{algorithm2e}

\usepackage{graphicx}
\usepackage{epstopdf}
\usepackage{multirow}
\usepackage{paralist}
\usepackage{amsmath,amsthm,amssymb}
\usepackage{setspace}
\usepackage{float}
\usepackage{morefloats}
\usepackage[]{algorithm2e}
\usepackage{mdwlist}
\usepackage{booktabs} 
\usepackage{amsfonts}
\usepackage{paralist}
\usepackage{caption}

\newcommand{\be}{\begin{equation}}
\newcommand{\ee}{\end{equation}}
\newcommand{\highlighttext}[1]{#1}

\newcommand{\RNum}[1]{\uppercase\expandafter{\romannumeral #1\relax}}
\theoremstyle{remark}

\usepackage{lineno,hyperref}
\modulolinenumbers[1]
\usepackage{geometry}
\usepackage{subfig}
\usepackage{color}
\usepackage[export]{adjustbox}
\usepackage{listings}
\usepackage{epstopdf} 
\usepackage{graphicx}
\begin{document}
\begin{frontmatter}
    \title{Spatiotemporal impacts of human activities and socio-demographics during the COVID-19 outbreak in the U.S.}
    \author[mymainaddress]{Lu Ling}
    \ead{ling58@purdue.edu}
    \author[bamaaddress]{Xinwu Qian}
    \ead{xinwu.qian@ua.edu}
    \author[mymainaddress]{Satish V. Ukkusuri}
    \ead{sukkusur@purdue.edu}
    \author[bamaaddress]{Shuocheng Guo}
    \ead{sguo18@ua.edu}

    \begin{abstract}
    Understanding influencing factors is essential for the surveillance and prevention of infectious diseases, and the factors are likely to vary spatially and temporally as the disease progresses. Taking daily cases and deaths data during the coronavirus disease 2019 (COVID-19) outbreak in the U.S. as a case study, we develop a mobility-augmented geographically and temporally weighted regression (M-GTWR) model to quantify the spatiotemporal impacts of social-demographic factors and human activities on the COVID-19 dynamics. Different from the base GTWR model, we incorporate a mobility-adjusted distance weight matrix where travel mobility is used in addition to the spatial adjacency to capture the correlations among local observations. The model residuals suggest that the proposed model achieves a substantial improvement over other benchmark methods in addressing the spatiotemporal nonstationarity. Our results reveal that the impacts of social-demographic and human activity variables present significant spatiotemporal heterogeneity. In particular, a 1\% increase in population density may lead to 0.63\% and 0.71\% more daily cases and deaths, and a 1\% increase in the mean commuting time may result in 0.22\% and 0.95\% increases in daily cases and deaths. Although increased human activities will, in general, intensify the disease outbreak, we report that the effects of grocery and pharmacy-related activities are insignificant in areas with high population density. And activities at the workplace and public transit are found to either increase or decrease the number of cases and deaths, depending on particular locations. The results of our study establish a quantitative framework for identifying influencing factors during a disease outbreak, and the obtained insights may have significant implications in guiding the policy-making against infectious diseases. 
    \end{abstract}
    
    
\end{frontmatter}


\section{Introduction}
The coronavirus disease 2019 (COVID-19) has infected 22 million people and caused at least 375 thousand deaths~\citep{covid19global} in the U.S. since its initial outbreak in February 2020. The estimated mortality rate is 8.1 times higher than seasonal influenza, according to the Centers for Disease Control and Prevention (CDC)~\citep{livingston2020coronavirus}. The pandemic increased the unemployment rate~\citep{swasey2020staggering} and is continuing to have a significant impact on the economy~\citep{coibion2020labor}. In light of the severe consequences from the COVID-19 outbreak, different public authorities quickly responded to the outbreak through various strategies, including the declaration of emergency, travel restrictions, city lock-down, and enforcing social distancing~\citep{brzezinski2020covid}. If properly followed and executed, these measures serve as the crucial first steps to limit physical contact and mitigate the extent of the outbreak before a vaccine is available. Nevertheless, under similar mitigation measures, significant differences are observed in the number of reported infections and the mortality rate across the U.S.~\citep{dong2020interactive}. This motivates us to explore the underlying factors that result in the heterogeneous disease dynamics for assisting the disease mitigation policies in the remaining phase of the COVID-19, and for better preparing against future risks of unknown infectious diseases.

As mentioned in the WHO study for the 2009 H1N1 pandemic~\citep{mateus2014effectiveness}, in addition to the pathological variables, the extent of the disease outbreak may be attributed to various non-epidemiological factors, including mobility level, social-demographics, pre-existing conditions of the population~\citep{christensen2010disease}, quality of health services, travel patterns, social network~\citep{eubank2004modelling,pinshi2020uncertainty,9043580,pinshi2020uncertainty}, ecological factors~\citep{ficetola2020climate,mateus2014effectiveness}, etc. But our knowledge on the precise impacts of these factors is very limited, primarily due to the lack of data that may enable the nexus between disease dynamics and the possible contributing factors. With recent advances in ubiquitous computing and epidemiology and the wide adoption of smartphones in the past decade, we are now able to monitor human activities at a fine spatiotemporal level and overlay such dynamics with high-resolution trajectories of disease outbreaks. This, together with the available data on socioeconomic, demographics and historical daily commuting patterns, provides an unprecedented opportunity to scrutinize the impacts of non-epidemiological factors and comprehensively evaluate how these factors drive the fate of the disease outbreak across the U.S. 

Existing studies have related social-demographic characteristics and human mobility with the spread of the COVID-19. The social-demographic structure of the population is demonstrated to have a significant effect on the fatality rate. An early study in China~\citep{novel2020epidemiological} suggested that people with age greater than 80 years older has the highest fatality rate of 14.8\%, and similar findings were obtained from studies in other countries~\citep{dowd2020demographic,sannigrahi2020overall}. In addition, a study~\citep{almagro2020differential} in New York City revealed the existence of racial disparities among the Whites, the blacks, the Asians and the Hispanics in the COVID-19 outbreak. In particular, nearly 20\% of the U.S. counties had a disproportionate Black population~\citep{millett2020assessing}, and they accounted for 52\% of the confirmed cases and 58\% of the deaths nationally. Except for demographic factors, the social and economic factors are also found to affect the fate of the COVID-19 outbreak. The study~\citep{atchison2020perceptions} suggested that households with the lowest income level are six times less likely to be able to work from home (WFH) and three times less likely to be able to self-isolate in the U.K. during the COVID-19. Besides, ~\cite{stojkoski2020socio} mentioned that the high income population is more resilient to being infected by the COVID-19. Finally, extensive efforts have shown that human activities and mobility dynamics are dominating factors that facilitate the spread of infectious diseases~\citep{wesolowski2012quantifying,vazquez2013using,qian2020scaling,qian2020modeling}. Studies suggested both information propagation and human mobility patterns co-affect the epidemic propagation~\citep{wang2017interplay,qian2020modeling2,yabe2020non}. Nevertheless,~\cite{lima2015disease} provided evidence that restricting mobility may not eliminate the diseases. And~\cite{bajardi2011human} recommended that stricter regimes of travel reduction would have led to a delayed outbreak of two weeks based on the study of the 2009 H1N1 pandemic.

\highlighttext{The aforementioned studies highlighted the significant roles played by mobility-related and social-demographic factors in the disease spreading process. Nevertheless, few studies examined the collective impacts of non-epidemiological factors on the spatiotemporal dynamics of infectious disease. In addition, the impacts of these influencing factors were primarily assumed to be stationary over time and space in the existing literature. The lack of consideration of these aspects will fail to fully reveal the interdependencies among modeling determinants and may result in biased model estimations. To address the issues, in this study, we introduce a quantitative approach, named mobility-augmented geographically and temporally weighted regression model (M-GTWR), to investigate the heterogeneous effects of non-epidemiological factors on the spreading dynamics of the COVID-19. The M-GTWR model can examine the heteroscedasticity of weekly average daily confirmed cases (WADC) and weekly average daily deaths (WADD) during the COVID-19 outbreak and quantify the spatiotemporal effects of the social-demographic characteristics in U.S. counties. By relating daily activity data and inter-county transportation data with the propagation dynamics of the COVID-19, the model confirms that the relationship between the spread of diseases and human activities is spatiotemporally heterogeneous. We also measure how mobility and activity dynamics affect the disease propagation. Our results suggest that counties with a high percentage of black population, a high household income level, a low education level, and a high unemployment rate are associated with more WADC and WADD. Moreover, the impact of human activity dynamics is found to differ spatially. Grocery and pharmacy activities only show positive and statistically significant effects on the COVID-19 cases in rural counties with a low income level, and effects of the public transit activities are tightly related to the WFH and reopening strategies.}

The outline of this study is as follows. The ‘Data Preparation’ section describes the study area and data preparation. The ‘Methodology’ describes the geographically and temporally weighted regression model. The ‘Model Calibration’ section calibrates the model specification. The `Results' section elaborates on estimated results from the model. Finally, the ‘Conclusion’ section includes key insights. 
\section{Data Preparation}
\subsection{Study area}
\begin{figure}[!h]
\centering
\includegraphics[width=.7\linewidth,trim={4.5cm 7cm 2.0cm 6cm},clip]{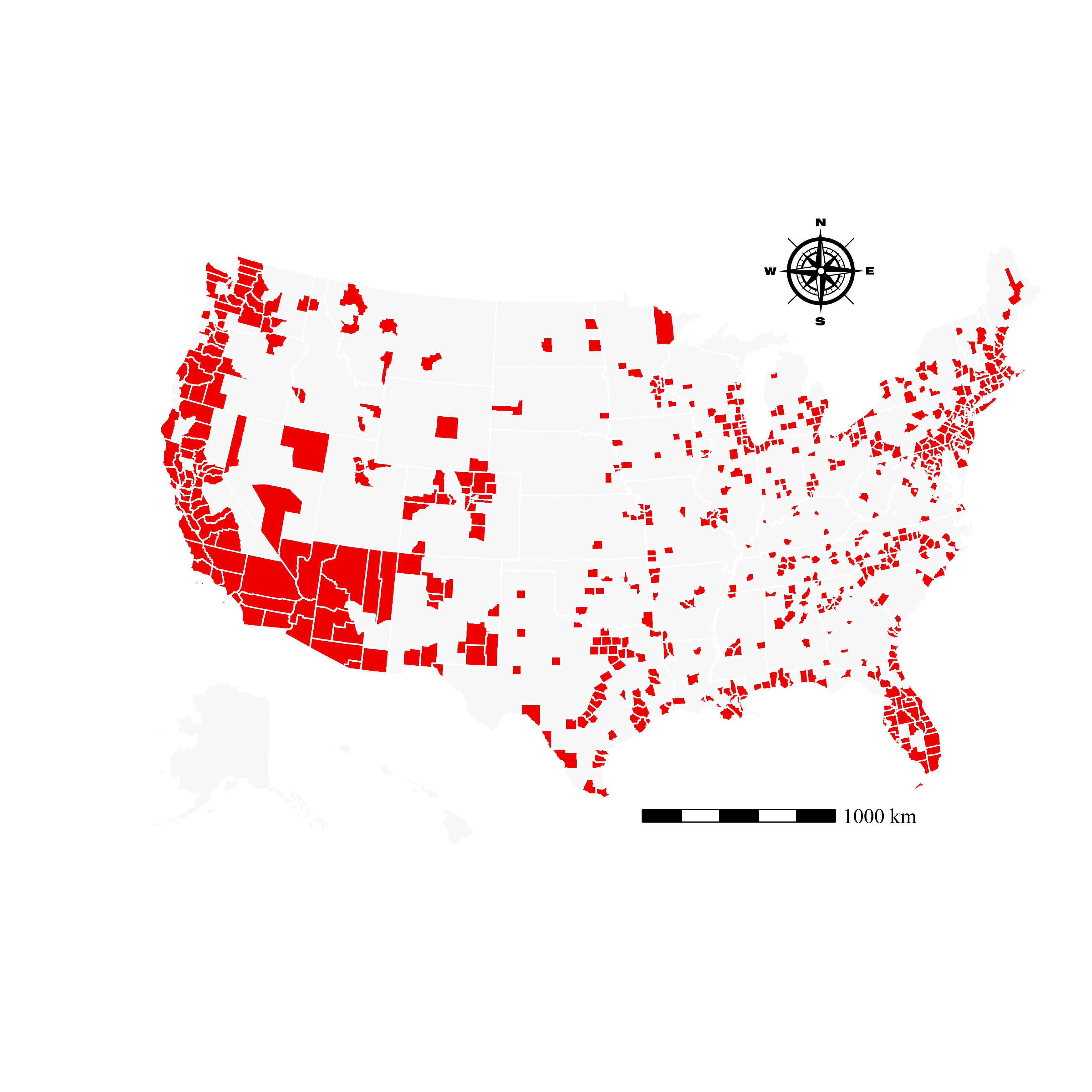}
\caption{The spatial distribution of selected counties (in red) in the U.S.}
\label{fig:targeted_county}
\end{figure}

 We investigate the COVID-19 dynamics at the U.S. county-level. There are 3141 counties in the U.S., and the counties present a significant variation of reported daily cases. To ensure that the disease dynamics are statistically meaningful, we target the counties with at least 100 confirmed cases from March 23, 2020 to December 13, 2020. In addition, counties with incomplete data are also removed. Finally, we keep the counties within 48 contiguous states. The preprocessing results in 699 selected counties that cover both metropolitan areas (592) and non-metropolitan areas (107) according to the definition in the rural-urban commuting area (RUCA) code~\citep{morrill1999metropolitan}. And the spatial distribution of the study areas (red-colored counties) is shown in Figure~\ref{fig:targeted_county}. The selected counties cover $79\%$ of the total U.S. population. We report that the processed data provides a reasonable scale to understand the non-epidemiological determinants of the COVID-19 propagation and linked with the underlying effects of the social-demographics characteristics and human activities on disease propagation in the metropolitan counties in the U.S.  

\subsection{Dependent variables}
\begin{figure}[!h]
\centering
\includegraphics[width=\linewidth]{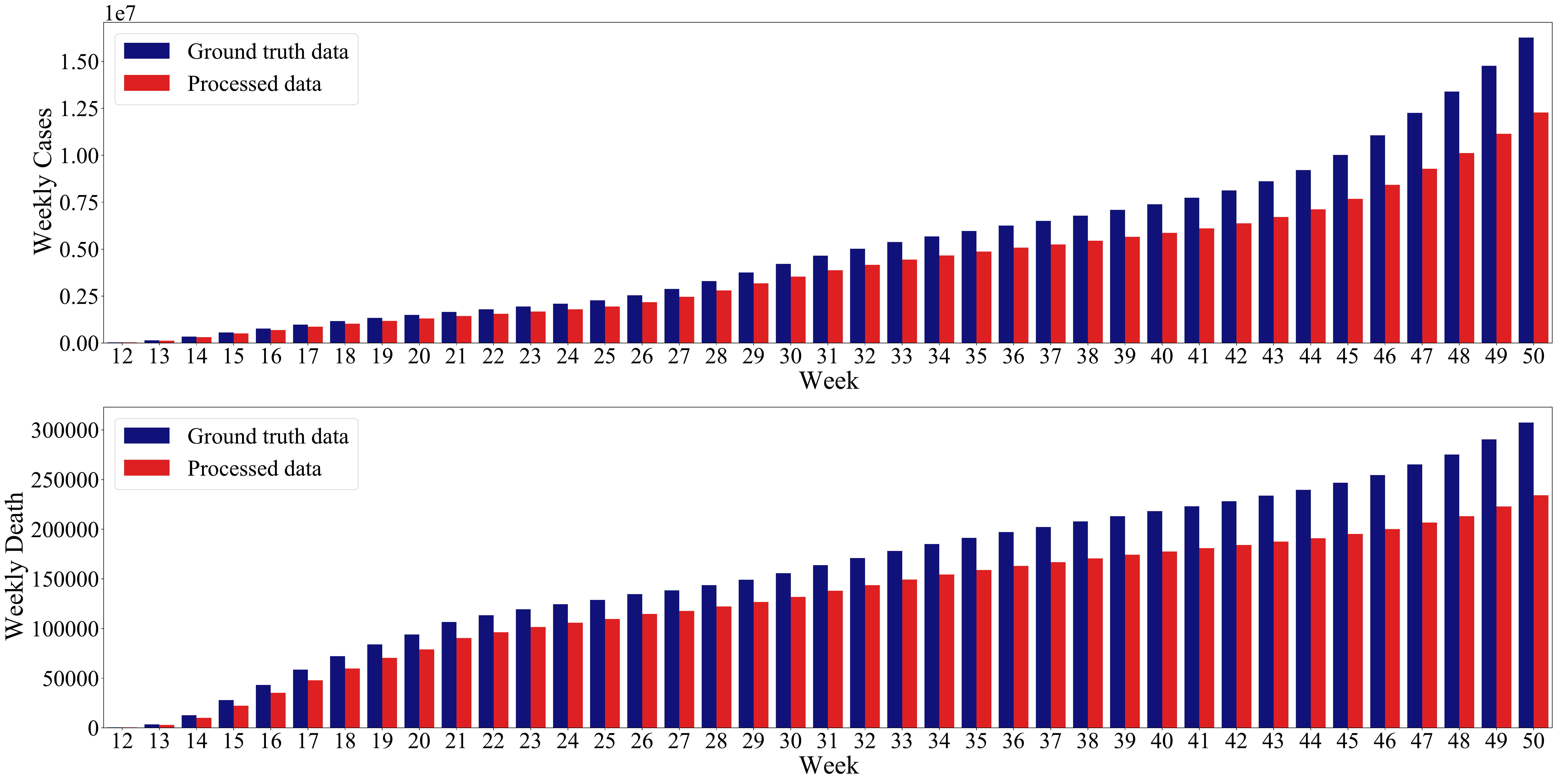}
\caption{The number of WADC and WADD between processed counties and the ground truth counties}
\label{fig:case-death}
\end{figure}

The dependent variables used in the study are the number of WADC and the number of WADD at the county level, which are obtained from the Center for Systems Science and Engineering at the Johns Hopkins University~\citep{dong2020interactive}. We use the WADC rather than daily cases to smooth the daily fluctuations in the data. As suggested in other studies~\citep{novel2020epidemiological}, there is a time delay between the date when an individual was actually infected and the reported date, which is usually two weeks for the COVID-19. On the other hand, human activity dynamics were reflective of the reporting date. To ensure the consistency between the disease and mobility dynamics, we apply a two-week delay to the human activity data to match the disease data. We report that the number of confirmed cases and deaths in the selected counties accounted for 75.5\% of the total cases and 76.1\% of total deaths during the study period in the U.S, as shown in Figure~\ref{fig:case-death}. These indicate that the dependent variables in the selected counties are representative of the general disease dynamics in the metropolitan areas in the U.S. 

\subsection{Explanatory variables}
The explanatory variables used in our study contain both the dynamic variables that changed over time as the disease progressed and the static variables that remained constant during the study period. Table~\ref{tab:covid-19} summarizes the selected variables and their summary statistics. 

\subsubsection{Dynamic variables}
\begin{figure}[!h]
\centering
\includegraphics[width=\linewidth, trim={0 1.3cm 0 0}]{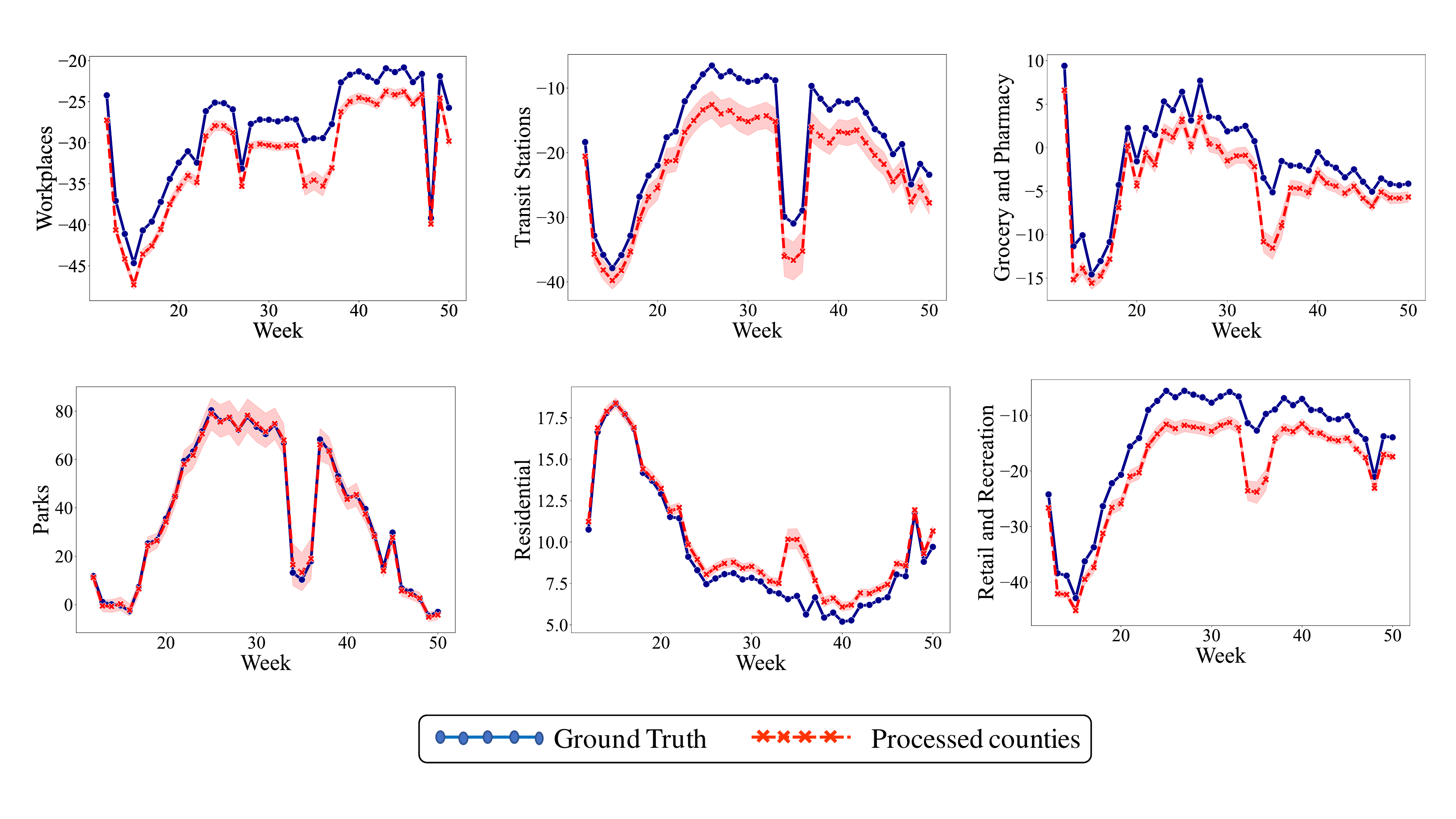}
\caption{The dynamic variables of the processed county and ground truth}
\label{fig:dynamic}
\end{figure}

The dynamic variables in the study are obtained from the Google Mobility Report\citep{GTOS}. The Google Mobility Report describes the change of daily activities in terms of recreation activity(RRPC), park activity(PAPC), residential activity(RAPC), grocery and pharmacy activity(GPPC), transit activity(TSPC), and workplace activity(WOPC) from the baseline value. The baseline value is the median value for the corresponding day of the week during January 3, 2020 and February 6, 2020. The dataset demonstrates the changes in visiting frequency of a particular activity category at individual counties and is also indicative of the activity intensity on the corresponding day of the week. To be consistent with the dependent variables, we calculated the average daily mobility in a week based on the daily mobility data. Then, we visualize the mean and standard deviation of the dynamic variables in the selected counties and its comparison with the ground truth data in Figure~\ref{fig:dynamic}. Similar to the dependent variables, we observe that the dynamic variables in selected counties also resemble the trends of the entire population.  

\begin{figure}[!h]
\centering
\includegraphics[width=.7\textwidth]{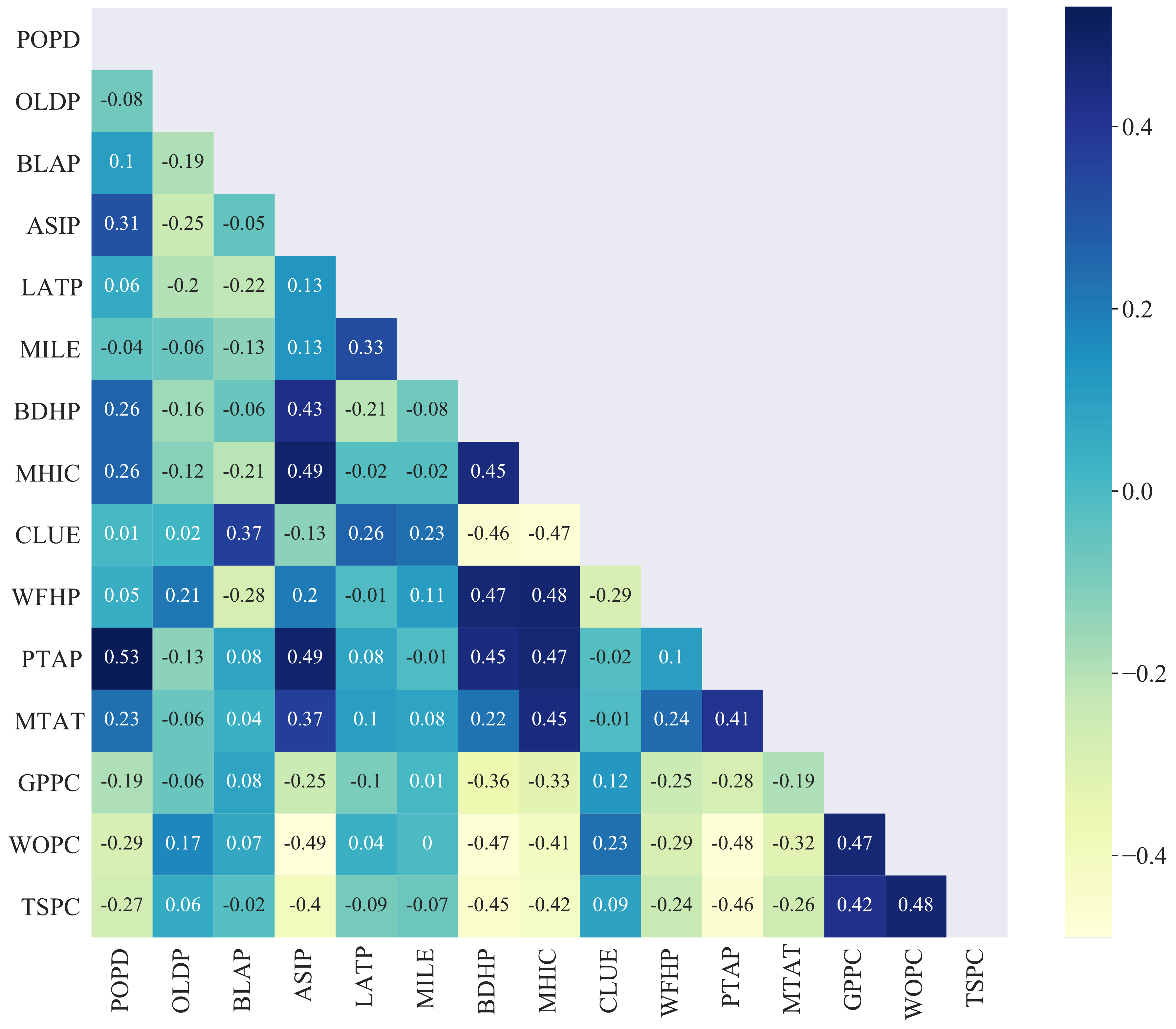}
\caption{Pearson product-moment correlation coefficient for explanatory variables}
\label{fig:corr}
\end{figure}

To eliminate the estimated bias from the association effects and multicollinearity issue, we tested the Pearson correlation and variance inflation factor(VIF) among dynamic variables in Figure~\ref{fig:corr}. It is observed that the Pearson correlations of the GPPC, TSPC, and WOPC are less than 0.6 and the VIF values are less than 10. However, the Pearson correlations among the RRPC, PAPC, and RAPC are near 0.7, which indicates a high correlation among these variables. Therefore, we selected GPPC, TSPC, and WOPC as model input. 

\subsubsection{Static variables}
In this study, the static variables include the county-level demographic factors, socioeconomic factors, and travel-related information. The demographic factors include the population, older population, white population, black or African American, Asian population, Hispanic Latino population, and land area. They are collected from the U.S. Census Bureau’s MAF/TIGER Geodatabases ~\citep{UScensus} and 2016 American Community Survey (ACS)~\citep{ACS2016}. In light of the varying size of the selected counties, we calculated the population density(POPD) by dividing the total population by the land area. Besides, we measured the percentage of the population by race via dividing the population in each race category by the total population. The socioeconomic factors in our study include the percentage of the population having at least a bachelor degree(BDHP), mean household income(MHIC), and the citizen labor force unemployment rate(CLUE). They are collected from the 2016 American Community Survey (ACS)~\citep{ACS2016}. The socioeconomic factors serve as indirect measures to probe how people may respond to the preventative measures and the economic resilience of the community against the disease outbreaks. 

Aside from the demographic and socioeconomic variables, we also include several travel-related factors as human activity intensity measures before the COVID-19. The travel-related factors include travel mode factors and public road mileage(MILE). The travel mode factors in our study contain the percentage of the population working at home(WFHP), the percentage of the population taking public transit(PTAP), and the mean commuting time(MTAT). These variables are obtained from the 5-Year ACS statistic (ACS 2011 to 2015)~\citep{ACS2016fiveyears}. Except for the above variables, the MILE is another crucial travel-related factor to reflect the intensity of economic activities in the county. The public road is described as any road under the jurisdiction and is maintained by a public authority. We collected the MILE from the 2018 Public Road Geodatabase~\citep{UScensus2018}. 

\begin{figure}[!h]
\centering
\includegraphics[width=\textwidth,trim={0cm, 0cm, 0cm, 1cm},clip]{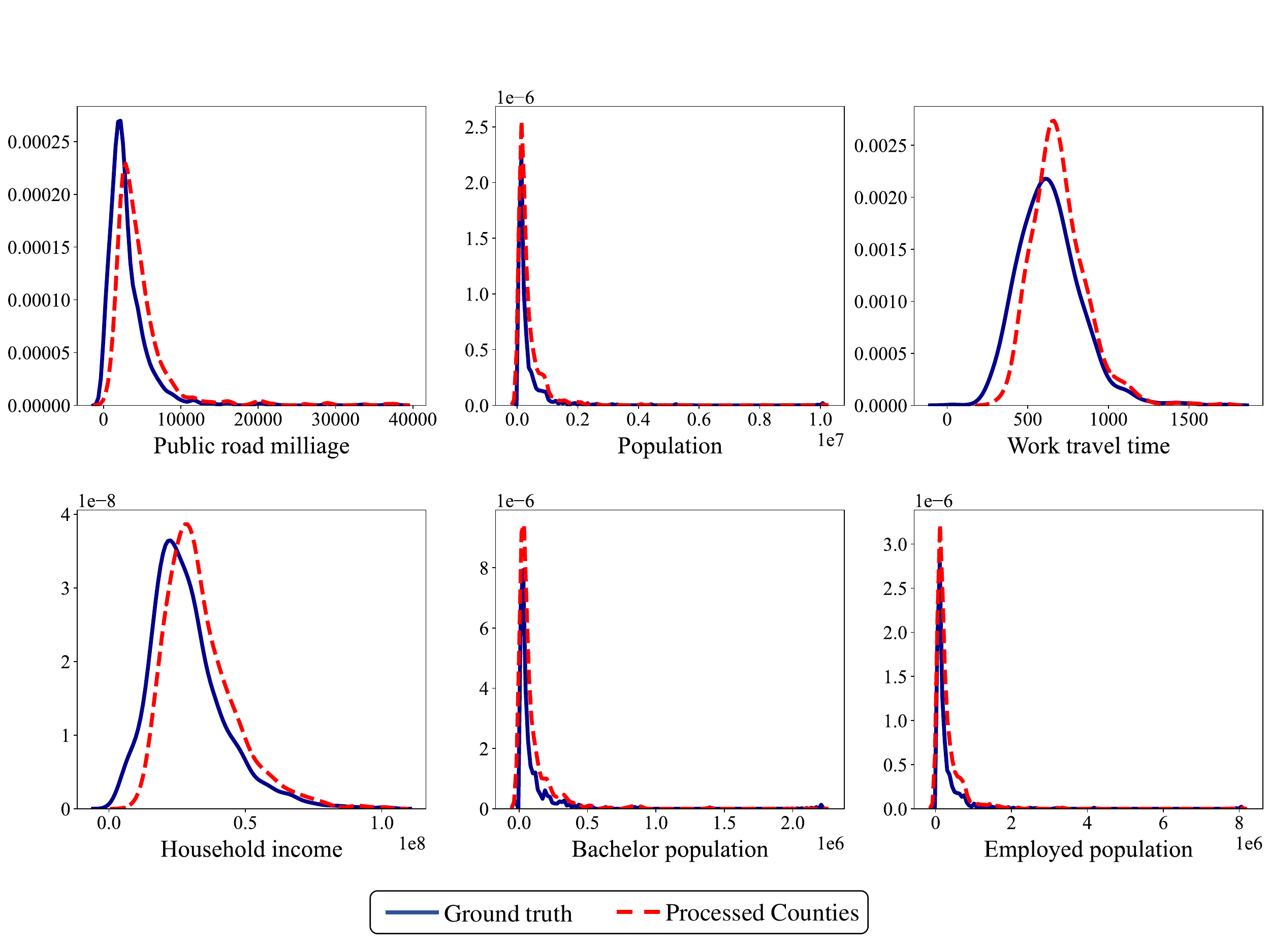}
\caption{The probability density function of the processed county and ground truth for independent variables}
\label{fig:static}
\end{figure}

The probability density function (PDF) of static variables in the selected counties and the comparisons to the ground can be seen in Figure~\ref{fig:static}, which further confirms the representativeness of the processed data. In addition, Figure~\ref{fig:corr} presents the Pearson correlation coefficients of the static factors, where most of the Pearson correlation coefficients of the static variables are less than 0.4. Some variables having correlations below 0.6 are also included in our model because they capture significant variations and provide non-overlapping effects for the dependent variables (we take the MILE as an explanatory factor instead of the MILE density to avoid the high correlation between MILE density and POPD).

The traffic flow variables include the road commuting flow and airline flow, which are gathered to describe the real travel connections among counties during the COVID-19 outbreak. The road commuting flow among counties is collected from the average 5-Year ACS statistic at the county-level~\citep{ACS2016fiveyears} and describes the traffic connections between residence counties and workplace counties. And the geographical airline passenger flow among airports is provided by the U.S. Department of Transportation and Bureau~\cite{data1999bureau}. The Bureau of Transportation Statistics offers quarterly airline and airport origination and destination survey (DB1B) within the U.S, which is a 10\% sample of airline tickets (passengers) from reporting carriers. The DB1B has the records of airline passenger volumes in U.S. airports, but it does not contain the information of airline flow among counties. To obtain the airline flow, we apply the airline traffic flow assignment method based on the origination and destination airport passenger volumes from DB1B.

\begin{longtable}{p{3.8cm}p{7.0cm}p{1.3cm}p{1.3cm}p{1.0cm}}
\caption{Deﬁnition and descriptive statistics of explanatory variables}\label{tab:covid-19}\\
\hline
Variables & Definitions & Mean &S.D.& VIF\\
\hline
\endhead
\hline
\endfoot
\textbf{\emph{Demographic}} &&&&\\
POPD & Population density (per square km) & 1210.83 & 4374.67&2.53\\
OLDP & The percentage of older population (\%) & 15.56&4.40&1.65\\
BLAP & The percentage of Black or African American percentage (\%) &12.09 &12.95&2.29\\
ASIP & The percentage of Asian percentage (\%) & 3.89&4.48&2.14\\
LATP & The percentage of Hispanic Latino percentage (\%) & 15.53 & 16.31 & 2.07\\

\textbf{\emph{Socioeconomic}} &&&&\\
BDHP & Percentage of population having at least bachelor degree (\%) & 31.26 &10.71&7.11\\
MHIC & Yearly mean household income (\$) & 82835.56 &21100.91&6.68\\
CLUE & Citizen labor force unemployment rate (\%) & 5.91&1.81&2.45\\

\textbf{\emph{Travel}} &&&&\\
MILE & Mileage (meter) &5238.21 &5128.13&1.32\\
WFHP & Percentage of population work at home (\%) & 4.83&1.88&2.21\\
PTAP & Percentage of population taking public transit (\%) & 3.09 & 6.86 &2.09\\
MTAT & Mean commuting time (minutes) & 24.94&5.01&2.60\\

\textbf{\emph{Human Activity}}&&&&\\
GPPC & Percentage of grocery and pharmacy mobility change from baseline (\%) & -3.03&10.80&3.27\\
WOPC & Percentage of workplace mobility change from baseline (\%) &-27.70&14.49&4.74\\
TSPC & Percentage of transit stations mobility change from baseline (\%) & -19.06&23.27&2.24\\
\end{longtable}

\section{Methodology}

\subsection{Mobility-augmented geographically and temporally weighted regression model}
In light of the heterogeneous disease dynamics in the U.S., conventional global regression techniques are no longer appropriate by assuming that all determinants are stationary over space and time. The geographically and temporally weighted regression model (GTWR)\citep{huang2010GTWR} is an effective method to account for the spatial and temporal non-stationarity issues and provide more interpretable estimations for the influencing factors during the COVID-19 pandemic. Considering that we have a series of $N$ observations $(Y_1,X_1),(Y_2,X_2),...,(Y_N,X_N)$ over time and space, each at location $(u_i,v_i)$ and at time $t_i$, the GTWR model can be formulated as follows:
\begin{equation}
Y_i = \beta_0 (u_i, v_i, t_i) + \sum_{k=1} ^ n \beta_{k} (u_i, v_i, t_i) X_{ik} + \epsilon_i 
\end{equation}

where $Y_i$ refers to the dependent variable of the $i$th observation, and $X_{ik}$ is the $k$th explanatory variable of observation $i$. $\epsilon_i$ is the error term that follows the normal distribution with zero mean and constant variance. $\beta_0 (u_i, v_i, t_i)$ is the intercept, and $\beta_{k} (u_i, v_i, t_i)$ is the regression parameter between $Y_i$ and $X_{ik}$.

Specifically, the regression coefficients $\hat{\beta}(u_i, v_i, t_i)$ are obtained as:

\begin{equation}\hat{\beta}(u_i, v_i, t_i) = [X^T W(u_i, v_i, t_i) X ]^{-1} X^T W(u_i, v_i, t_i) Y  \end{equation}

where $W(u_i, v_i, t_i)$ is the spatiotemporal weight matrix.

In the GTWR model, the weight between two observations is estimated solely based on the spatial distance and the time gap. Nevertheless, the correlation between two spatial locations shall not only be measured based on their geographic distance but mobility connections from the infectious disease perspective. Therefore, the M-GTWR is developed in our study. In addition to the great circle distance-based weight matrix, we also incorporate the mobility-based weight (the components include airline volume and commuting volume) into consideration to improve the baseline GTWR model~\citep{huang2010GTWR,wu2014IGTWR}. The new weight matrix is considered as the weighted combination of the distance weight matrix and the mobility volume weight matrix. More importantly, to reduce the fluctuation of the components in the mobility-augmented weight function $d^{M} _{ij}$ between node $i$ and $j$, the standardized form of each component is proposed in this study. The standard deviation form of the elements in the mobility-augmented weight matrix is:

\begin{equation}
  d^{M} _{ij} = \frac{d_{ij} ^ {S} }{\sigma\left( {d_{ij}^{S}}\right)} + \tau_{air}\exp\left[- \frac{N^{air}_{ij}}{\sigma\left( N^{air}_{ij} \right)} \right ] + \tau_{commuting}\exp\left[- \frac{N^{commuting}_{ij}}{\sigma\left( N^{commuting}_{ij} \right)} \right ]
\end{equation}

where $\tau_{air}$ and $\tau_{commuting}$ represent the airline volume and commuting volume, respectively; $N_a$ denotes the real airline volumes; $N_c$ denotes the real commuting volumes; $\sigma\left( {d_{ij}^{S}}\right)$, $\sigma\left( N^{air}_{ij} \right)$, and $\sigma\left( N^{commuting}_{ij} \right)$ are the prior bandwidth to standardize each component in the mobility-augmented weight matrix.

The procedure to generate the mobility-augmented weight matrix by combining the temporal distance matrix is shown below. Note that the geographically weighted regression (GWR) model with mobility-augmented weight matrix (M-GWR) can be achieved when $\lambda =1$.

\begin{equation}
   d_{ij} ^ {MST} =   \lambda d_{ij} ^{M} + (1- \lambda) d_{ij} ^T +2 \sqrt{\lambda (1- \lambda) d_{ij} ^{M} d_{ij} ^T } 
\end{equation}

Similarly, the Gaussian and Exponential kernels are employed for the M-GTWR model. The spatiotemporal weight function for each kernel can be constructed as follows:

\begin{equation}
  W_{ij}= \exp{ \left[ -\frac{1}{2} \left( \frac{d_{ij} ^{MST}}{h^{MST}} \right)^2 \right]}
\end{equation}

\begin{equation}
   W_{ij}= \exp{  \left( -\frac{d_{ij} ^{MST}}{h^{MST}} \right)}
\end{equation}

We apply the cross-validation (CV) procedure to select the optimal bandwidth $h^{MST}$ for M-GTWR~\citep{fotheringham2003geographically}, and choose the spatiotemporal effect parameter $\lambda$~\citep{huang2010GTWR}, and calibrate parameters ($\tau_{air},~\tau{commuting}$). Besides, we use the corrected Akaike Information Criterion (AICc) as the performance measurement to calibrate the trade-off between goodness of fit and degrees of freedom~\citep{fotheringham2003geographically}. Finally, we verified the effectiveness of the M-GTWR model based on the analysis of variance (ANOVA) method.

\subsubsection{Airline traffic assignment}
Since the airline passenger volume has potential seasonal fluctuation, we used the DB1B market records of the first quarter of 2019 to infer the airline travel connections in the first quarter of 2020 (when the COVID-19 is started) in this study. The dataset contains 420 original airports and 419 destination airports, which server 396 original cities and 394 destination cities in The U.S. To obtain the airline flow among the selected counties, we first assign the number of passengers from the origin airport to the nearby counties that are within the radiation range using the distance-based gravity model~\citep{balcan2009multiscale} as follows:
\begin{equation}
    w_{ij}=C\times{\frac{(P_i)^\alpha\times(P_j)^\gamma}{f(d_{ij)}}}
\end{equation}

In the equation, $C$ is a proportionality constant, and $\alpha$ and $\gamma$ tune the dependence related to the population size of each county.  The distributed weight $w_{ij}$ is positively related to the product of the population of served county $P_i$ and the population of the airport located county $P_j$, and negatively related to the distance $d_{ij}$ between the two counties. And $f(d_{ij})$ is a distance-dependent functional form, which assumes to be an exponential law for the dis-attraction between two counties and is defined as: 
\begin{equation}
    f(d_{ij})=e^{(\beta d_{ij})}
\end{equation}

According to~\cite{balcan2009multiscale}, the empirical PDF of the connected airline volume reaches the summit when the distance between the two areas is around 250km and decays exponentially afterward. The parameters $\alpha$, $\gamma$, and $\beta$ used in the study are, therefore, obtained from their statistical analysis: $\alpha$ is 0.46, $\gamma$ is 0.64, and $\beta$ is 0.0122 when the distance is less than 
or equal to $300 km$, and $\alpha$ is 0.35 and $\gamma$ is 0.75 when the distance is great than $300 km$.

The assigned air travel demand $PD_i$ of county $i$ is $PD_i=TD_j w_{ij}$, where $TD_j$ is the total original passengers volume of airport $j$. After distributing the passengers demand of served county in The U.S., we then conduct a similar approach to get the assignment weight $w^{'}_{i^{'}j}$ between the destination airport $i^{'}$ and the county $j$. Finally, the airline passengers' demand from county $i$ to county $j$ is calculated as 
\begin{equation}
    PD_{ij}=PD_i\times w^{'}_{i^{'}j} 
\end{equation}

\section{Model Calibration}

\subsection{Spatial autocorrelation and heterogeneity test}
To understand the effects of influencing factors on the COVID-19 propagation based on the GTWR model, we need to first assess whether there are significant spatial nonstationarity and autocorrelation of the dependent variables over the study period. In this study, we apply the Breusch-Pagan(BP) test to examine the spatial heterogeneity of the dependent variable. The BP test is a classic approach used to detect spatial heterogeneity~\citep{nurhayati2019robust}. The null hypothesis in the BP test is that the error variables are equal in all areas: $\sigma^2_1=\sigma^2_2=\sigma^2_i=\dots=\sigma^2$. The alternative hypothesis is that there should be at least one location $i$, such that $\sigma^2_i \neq \sigma^2$. Besides, we use the adjusted Moran's I test to examine the spatial autocorrelation of the dependent variable that varies from time, which is proposed by~\cite{gao2019measuring}. The results for the significance of the spatial heterogeneity and autocorrelation are shown in table~\ref{tab:test heterogenity} and table~\ref{tab:test auto-correlation}.

\begin{table}[!h]
\centering
\caption{Breusch-Pagan test for the spatial heterogeneity }
\label{tab:test heterogenity}
\begin{tabular}{lcccc}
\toprule
\multirow{2}{*}{Weeks}   &
\multicolumn{2}{c}{WADC} & \multicolumn{2}{c}{WADD} \\
\cmidrule(l){2-3} \cmidrule(l){4-5}
& BP value &  p-value & BP value & p-value   \\
\hline

All week&332.77&$<$2.2e-16 ***&85.65 & 8.648e-11 ***\\
\hline

\bottomrule

\multicolumn{5}{C{0.6\linewidth}}{\footnotesize Note: .$\sim p < 0.1$; * $\sim p<0.05
$; **$\sim p<0.01$; ***$\sim p<0.001$}
\end{tabular}

\end{table}

\begin{table}[!h]
\centering
\caption{Adjusted Moran’s I test for spatial auto-correlation}
\label{tab:test auto-correlation}
\begin{tabular}{lcccc}
\toprule
Dependent variables & Adjusted Moran’s I &  Z scores & p-value  \\
\hline
WADC& 0.573&445.11 & ***\\
WADD & 0.546&424.51 & ***\\
\hline

\bottomrule
\end{tabular}
\end{table}

From the test results, the BP test rejects the null hypothesis of homoscedasticity for WADC and WADD in the 38 weeks period (significant at 0.001 level). In addition, the results of the adjusted Moran's I test demonstrates the spatial misspecification of dependent variables. Moreover, since the Z-score values are positive in the number of WADC and WADC, it implies that the spatial distribution of counties with WADC and WADD is more likely to be spatially clustered.

\subsection{Mobility-augmented weight matrix calibration}

\begin{figure}[!h]
\centering
\includegraphics[width=0.8\linewidth]{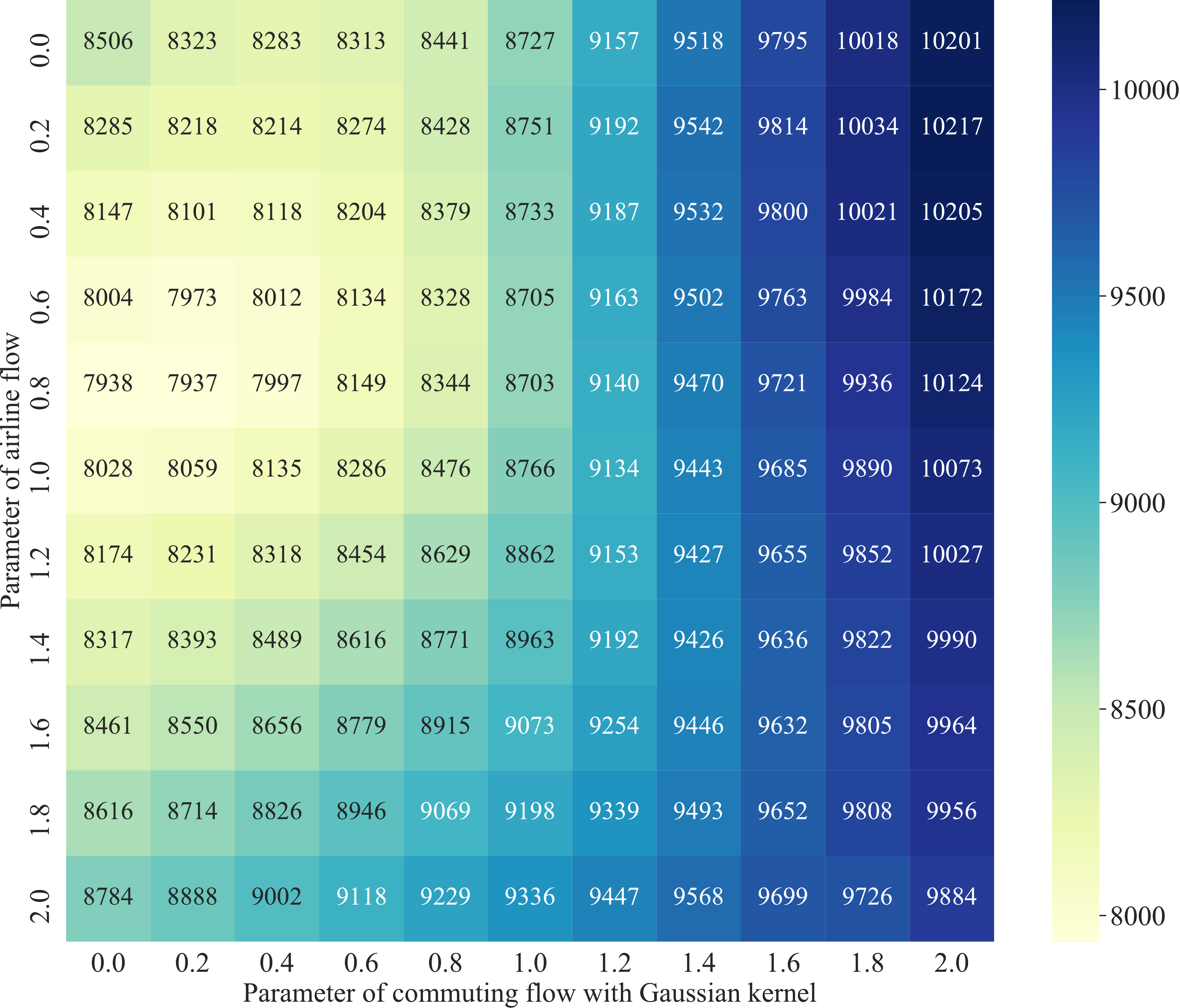}
\caption{The parameter calibration in the WADC model}
\label{fig:case_para}
\end{figure}

The mobility-augmented weight matrix in the M-GTWR model is incorporated by the standardized form of the great circle distance with the travel flow connection (airline flow and commuting flow). To calibrate the parameters($\tau_{air},~\tau_{commuting}$), we conducted the cross-validation (CV) to verify the performances of different parameter combinations. The CV procedure is described as follows:
\begin{enumerate}
    \item We first apply the CV method to calibrate the parameters for the best combination of the weight matrix based on the AICc value. The results are shown in Figure~\ref{fig:case_para}.
    \item We next compare the Gaussian kernel with the Exponential kernel functions for the model. 
    \item We then find the parameter $\lambda$ with the CV based on the optimal kernel selected from the second step.
\end{enumerate}

\begin{figure}[!h]
\centering
\includegraphics[width=0.7\linewidth]{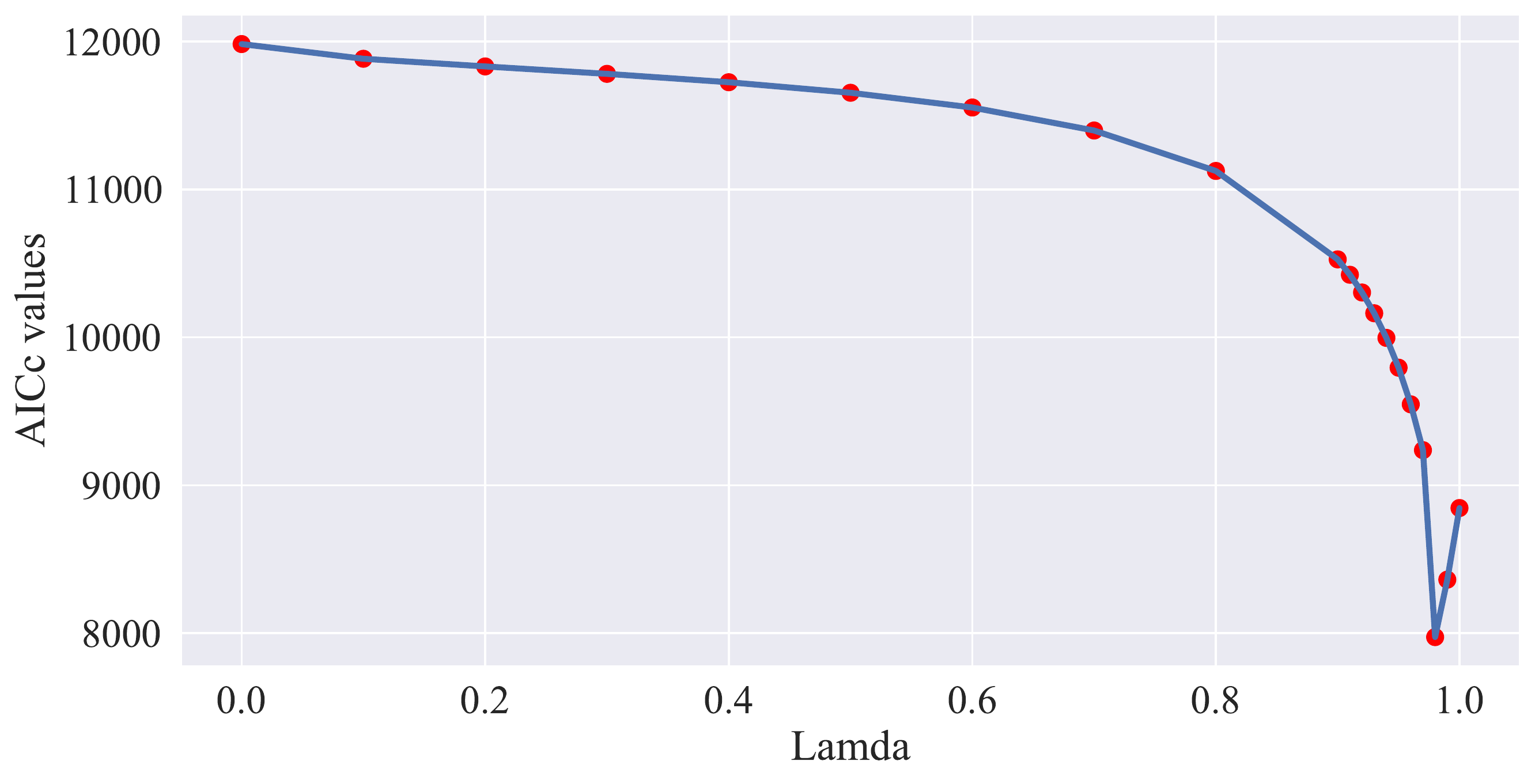}
\caption{The performance of Lambda in the WADC model}
\label{fig:case_death_lamda}
\end{figure}

Based on the results, we find that the optimal parameters setting for the combination of the standardized travel connection is 0.8 for the airline flow and 0.2 for the commuting flow in the WADC model. The corresponding AICc value is 7937 for the Gaussian kernel and 7963 for the Exponential kernel. As for the WADD model, the parameters are 0.8 for airline flow and 0.4 for commuting flow, where the corresponding AICc value is 7395 for the Gaussian kernel and 10634 for the Exponential kernel. The performances suggest that the Gaussian kernel outperforms the Exponential kernel in all combinations. Therefore, the Gaussian kernel is selected as the final weight matrix kernel function. The final weight matrix for the WADC and WADD models are:

\begin{equation}
  d^{M} _{ij} = \frac{d_{ij} ^ {S} }{\sigma\left( {d_{ij}^{S}}\right)} + 0.8\times\exp\left[- \frac{N^{air}_{ij}}{\sigma\left( N^{air}_{ij} \right)} \right ] + 0.2\times\exp\left[- \frac{N^{commuting}_{ij}}{\sigma\left( N^{commuting}_{ij} \right)} \right ]
\end{equation}

\begin{equation}
  d^{M} _{ij} = \frac{d_{ij} ^ {S} }{\sigma\left( {d_{ij}^{S}}\right)} + 0.8\times\exp\left[- \frac{N^{air}_{ij}}{\sigma\left( N^{air}_{ij} \right)} \right ] + 0.4\times\exp\left[- \frac{N^{commuting}_{ij}}{\sigma\left( N^{commuting}_{ij} \right)} \right ]
\end{equation}

Based on the Gaussian kernel function selected from the second step, we then conducted the CV approach to calibrate the spatiotemporal distance parameter $\lambda$. The result of the WADC model can be seen in the Figure~\ref{fig:case_death_lamda}. It shows that the optimal $\lambda$ is equal to 0.98 based on the comparison of the AICc values, which means the spatial effect is the dominant effect in the spatiotemporal relationship.

\subsection{Model comparisons}
After finalizing the modeling parameters, we then evaluate whether the proposed M-GTWR model is superior than other benchmarks in characterizing the spatial and temporal variations of the dependent variables and offering better explanatory power for the COVID-19 case. The selected benchmarks include the base GTWR model, the M-GWR and GWR model that only consider spatial heterogeneity, and the OLS model that assumes stationarity. We use the ANOVA for comparing the improvements in the residual reduction among the candidate models. This approach is also adopted by~\cite{huang2010GTWR} for model comparisons in studying the spatiotemporal variations of real estate prices. The results on the ANOVA test are summarized in Table~\ref{tab:anovacase}.

\begin{table}[!h]
\centering
\caption{ANOVA comparison between GWR and OLS models (WADC and WADD models)}
\label{tab:anovacase}
\resizebox{\textwidth}{!}{
\begin{tabular}{lccccc}
\toprule
Source of variance & RSS &DF&MS&F-test&P-value\\
\hline

OLS residuals                   & 32051 (28462)     & 15(15)        & 2136.7(1897.5)    &  &\\
GWR-basic residual              & 18988(12848)      & 15856(15721)  & 1.2(0.8)      & 1784(2322)&  ***(***)\\
M-GTWR residual                 & 18484(12647)      & 15949(15946)  & 1.2(0.8)      & 1844 (2392)& ***(***)\\
GTWR-basic residual residual    & 6168(8889)        & 1456(15749)   & 4.2(0.6)      & 504(3362)&  ***(***)\\
M-GTWR residual residual        & 2869(4789)        & 12025(12202)  & 0.2(0.4)      & 8956(4834)&  ***(***)\\
GWR-basic/OLS improvement       & 13063(15614)      & 1423(1557)    & 9.2(10.0)      & & \\
M-GTWR/OLS improvement          & 13567(15815)      & 1329(1332)    & 10.2(11.9)     &&\\
GTWR-basic/OLS improvement      & 25883(19573)      & 1306(1529)    & 19.8(12.8)     &&\\
M-GTWR/OLS improvement          & 29182(23673)      & 5253(5076)    & 5.6(4.7)      &&\\
M-GTWR/GWR-basic improvement    & 16119(8059)       & 3830(3519)    & 4.2(2.3)      &&\\
M-GTWR/M-GTWR improvement       & 15615(7858)       & 3924(3744)    & 4.0(2.1)     &&\\
M-GTWR/GTWR-basic improvement   & 3299(4100)        & 3947(3547)    & 0.8(1.2)      &&\\

\hline
\bottomrule
\multicolumn{6}{C{\linewidth}}{\footnotesize Note: RSS: the residual sum of squares; DF: the degree of freedom; MS: residual mean of squares. The ANOVA test results for the WADD model are in the parenthesis.}
\end{tabular}}

\end{table}





The statistics demonstrate the significance of the spatial and temporal nonstationarity in the study area over the period. And it is preferable to adopt the GWR-based model instead of the OLS model. Besides, the comparison between the GWR model and the GTWR model asserts the importance for considering the temporal nonstationarity of the data. Finally, we also observe that the including of the mobility-augmented weighting scheme achieves notable improvements in the modeling residual for both the GWR model and the GTWR model, and that the improvements is statistically significant following the ANOVA test. The results are consistent for the WADC and WADD models, and therefore support the superiority of the M-GTWR model over all other alternatives in representing the COVID-19 dynamics in the U.S. 

\section{Results}

The estimated results of the M-GTWR model are obtained in Table~\ref{tab:gtwrcase}. The M-GTWR model addresses the nonstationarity issue for the fundamentally heterogeneous and dynamic disease propagation and provides more efficient estimates by assuming the effects of the influencing factors are spatiotemporal heterogeneous. Compared with the global OLS model, the M-GTWR model improves AICc and adjusted $R^2$ from 59779.59 and 0.54 to 26309.13 and 0.94 for the WADC. As for the WADD, the corresponding AICc and adjusted $R^2$ are 34801.25 and 0.90 respectively.

\begin{table}[H]
\centering
\caption{Estimates of the M-GTWR model (WADC and WADD models)}
\label{tab:gtwrcase}
\begin{tabular}{lccccc}
\toprule
Variables & Min & Max &Median & Lower quartile &Upper quartile\\
\hline
Intercept& -73.19(-98.42) & 16.60(15.64) & -13.81(-24.62) & -28.58(-38.38) & -4.94(-13.55) \\
\textbf{\emph{Demographic}} &&&&&\\
Log.POPD            & 0.18(0.06) & 1.20(1.38) & 0.63(0.71) & 0.57(0.62) & 0.69(0.79) \\
OLDP                & -13.13(-11.69) & 8.96(15.35) & -1.30(4.51) & -4.05(2.27) & 0.85(6.51) \\
BLAP                & -0.05(-0.08) & 0.05(0.06) & 0.02(0.01) & 0.01(0.00) & 0.02(0.02) \\
ASIP                & -0.15(-0.19) & 0.12(0.09) & -0.03(-0.04) & -0.06(-0.07) & -0.01(-0.02) \\
LATP                & 0.02(-0.03) & 0.05(0.05) & 0.01(0.01) & 0.01(0.00) & 0.02(0.02) \\
\textbf{\emph{Socioeconomic}} &&&&&\\
CLUE                & -0.40(-0.27) & 0.35(0.45) & -0.05(0.01) & -0.09(-0.05) & 0.01(0.08) \\
BDHP                & -0.17(-0.19) & 0.07(0.11) & -0.01(-0.01) & -0.02(-0.03) & 0.00(0.02) \\
Log.MHIC            & -1.90(-3.16) & 6.71(9.97) & 0.83(1.17) & 0.13(0.02) & 1.86(2.26) \\
\textbf{\emph{Travel}} &&&&&\\
Log.MILE           & 0.44(-0.05) & 1.82(2.01) & 1.03(1.04) & 0.91(0.90) & 1.13(1.18) \\
WFHP               & -0.38(-0.64) & 0.27(0.15) & -0.07(-0.18) & -0.13(-0.23) & -0.03(-0.14) \\
PTAP               & -0.08(-0.08) & 0.17(0.19) & 0.01(0.02) & -0.01(0.01) & 0.02(0.04) \\
Log.MTAT           & -3.20(-5.36) & 3.69(4.16) & 0.22(0.95) & -0.44(0.16) & 0.88(1.71) \\
\textbf{\emph{Human Activity}} &&&&&\\
GPPC                & -0.06(-0.07) & 0.05(0.07) & -0.01(-0.01) & -0.02(-0.02) & 0.01(0.01) \\
WOPC                & -0.11(0.12) & 0.08(0.09) & -0.01(-0.02) & -0.03(-0.05) & 0.01(0.01) \\
TSPC                & -0.04(-0.03) & 0.02(0.03) & -0.01(-0.01) & -0.01(-0.01) & 0.01(0.01) \\
AIC&&&21356.98(30096.09)&&\\
AICc&&&26309.13(34801.25)&&\\
$R^2$&&&0.96(0.92) &&\\
Adjusted $R^2$&&&0.94(0.90)&&\\

\hline
\bottomrule
\multicolumn{6}{C{\linewidth}}{\footnotesize Note: The results for the WADD model are in the parenthesis.}
\end{tabular}
\end{table}

\begin{figure}[H]
\centering
\subfloat[][Coefficient,$12th$ week]{
\includegraphics[width=0.5\linewidth,trim={1.5cm 5.3cm 0cm 4cm},clip]{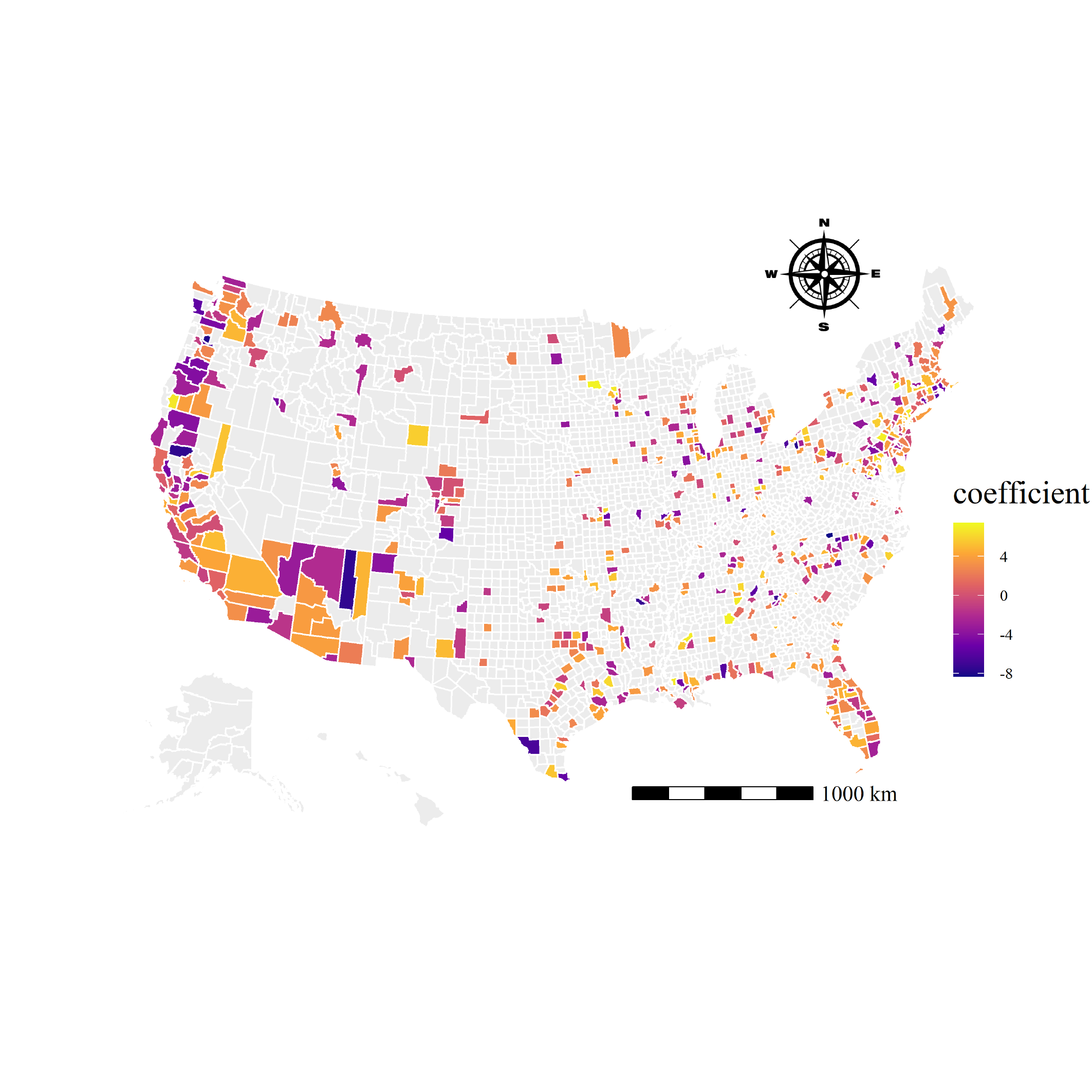}
\label{fig:caseolder12}}
\subfloat[][t statistic, $12th$ week]{
\includegraphics[width=0.5\linewidth,trim={1.5cm 5.5cm 0cm 4cm},clip]{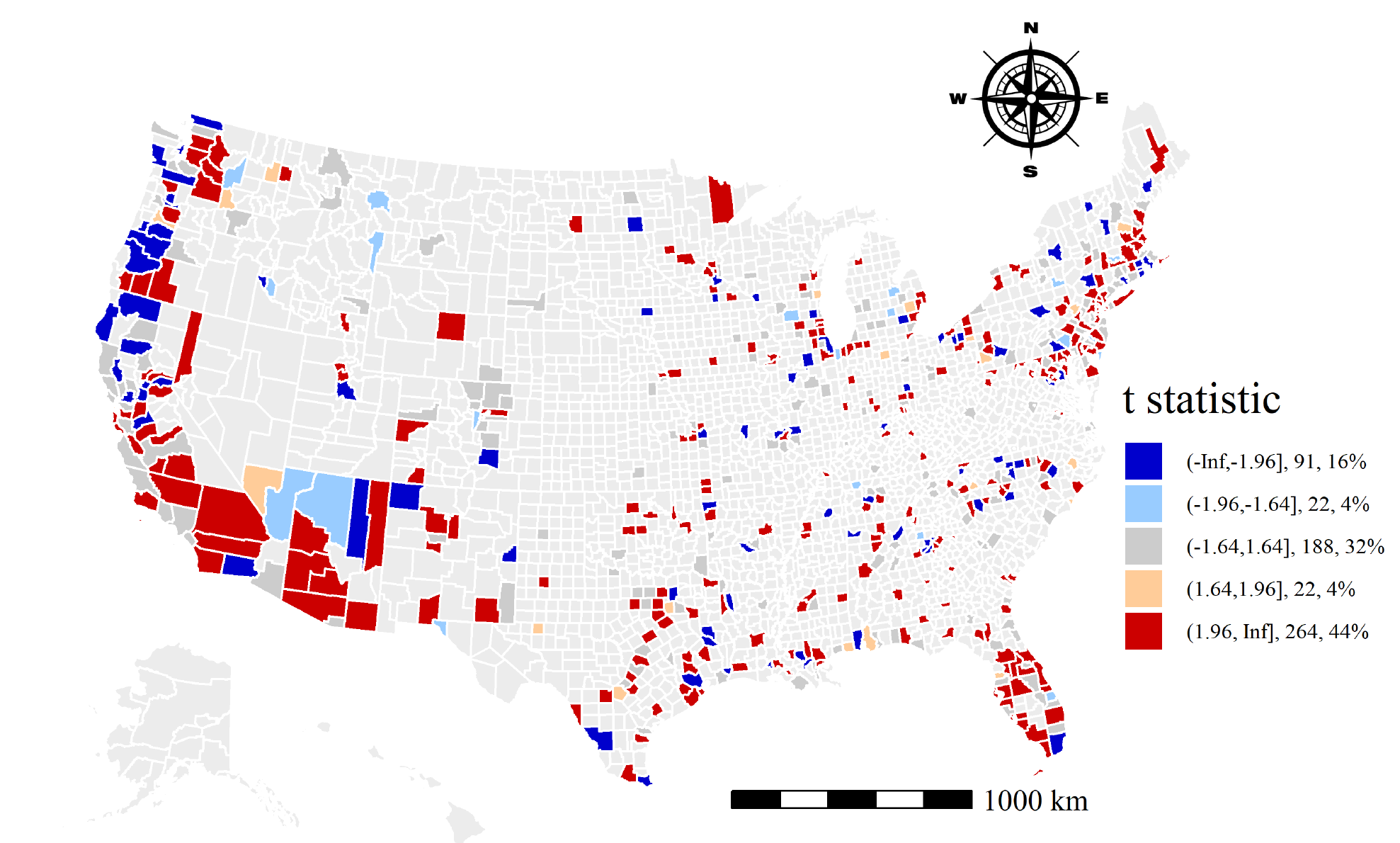}
\label{fig:caseolder12_tvalue}}
\qquad
\subfloat[][Coefficient, $30th$ week]{
\includegraphics[width=0.5\linewidth,trim={1.5cm 5.3cm 0cm 4cm},clip]{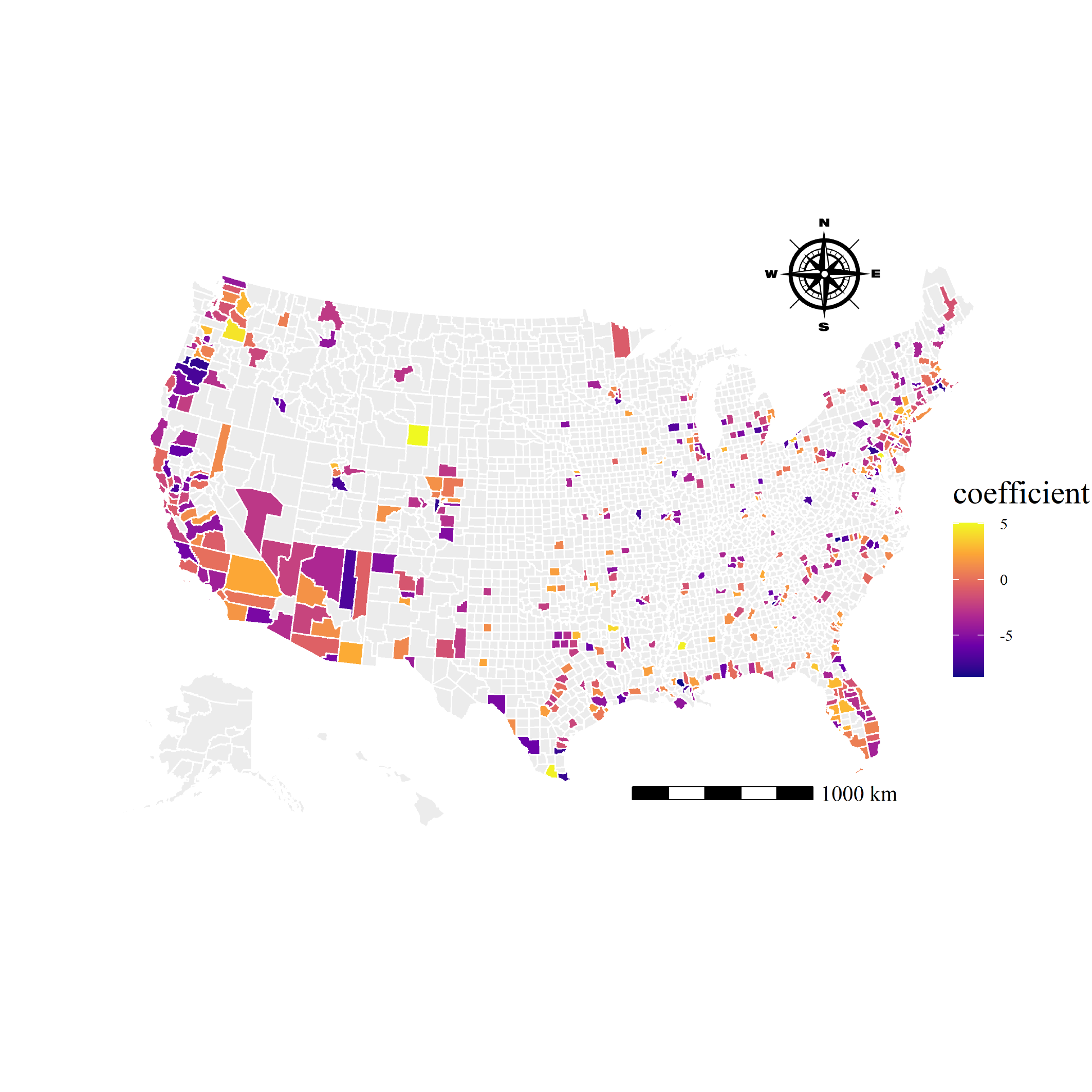}
\label{fig:caseolder30}}
\subfloat[][t statistic, $30th$ week]{
\includegraphics[width=0.5\linewidth,trim={1.5cm 5.5cm 0cm 4cm},clip]{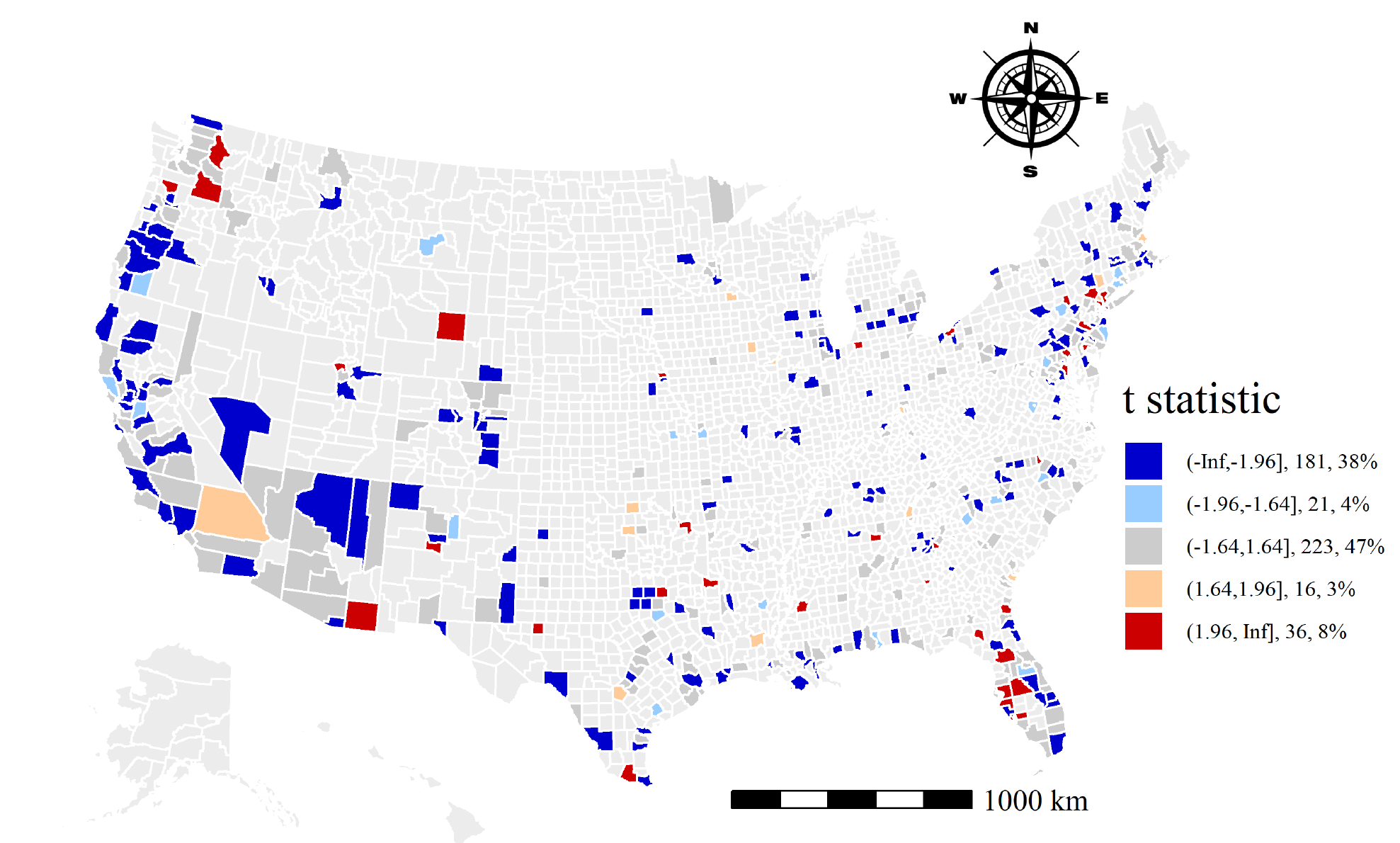}
\label{fig:caseolder30_tvalue}}
\qquad

\subfloat[][Coefficient, $48th$ week]{
\includegraphics[width=0.5\linewidth,trim={1.5cm 5.3cm 0cm 4cm},clip]{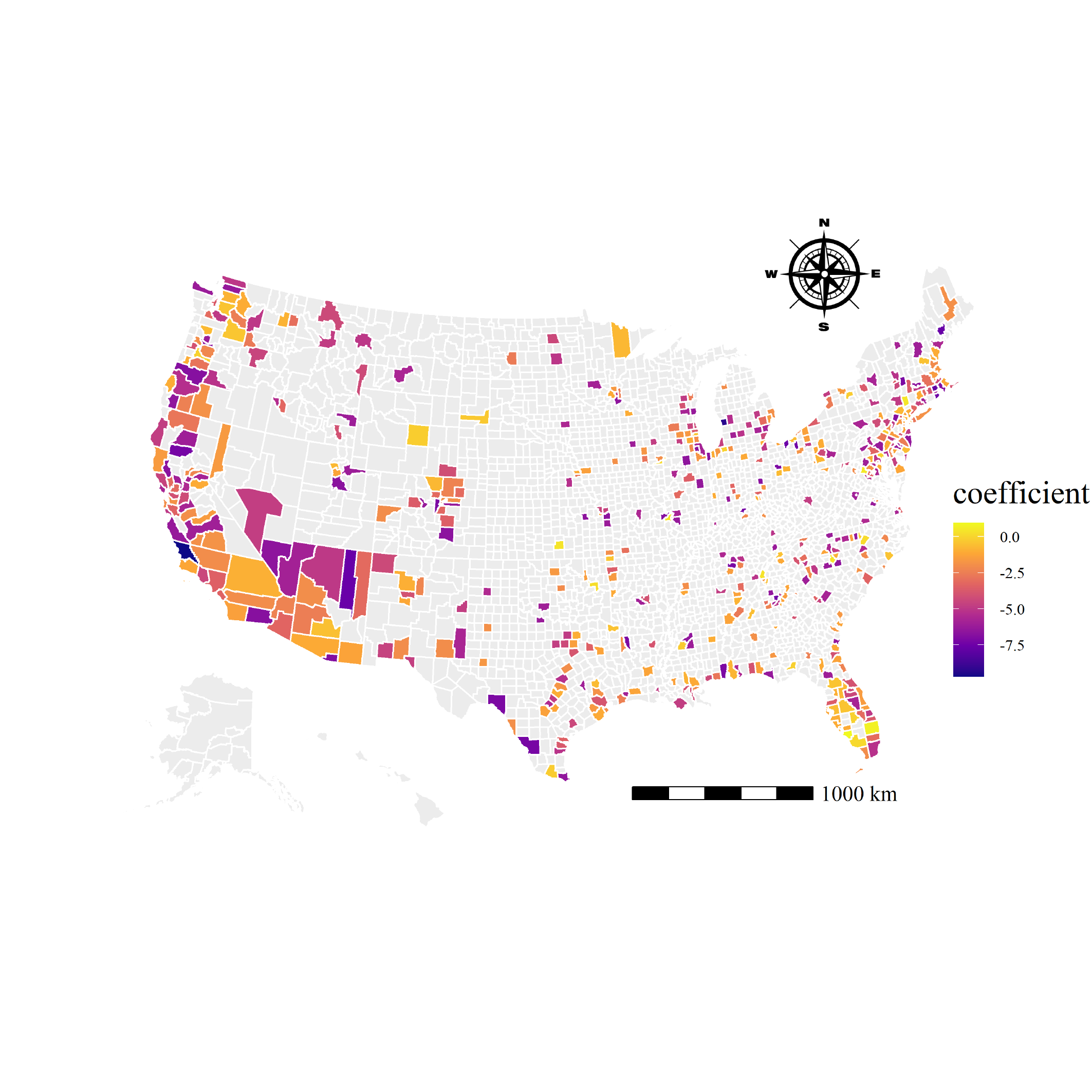}
\label{fig:caseolder48}}
\subfloat[][t statistic, $48th$ week]{
\includegraphics[width=0.5\linewidth,trim={1.5cm 5.5cm 0cm 4cm},clip]{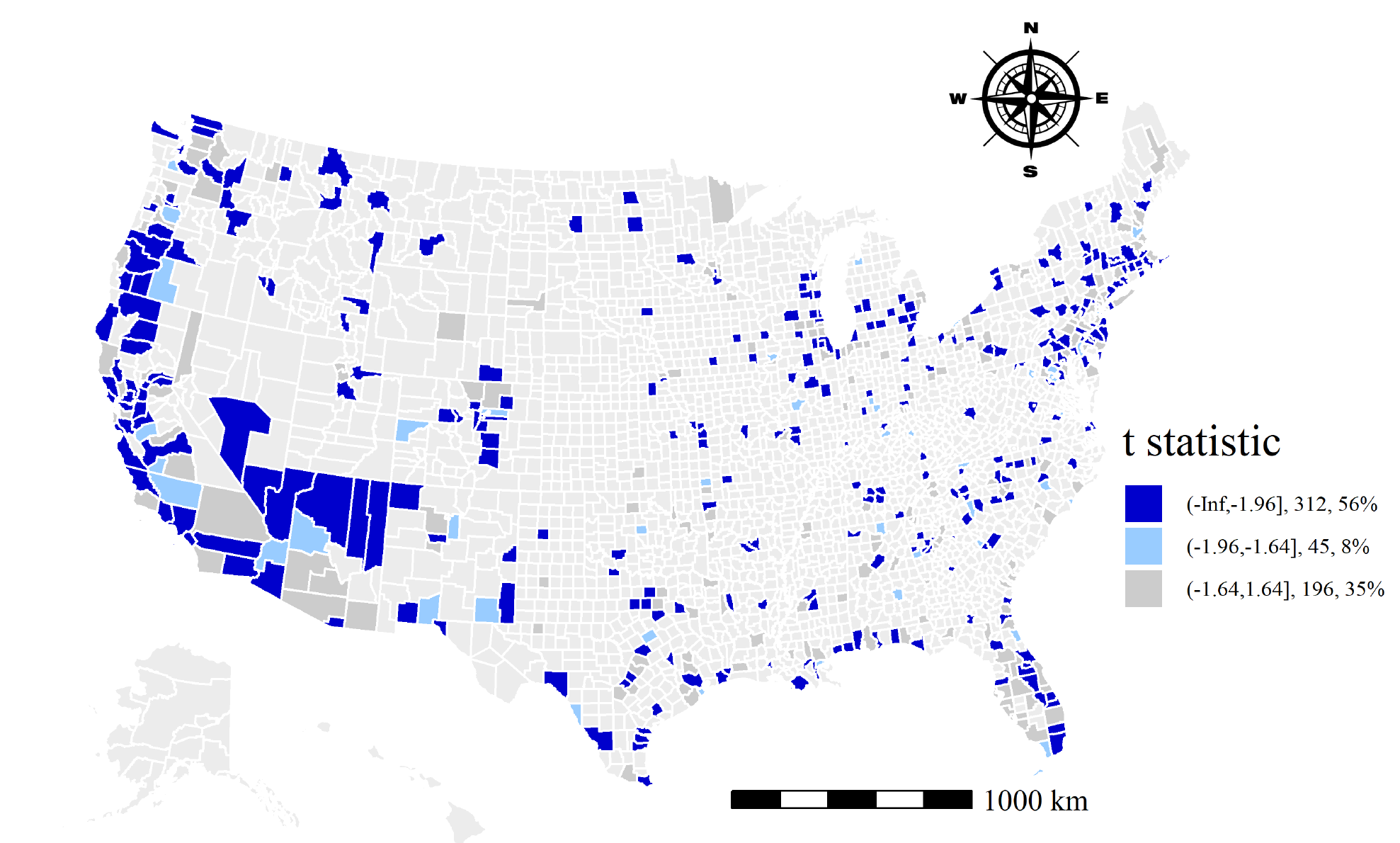}
\label{fig:caseolder48_tvalue}}
\qquad

\caption{The distribution of the effects of the OLDP in the WADC model}
\label{fig:olderfig}

\caption*{Note: the range of Z-scores, the number of counties in the range, and the percentage of counties in the range are calculated in the t statistic legend (all the figures below follow the same approach). }

\end{figure}  

 Instead of the static effects of the influencing factors estimated in the global OLS model, our findings show the effects of the factors vary among regions and time in both models. \highlighttext{For estimates of the demographic variables, the coefficient of POPD is positive and the estimates are statistically significant in the WADC and the WADD in the studied counties from the $12$th week and $48$th week. In particular, the median elasticity of the POPD shows that a 1\% increase in the POPD leads to 0.63\% increase in the the WADC and 0.71\% increase in the WADD on average. In Figure~\ref{fig:olderfig}, we visualize the coefficient and t-stats distributions for OLDP from three selected weeks ($12$th, $30$th, $48$th) based on the WADC model. The coefficients are observed to be positive and the estimates are statistically significant in around 50\% of the studied counties at the $12$th week. However, the corresponding coefficients among 34\% of these counties gradually shift to negative while the estimates remaining statistically significant at the $48$th week. One reason is that the older population are more vulnerable to the disease. This, along with the worse pandemic situation, indicates that the older population are increasingly cautious and adopting better preventative measures, which reduce their chances of being infected as the outbreak proceeded. In addition, among the studied counties that have insignificant effects at the $12$th week, 59\% of these counties has shifted to have negative coefficients and remaining statistically significant at the $48$th week, and the coefficients of the high population density counties (mean population density is 357 population$/km^{2}$) are more likely to shift to negative in the later stage than the relative low population density counties (mean population density is 288 population$/km^{2}$). That might because the high population density counties provide better guide for the older population in preventing the COVID-19. Finally, the coefficients are observed to be negative and statistically significant in around 64\% of the studied counties at the $48$th week.}

\begin{figure}[H]
\centering
\subfloat[][Asian, $12$th week, WADC model]{
\includegraphics[width=0.5\linewidth,trim={1.5cm 5.5cm 0cm 4.3cm},clip]{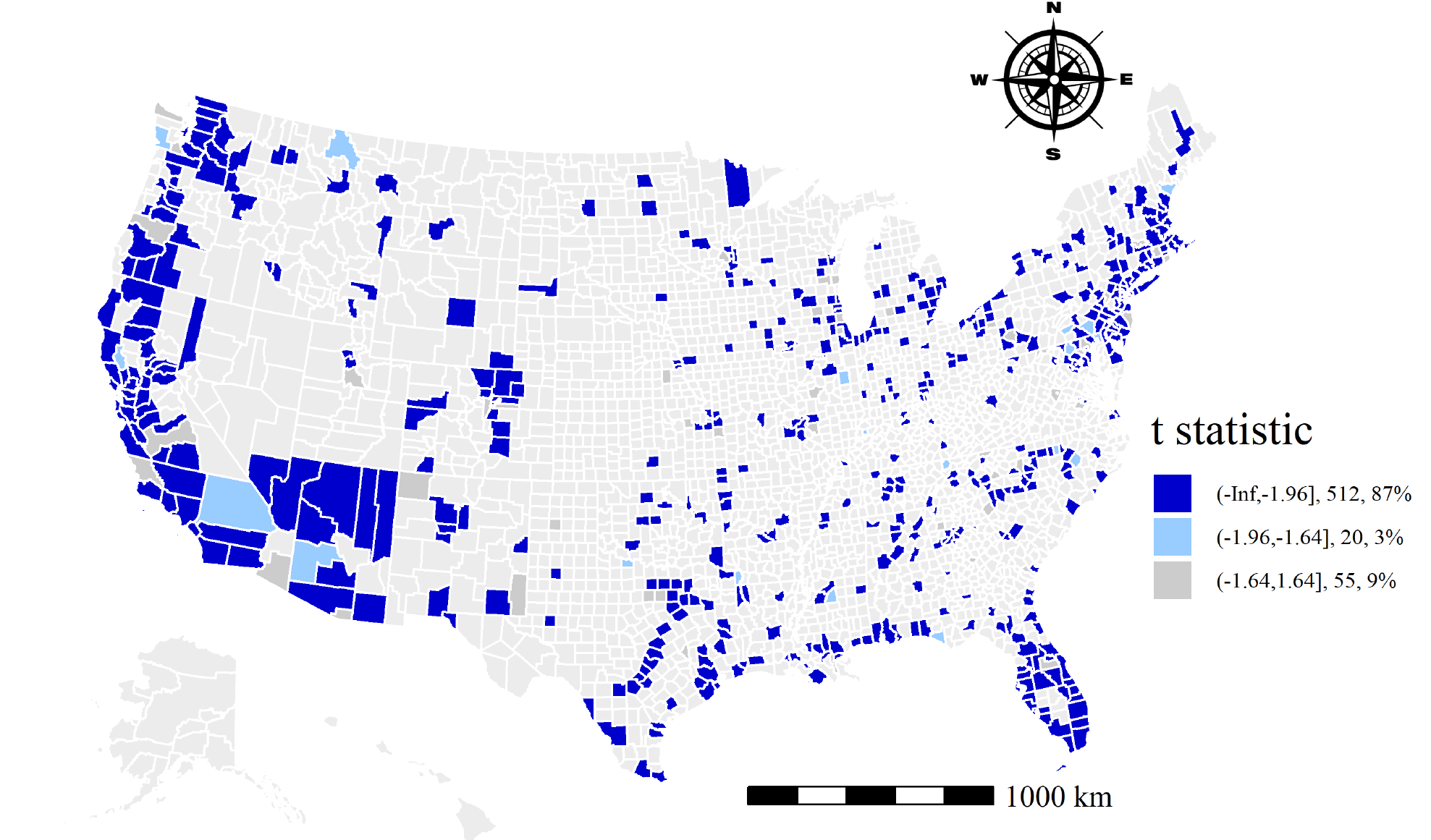}
\label{fig:caseasian12_tvalue}}
\subfloat[][Asian, $48$th week, WADC model]{
\includegraphics[width=0.5\linewidth,trim={1.5cm 5.5cm 0cm 4.3cm},clip]{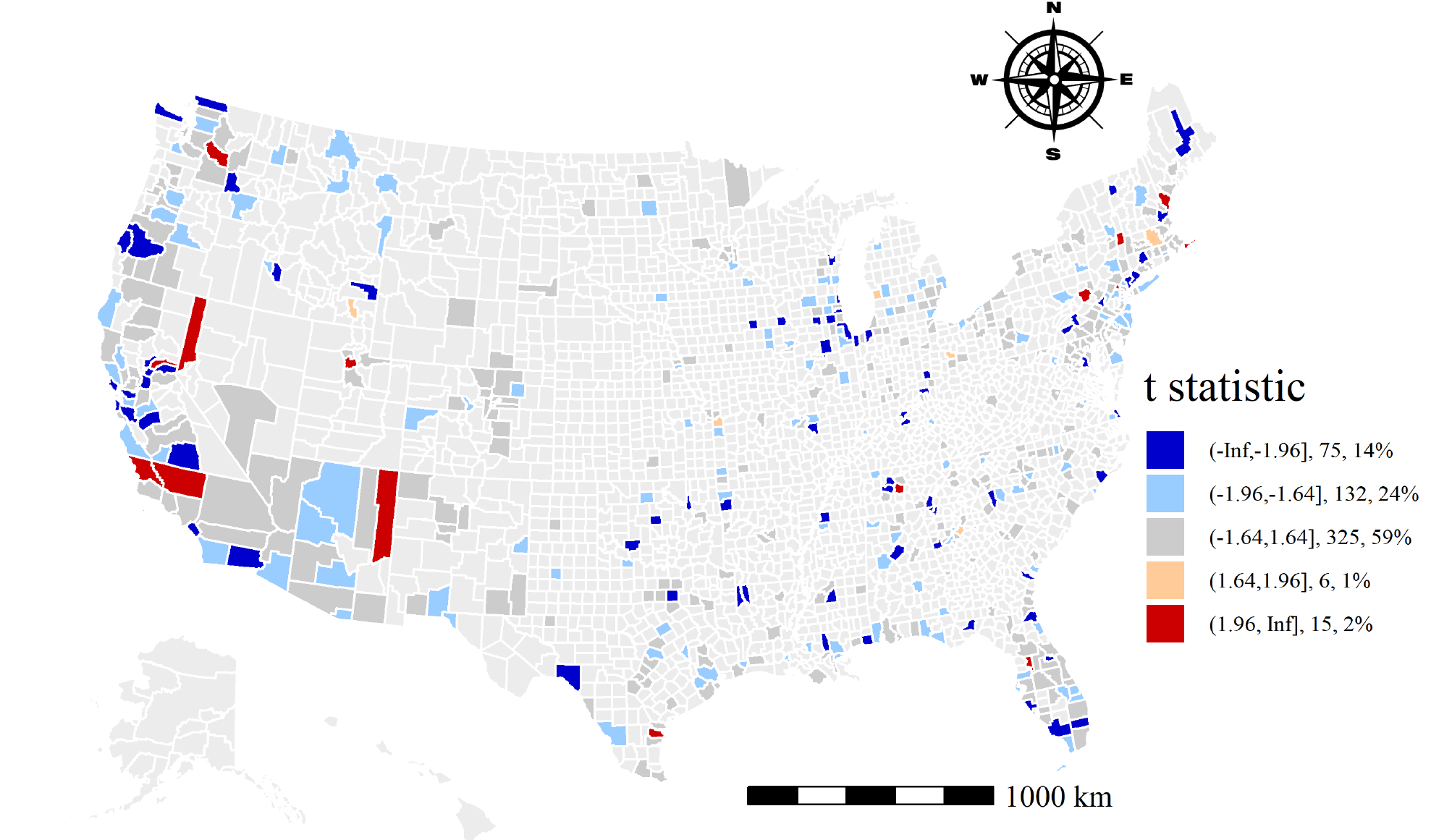}
\label{fig:caseasian48_tvalue}}
\qquad
\subfloat[][black, $12$th week, WADC model]{
\includegraphics[width=0.5\linewidth,trim={1.5cm 5.5cm 0cm 4.3cm},clip]{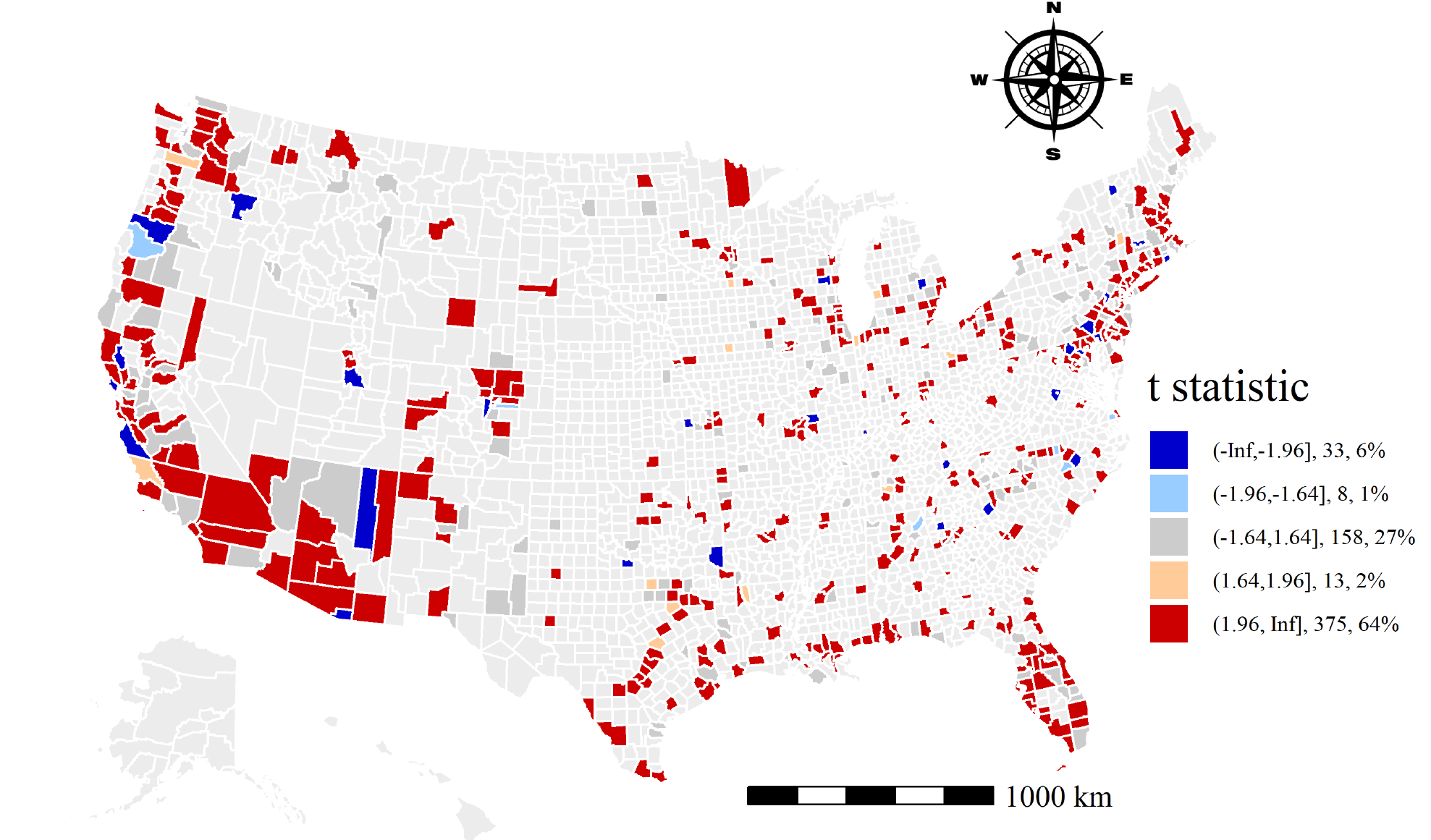}
\label{fig:caseblack12_tvalue}}
\subfloat[][black, $48$th week, WADC model]{
\includegraphics[width=0.5\linewidth,trim={1.5cm 5.5cm 0cm 4.3cm},clip]{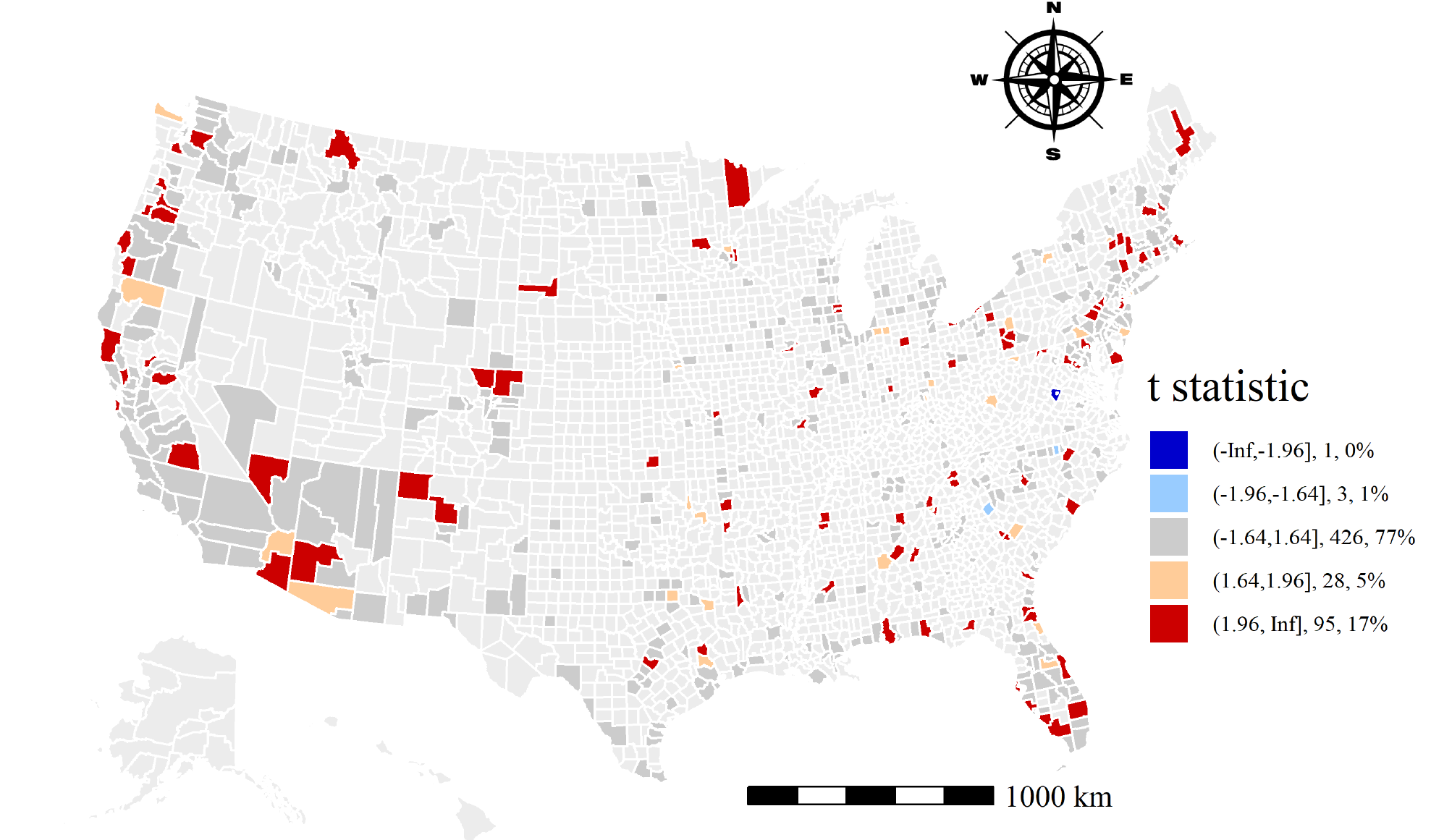}
\label{fig:caseblack48_tvalue}}
\qquad

\subfloat[][Asian, $12$th week, WADD model]{
\includegraphics[width=0.5\linewidth,trim={1.5cm 5.5cm 0cm 4.3cm},clip]{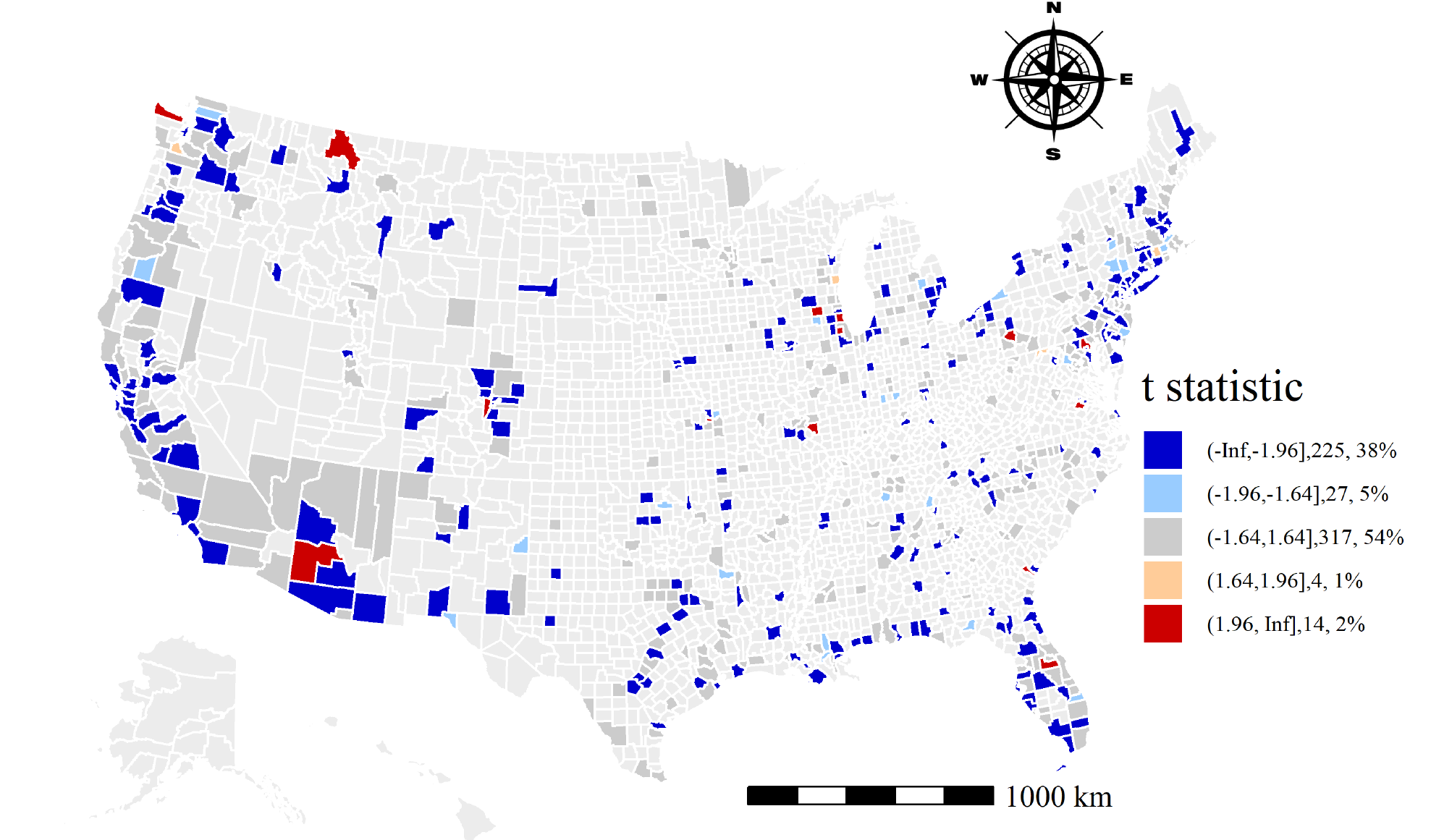}
\label{fig:deathAsian12_tvalue}}
\subfloat[][Asian, $48$th week, WADD model]{
\includegraphics[width=0.5\linewidth,trim={1.5cm 5.5cm 0cm 4.3cm},clip]{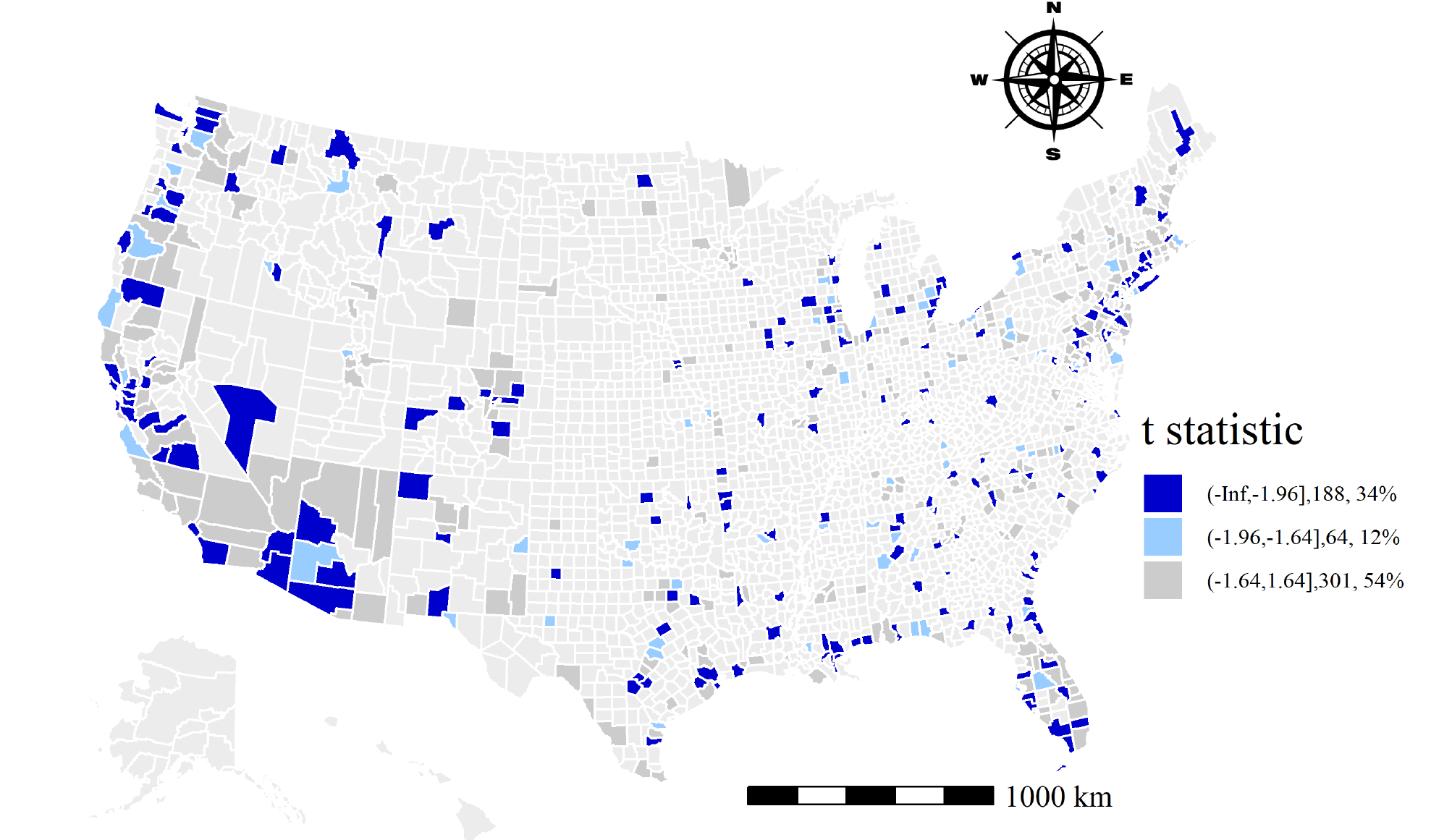}
\label{fig:deathAsian48_tvalue}}
\qquad
\subfloat[][black, $12$th week, WADD model]{
\includegraphics[width=0.5\linewidth,trim={1.5cm 5.5cm 0cm 4.3cm},clip]{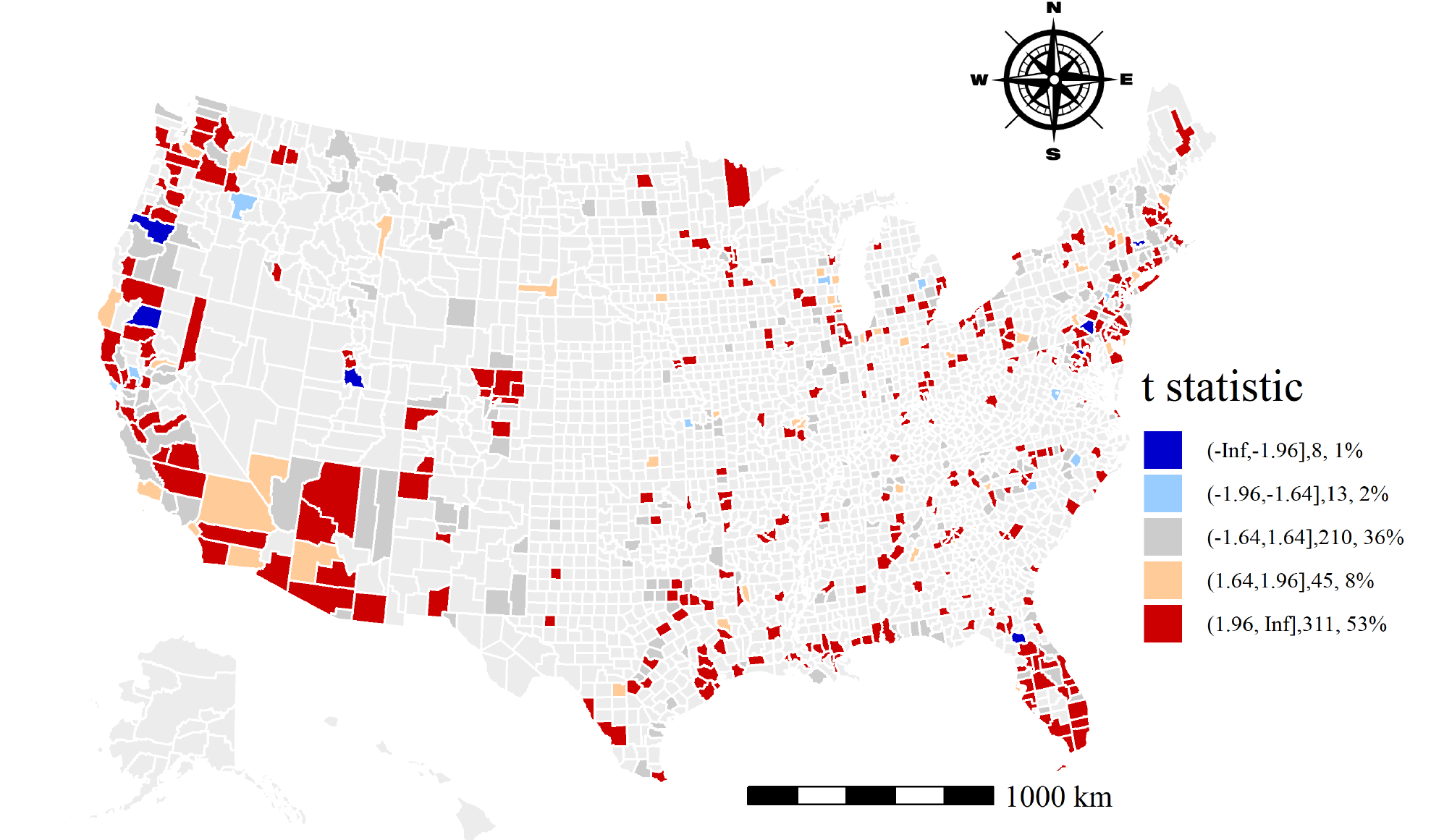}
\label{fig:deathblack12_tvalue}}
\subfloat[][black, $48$th week, WADD model]{
\includegraphics[width=0.5\linewidth,trim={1.5cm 5.5cm 0cm 4.3cm},clip]{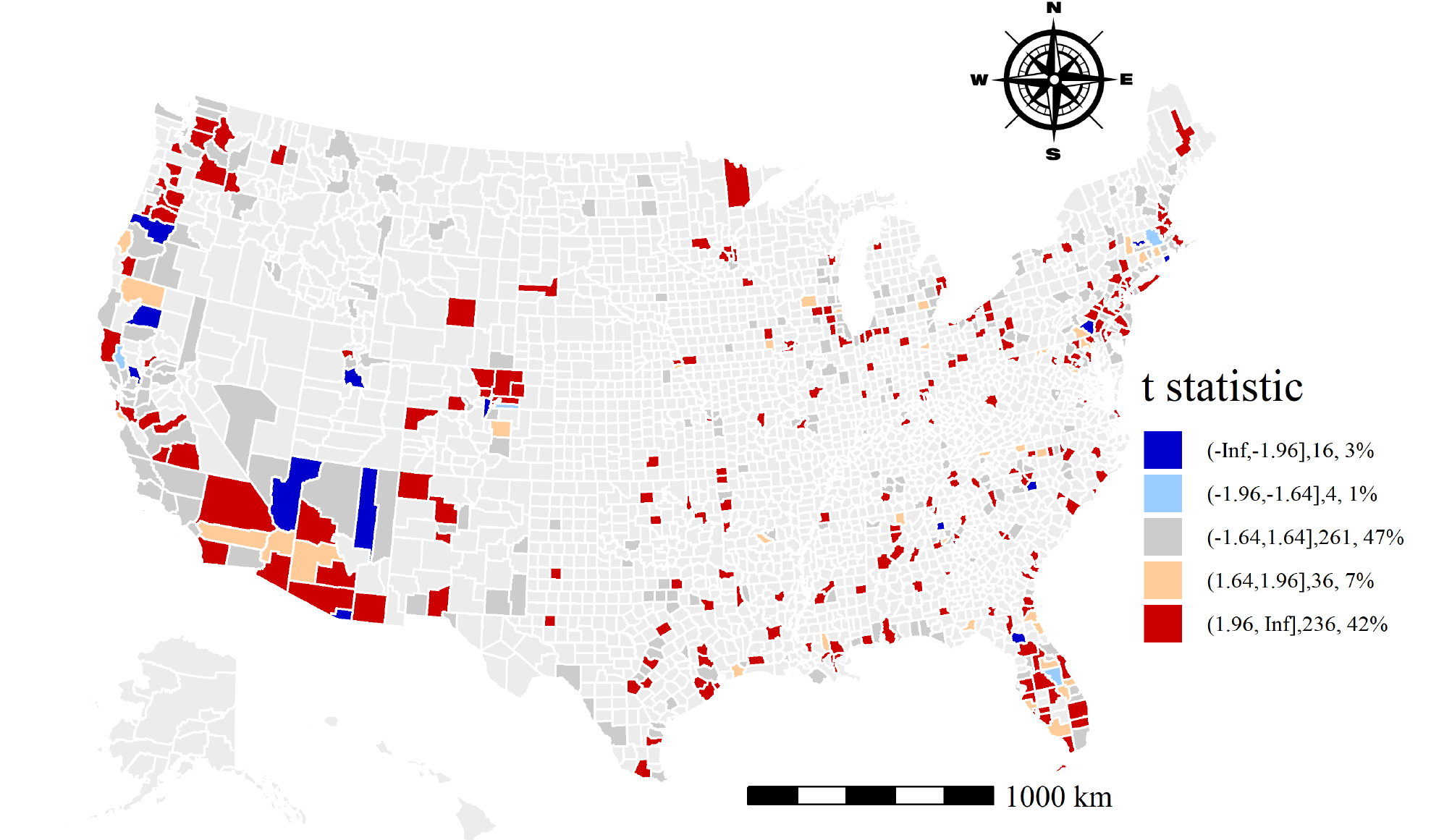}
\label{fig:deathblack48_tvalue}}
\qquad

\caption{The t statistic distribution of race type in two models}
\label{fig:globfig}
\end{figure}

 From the race perspective, the coefficient of the LATP is positive and the estimates are statistically significant in about 18\% of the counties in the $12$th week on the WADC and WADD models. Then, the coefficient of the LATP are positive and the estimates are statistically significant at over 90\% of counties in the $30$th week and $48$th week at both models. \highlighttext{On the other hand, as shown in Figures~\ref{fig:caseblack12_tvalue},  and~\ref{fig:caseblack48_tvalue}, while the BLAP is found to be a statistically significant factor in 70\% of the counties in the $12$th week for the WADC model, the corresponding value drastically reduces to 20\% of the counties in the $48$th week. Besides, the coefficients are found to be positive in over 50\% of the counties for the WADD as shown in Figure~\ref{fig:deathblack12_tvalue}, and ~\ref{fig:deathblack48_tvalue}. This finding is consistent with the previous survey~\citep{millett2020assessing}, where they asserted that the black communities are more vulnerable due to the spread of the COVID-19 with the lower coverage rate of health insurance.} However, As shown of the t-stat in Figure~\ref{fig:caseasian12_tvalue} and~\ref{fig:caseasian48_tvalue}, more than 40\% of counties are observed to have negative coefficient of ASIP in the WADC and the WADD models during the study period, which is more than the number of counties having positive coefficient, including the high population density cities and states (e.g., Californian, Seattle, and Florida). \highlighttext{As reported by the studies~\citep{abuelgasim2020covid,raisi2020greater} that the Asians have a higher infected rate by the COVID-19 due to the deficiency of Vitamin D and a higher incidence of coronary heart disease. However, the Asians might get warnings from their families and peers who experienced the earliest suffering of the COVID-19, which may help the Asian population to be more aware of the risk of the COVID-19 and may take better preventative actions against the disease in advance.} It yields similar results when we apply the normalization of the race population cross the studied counties.

\begin{figure}[H]
\centering
\subfloat[][Coefficient, $12$th week, WADC model]{
\includegraphics[width=0.5\linewidth, trim={1.5cm 5.35cm 0cm 4.3cm},clip]{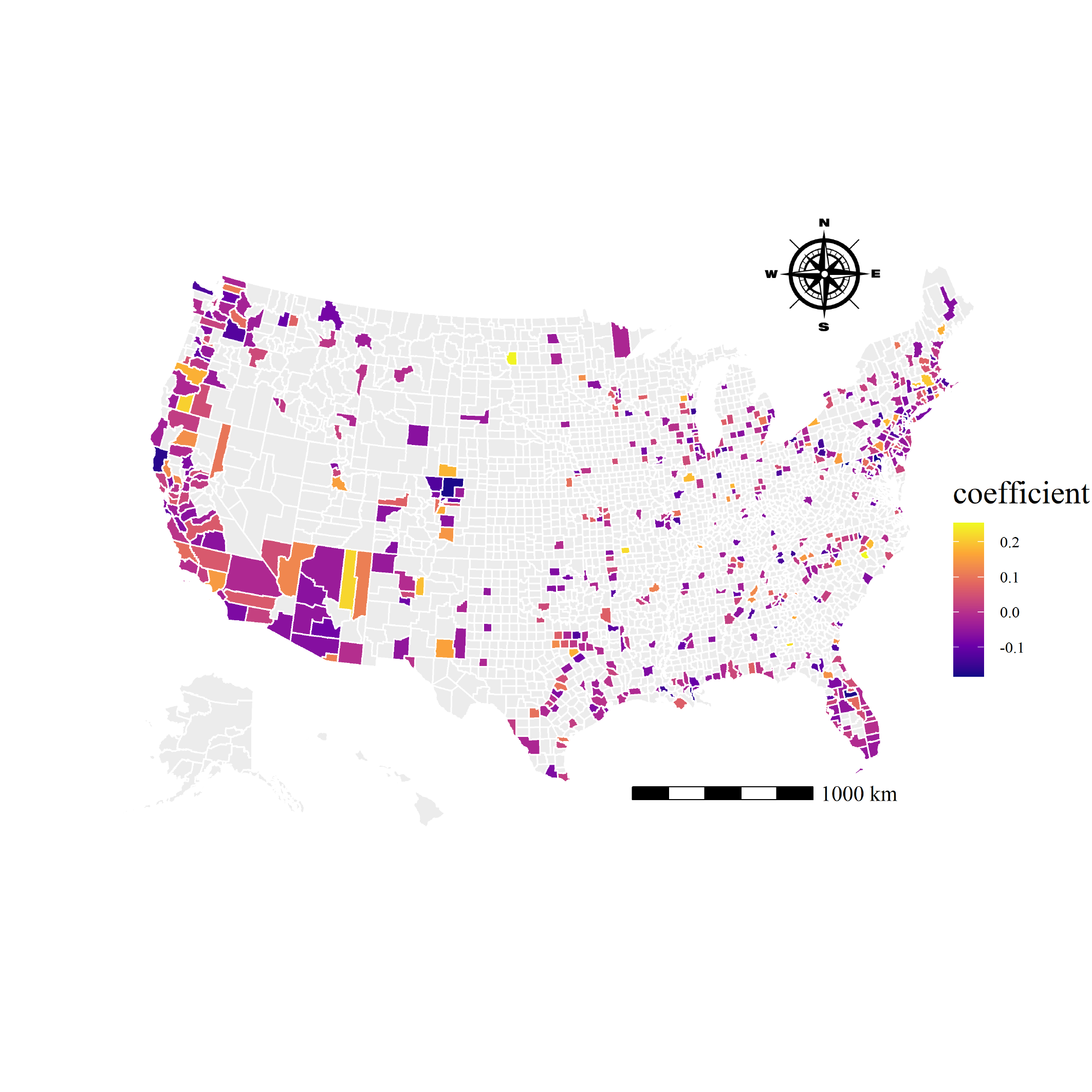}
\label{fig:caselabor12}}
\subfloat[][t statistic, $12$th week, WADC model]{
\includegraphics[width=0.5\linewidth, trim={1.5cm 5.5cm 0cm 4.3cm},clip]{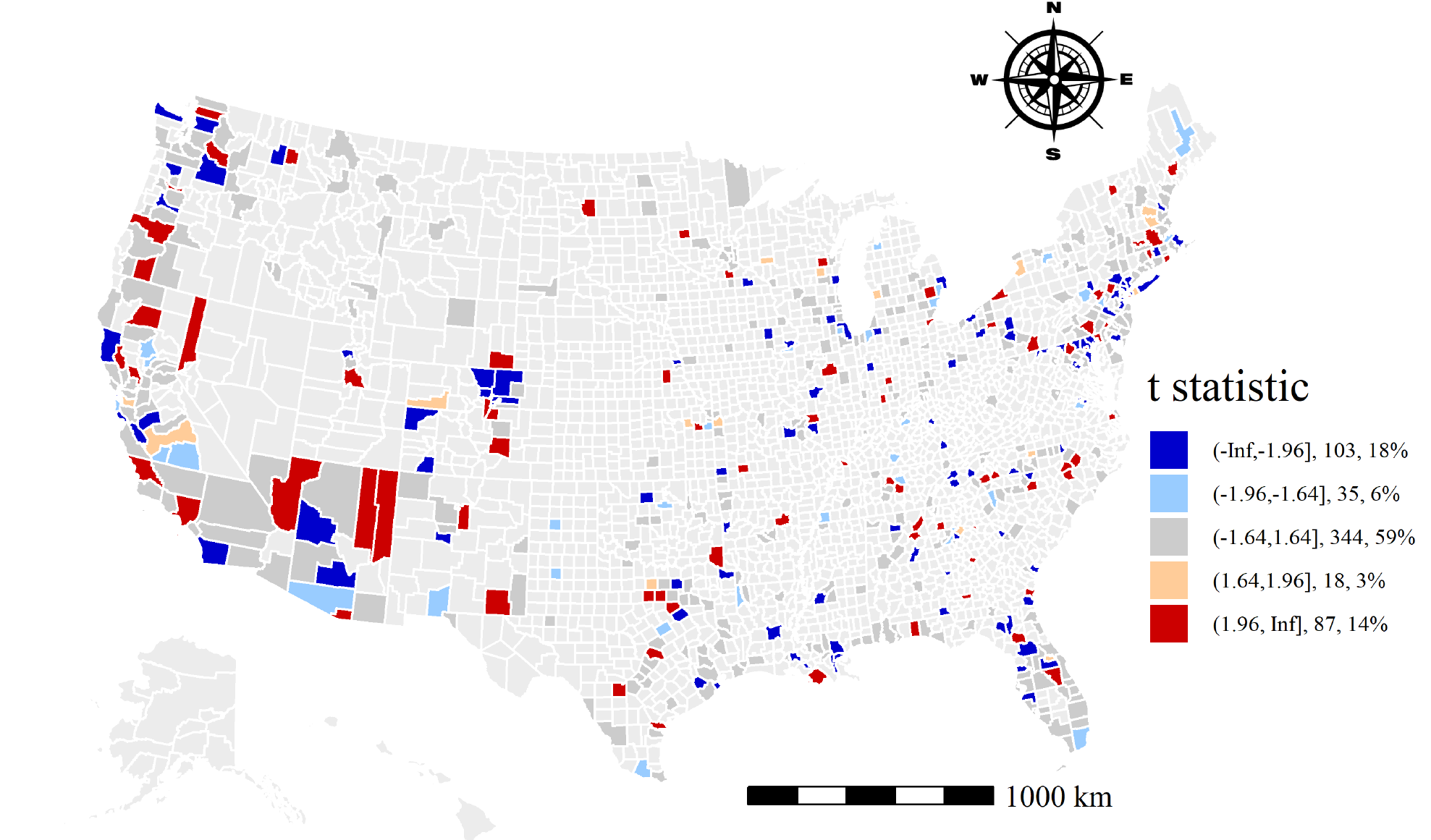}
\label{fig:caselabor12_tvalue}}
\qquad

\subfloat[][Coefficient, $12$th week, WADD model]{
\includegraphics[width=0.5\linewidth, trim={1.5cm 5.35cm 0cm 4.3cm},clip]{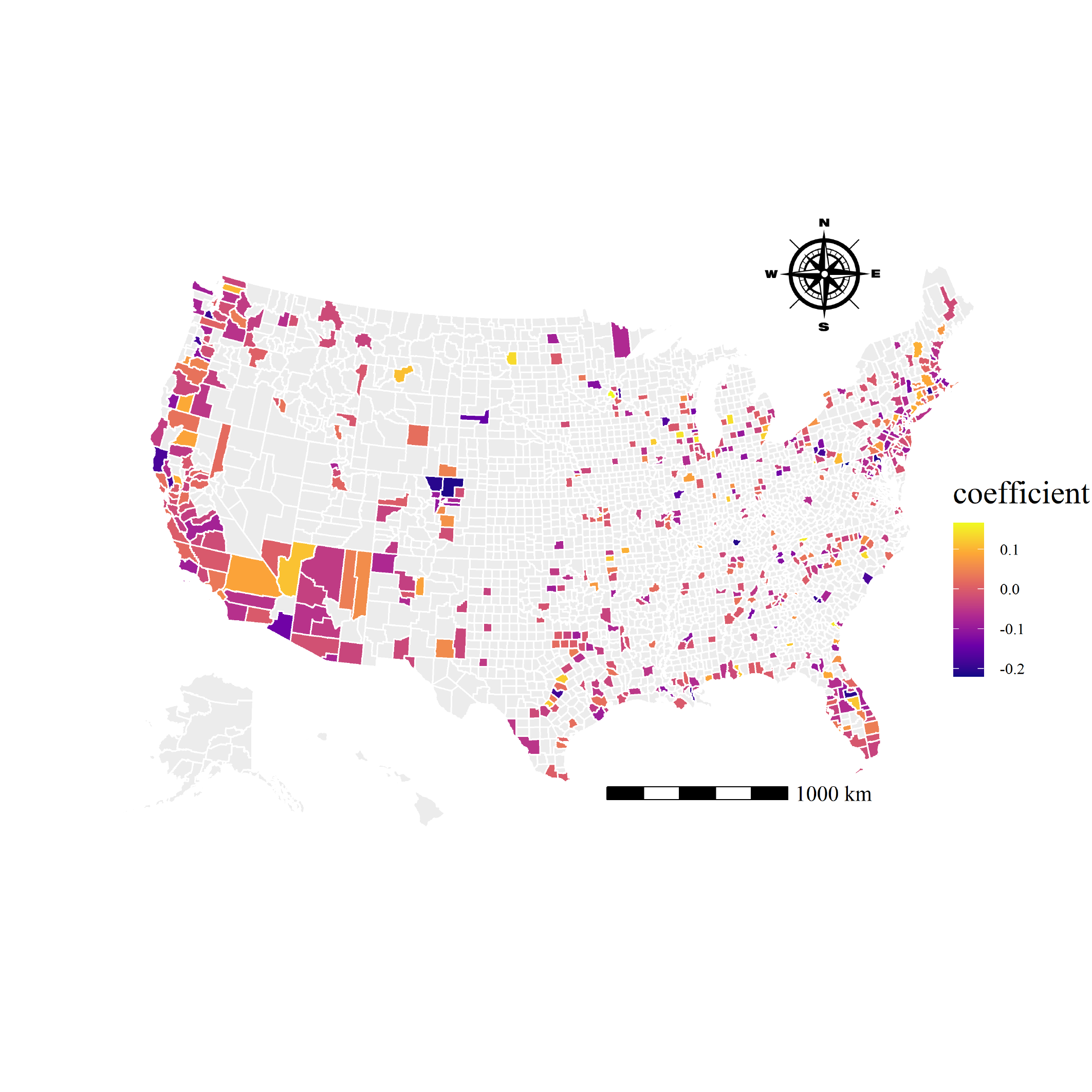}
\label{fig:deathlabor12}}
\subfloat[][t statistic, $12$th week, WADD model]{
\includegraphics[width=0.5\linewidth, trim={1.5cm 5.5cm 0cm 4.3cm},clip]{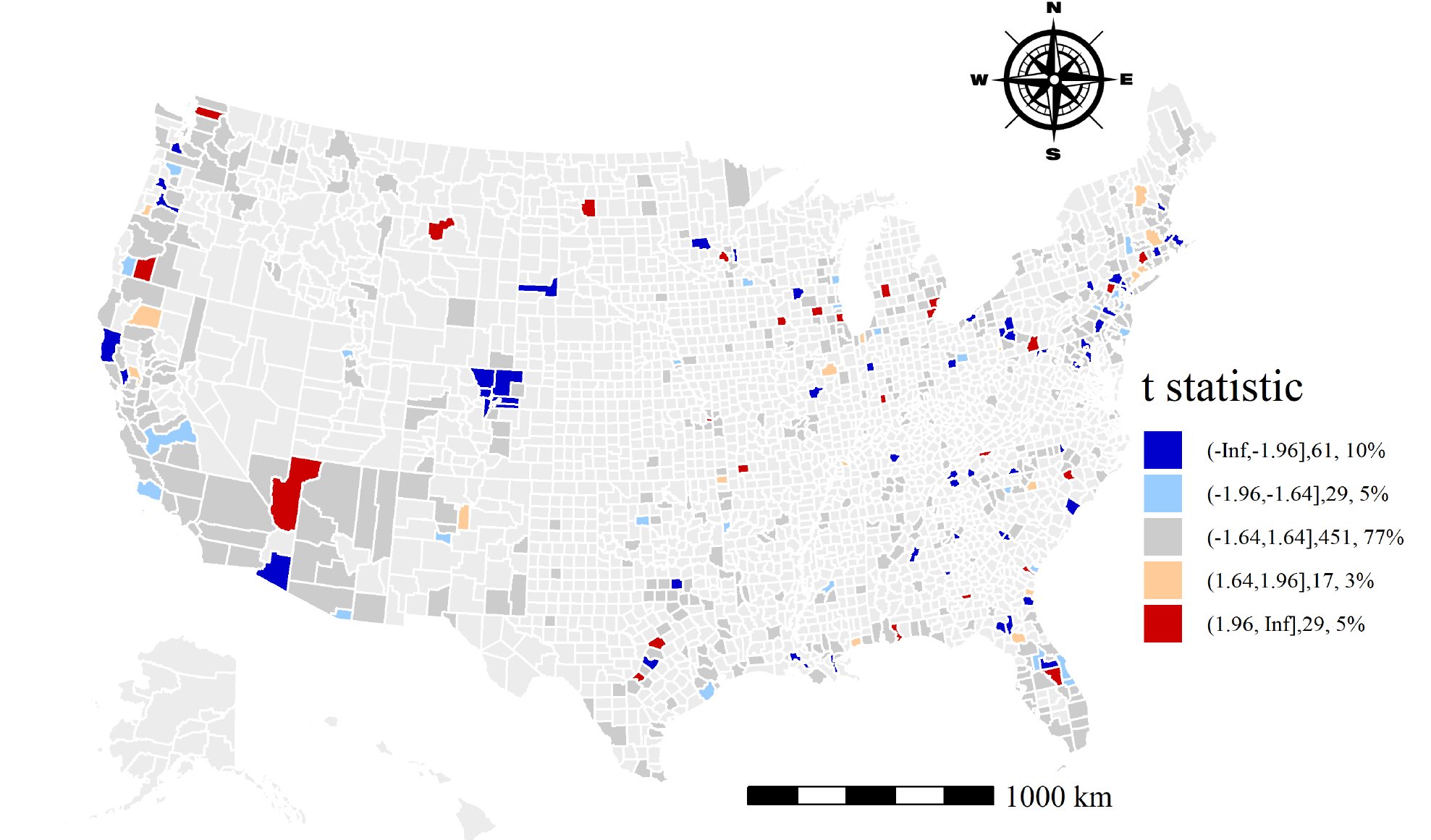}
\label{fig:deathlabor12_tvalue}}

\subfloat[][Coefficient, $48$th week, WADC model]{
\includegraphics[width=0.5\linewidth, trim={1.5cm 5.35cm 0cm 4.3cm},clip]{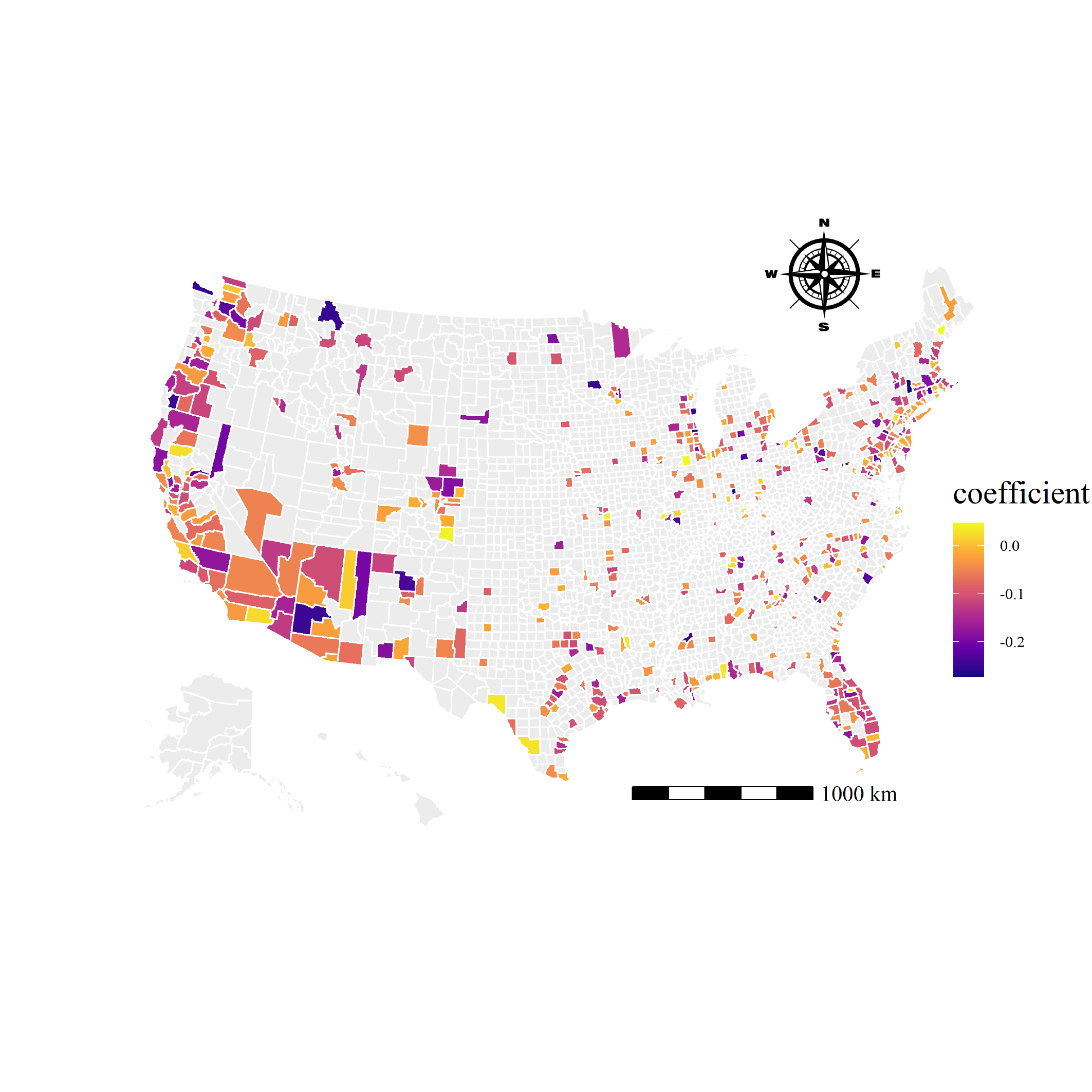}
\label{fig:caselabor48}}
\subfloat[][t statistic, $48$th week, WADC model]{
\includegraphics[width=0.5\linewidth, trim={1.5cm 5.5cm 0cm 4.3cm},clip]{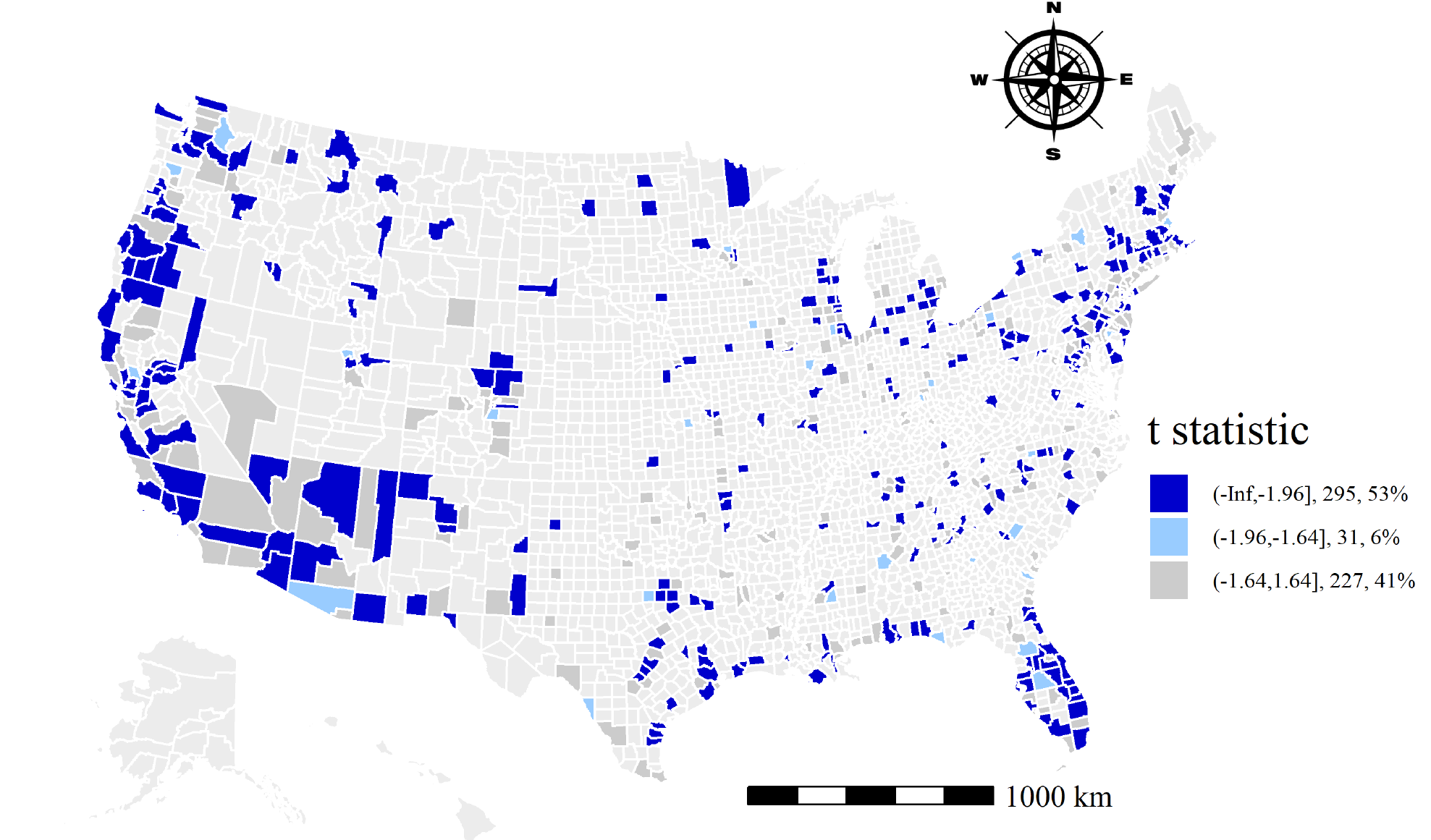}
\label{fig:caselabor48_tvalue}}

\subfloat[][Coefficient, $48$th week, WADD model]{
\includegraphics[width=0.5\linewidth, trim={1.5cm 5.35cm 0cm 4.3cm},clip]{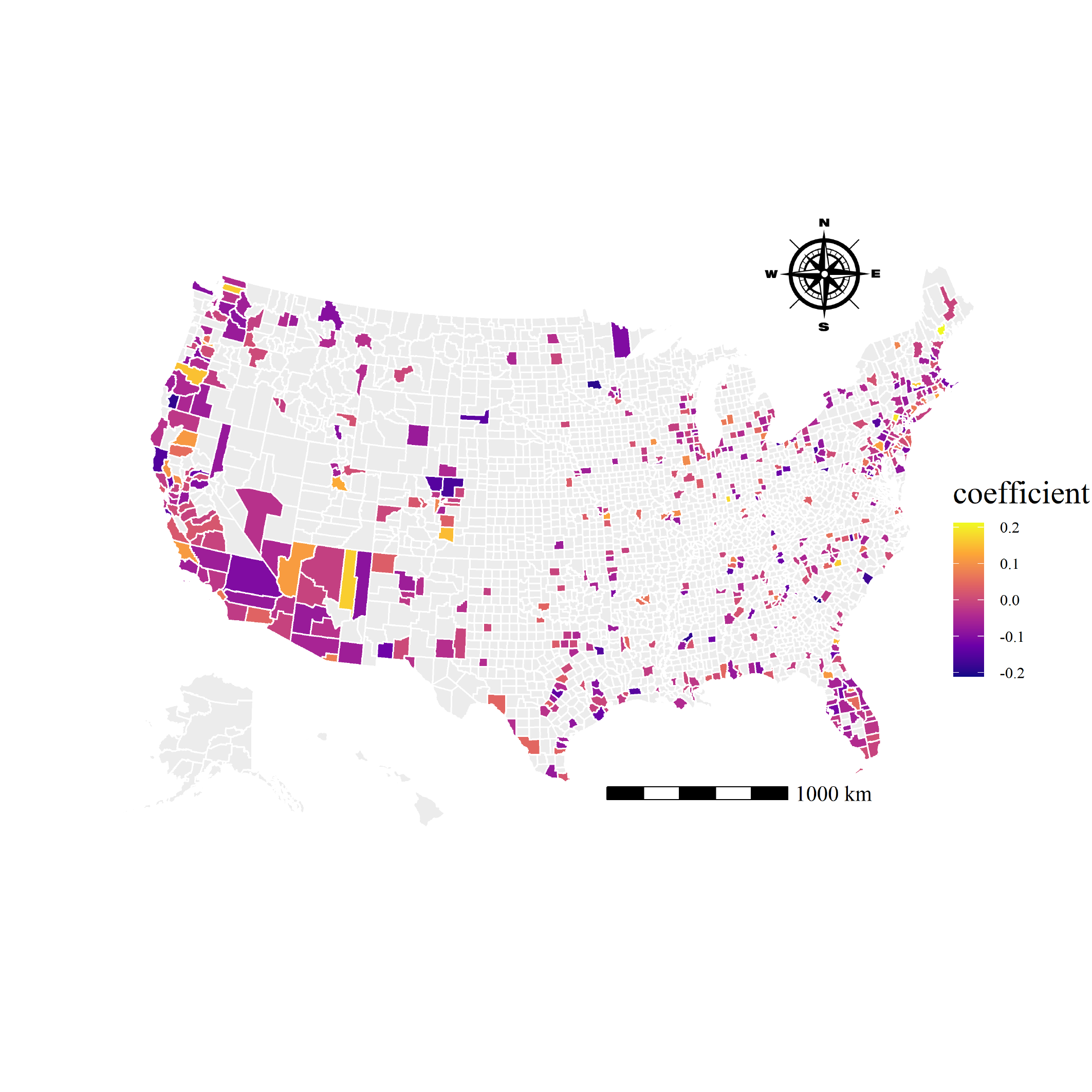}
\label{fig:deathlabor48}}
\subfloat[][t statistic, $48$th week, WADD model]{
\includegraphics[width=0.5\linewidth, trim={1.5cm 5.5cm 0cm 4.3cm},clip]{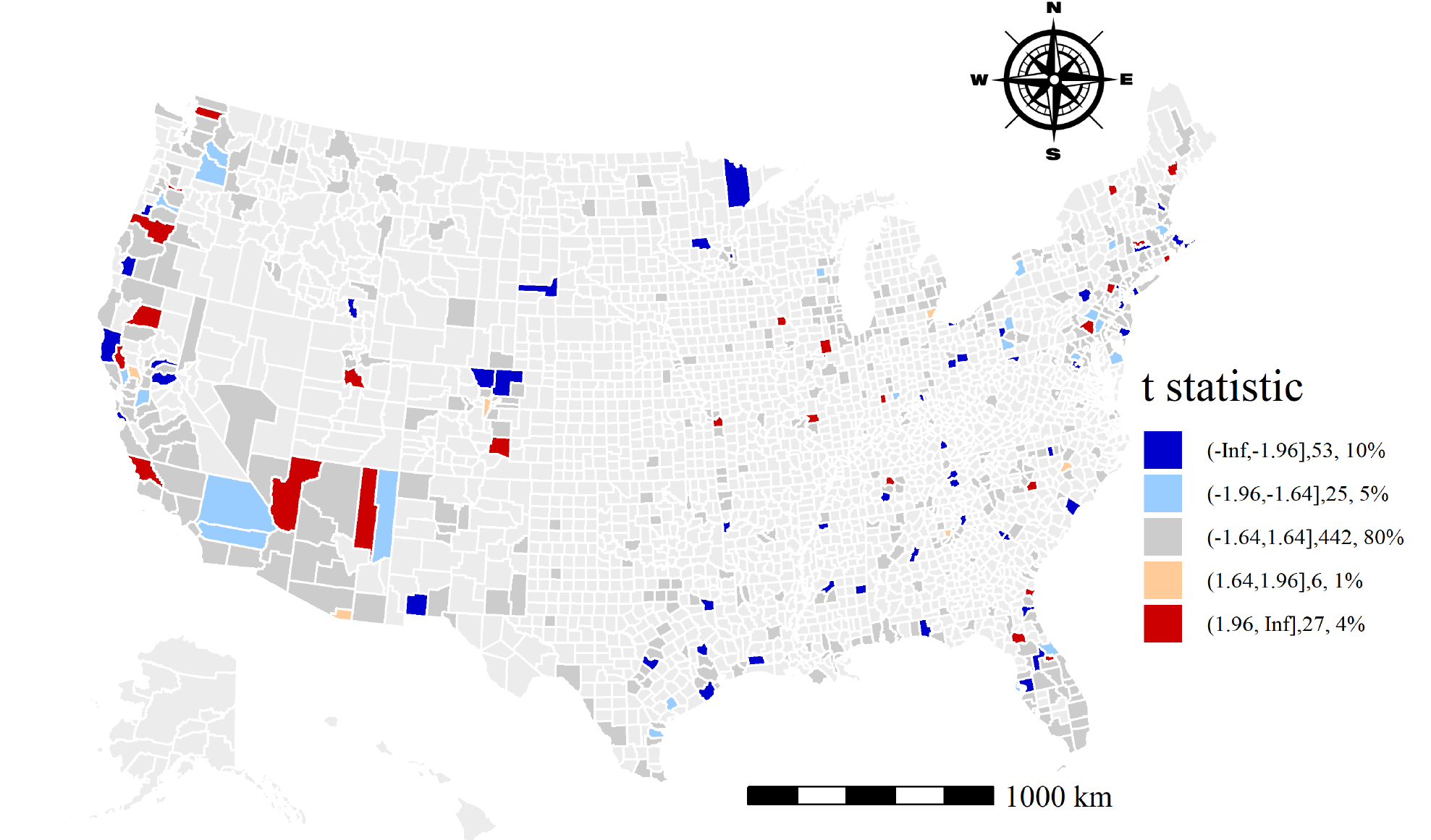}
\label{fig:deathlabor48_tvalue}}

\caption{The distribution of the effects of CLUE in two models}
\label{fig:labor}
\end{figure}

As for the effects of the socioeconomic variables, the estimates of the CLUE are statistically significant (t-stat $>$ 1.64 or t-stat $<$ -1.64) in around 50\% of the counties for the WADC model and around 25\% of the counties for the WADD model during the study period (see Figure~\ref{fig:labor}). In the $12$th week, the number of counties with a positive coefficient of the CLUE is more than the number of areas with negative effects on both models (24\% vs. 17\% for the WADC model, and 15\% vs. 8\% for the WADD model). The unemployed population with unstable(unsafe) workplaces and irregular social activity might intensify the disease propagation as suggested by the previous finding~\citep{poletto2013human}. More importantly, the high population density areas are more likely to have positive and statistically significant effects of the CLUE in the $12$th week (e.g., counties in California, Washington, Arizona,  Minnesota, and Florida). However, this situation has been changed for the WADC model in the $48$th week as shown in Figure~\ref{fig:caselabor48_tvalue}. That may be related to the effectiveness of the shelter policy and the the COVID subsidies, which mitigated the sufferings of the unemployment population in purchasing daily needs and helped reduce their daily activity levels. Similarly, the study~\citep{borkowski2020lockdowned} also indicates the occupation is the key factor affecting travel time change and infected rate. \highlighttext{For the education effects, the number of areas with negative coefficient (t-stat $<$ -1.64 ) of BDHP is more than the number of areas with positive coefficient (t-stat $>$ 1.64) in the WADC model (about 30\% vs. 10\%) and the WADD model (about 40\% vs. 20\%) during the study period. The reason might be the highly educated population have a higher awareness of the risks of the COVID-19 and are more acceptable to the public suggestions for preventing the COVID-19. On the other hand, the median elasticity indicates 1\% increase in the MHIC results in 0.83\% increase in WADC and 1.17\% increase in WADD. The reason might be that commercial activity and business communication are more active in counties with high MHIC.} 
 
\begin{figure}[H]
\centering
\subfloat[][$12$th week, WADC model]{
\includegraphics[width=0.5\linewidth, trim={1.5cm 5.5cm 0cm 4.3cm},clip]{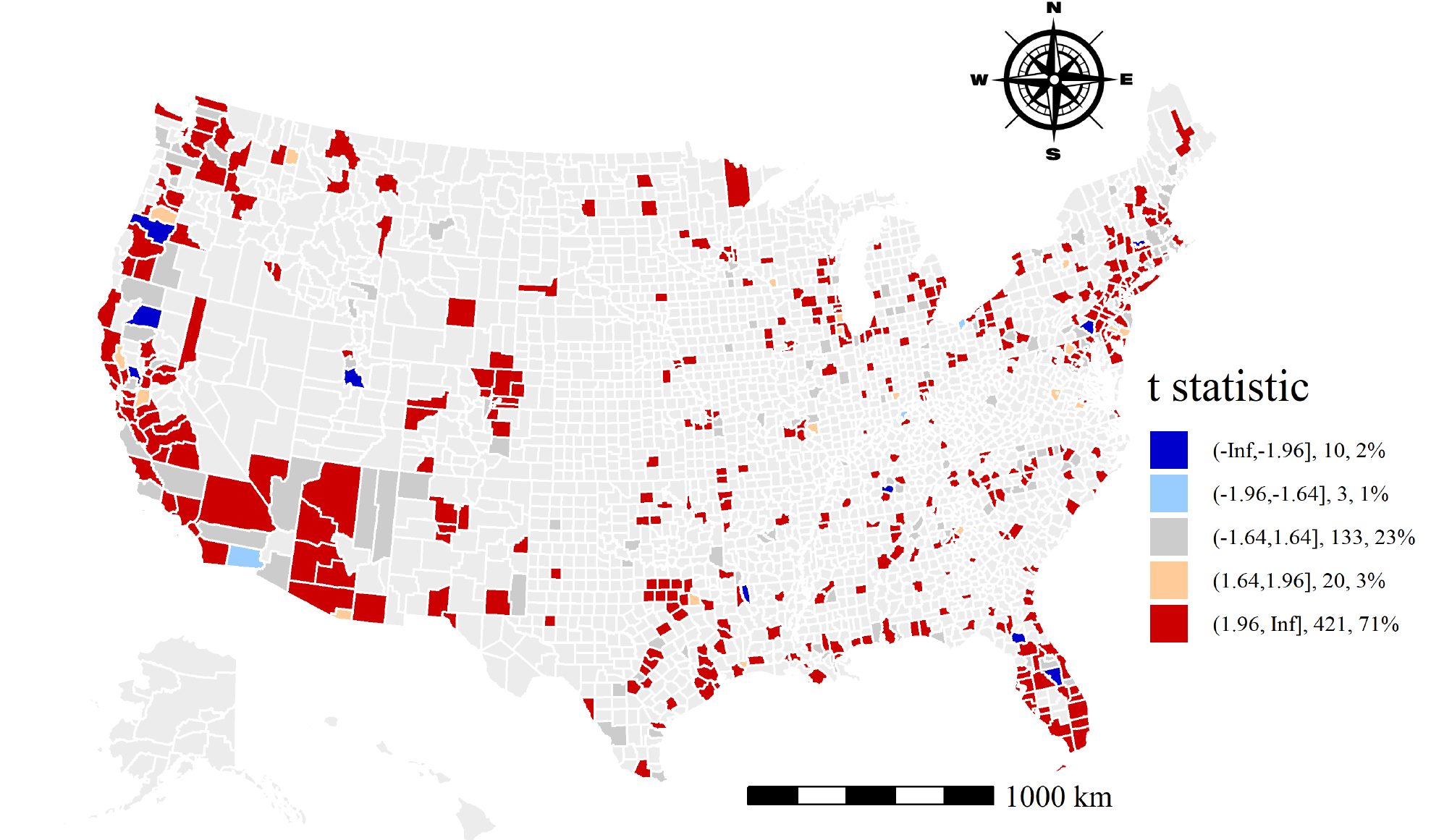}
\label{fig:casepubtran12_tvalue}}
\subfloat[][$48$th week, WADC model]{
\includegraphics[width=0.5\linewidth, trim={1.5cm 5.5cm 0cm 4.3cm},clip]{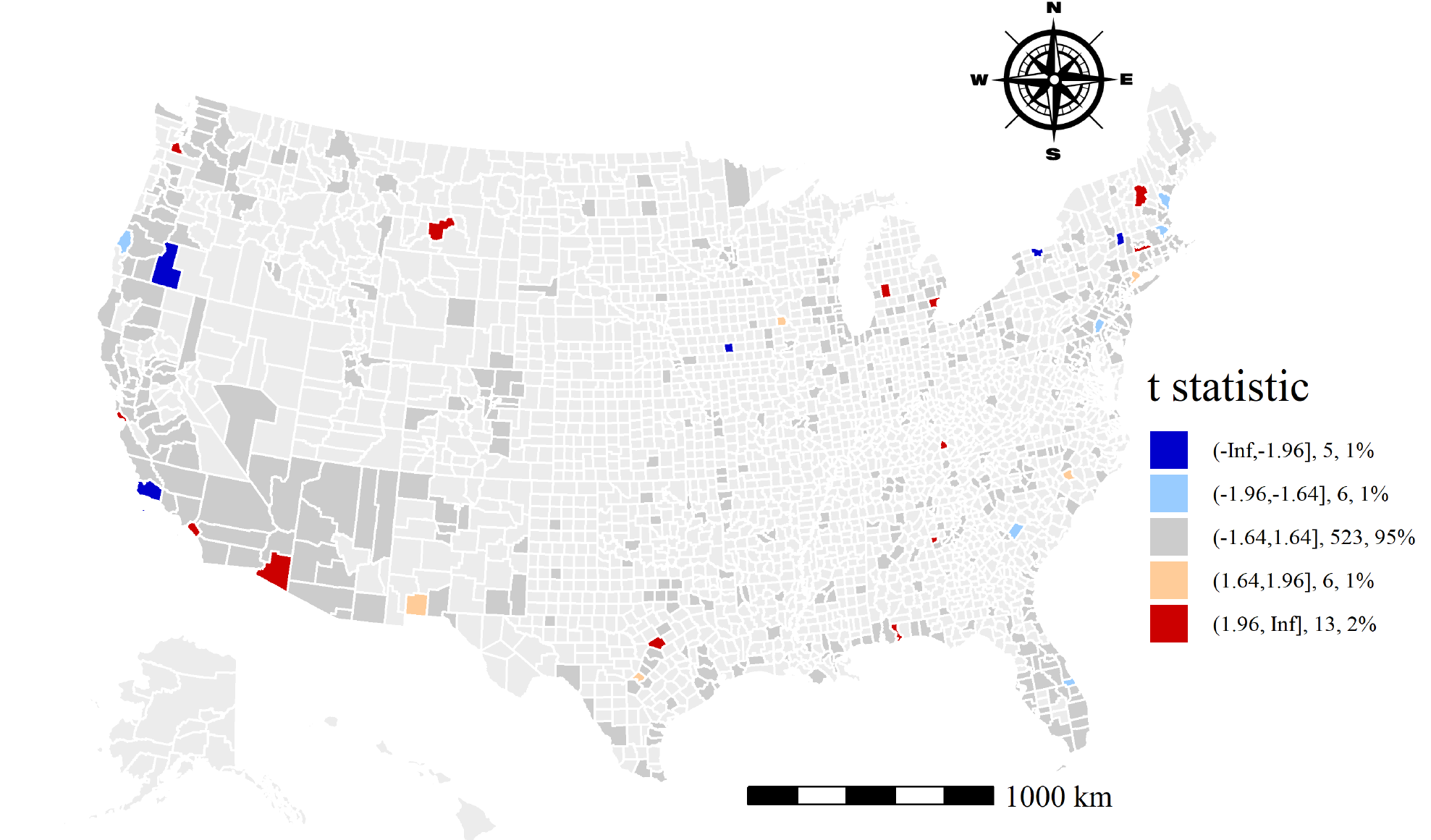}
\label{fig:casepubtran48_tvalue}}
\qquad

\subfloat[][$12$th week, WADD model]{
\includegraphics[width=0.5\linewidth, trim={1.5cm 5.5cm 0cm 4.3cm},clip]{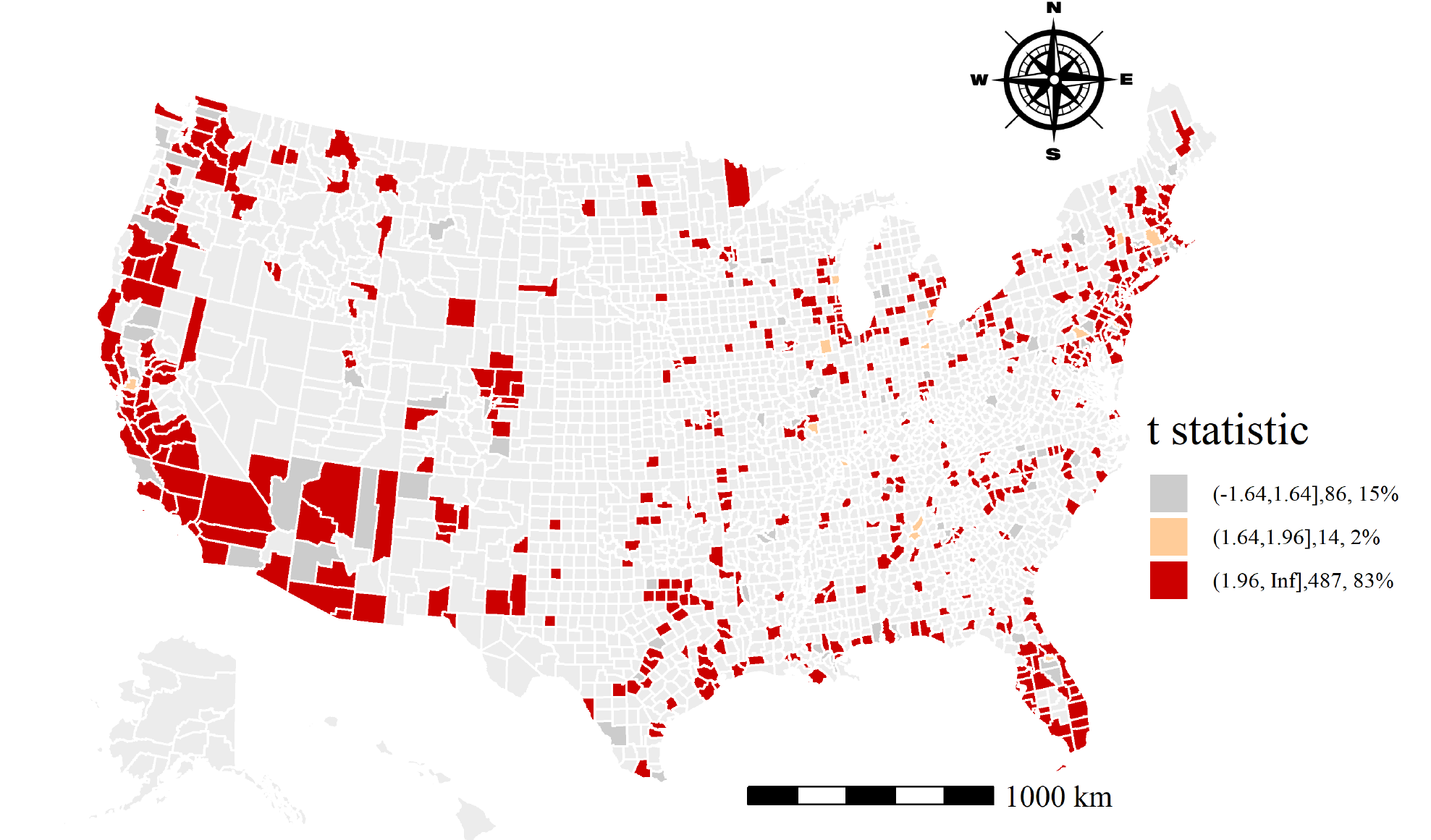}
\label{fig:deathpubtran12_tvalue}}
\subfloat[][$48$th week, WADD model]{
\includegraphics[width=0.5\linewidth, trim={1.5cm 5.5cm 0cm 4.3cm},clip]{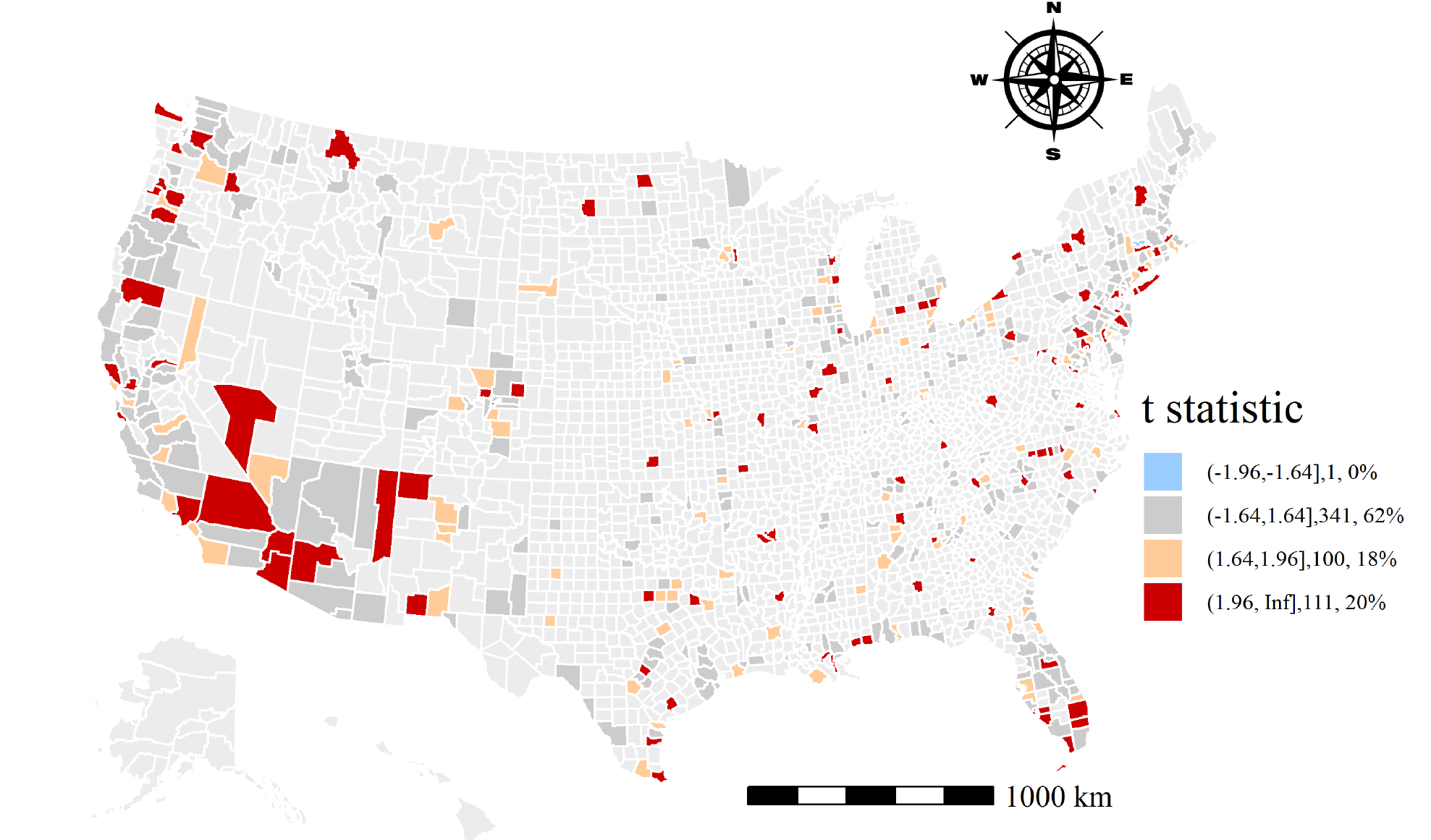}
\label{fig:deathpubtran48_tvalue}}

\caption{The t statistic distribution of PTAP at two models}
\label{fig:pubtrans}
\end{figure}

\begin{figure}[H]
\centering
\subfloat[][$12$th week, WADC model]{
\includegraphics[width=0.5\linewidth, trim={1.5cm 5.5cm 0cm 4.3cm},clip]{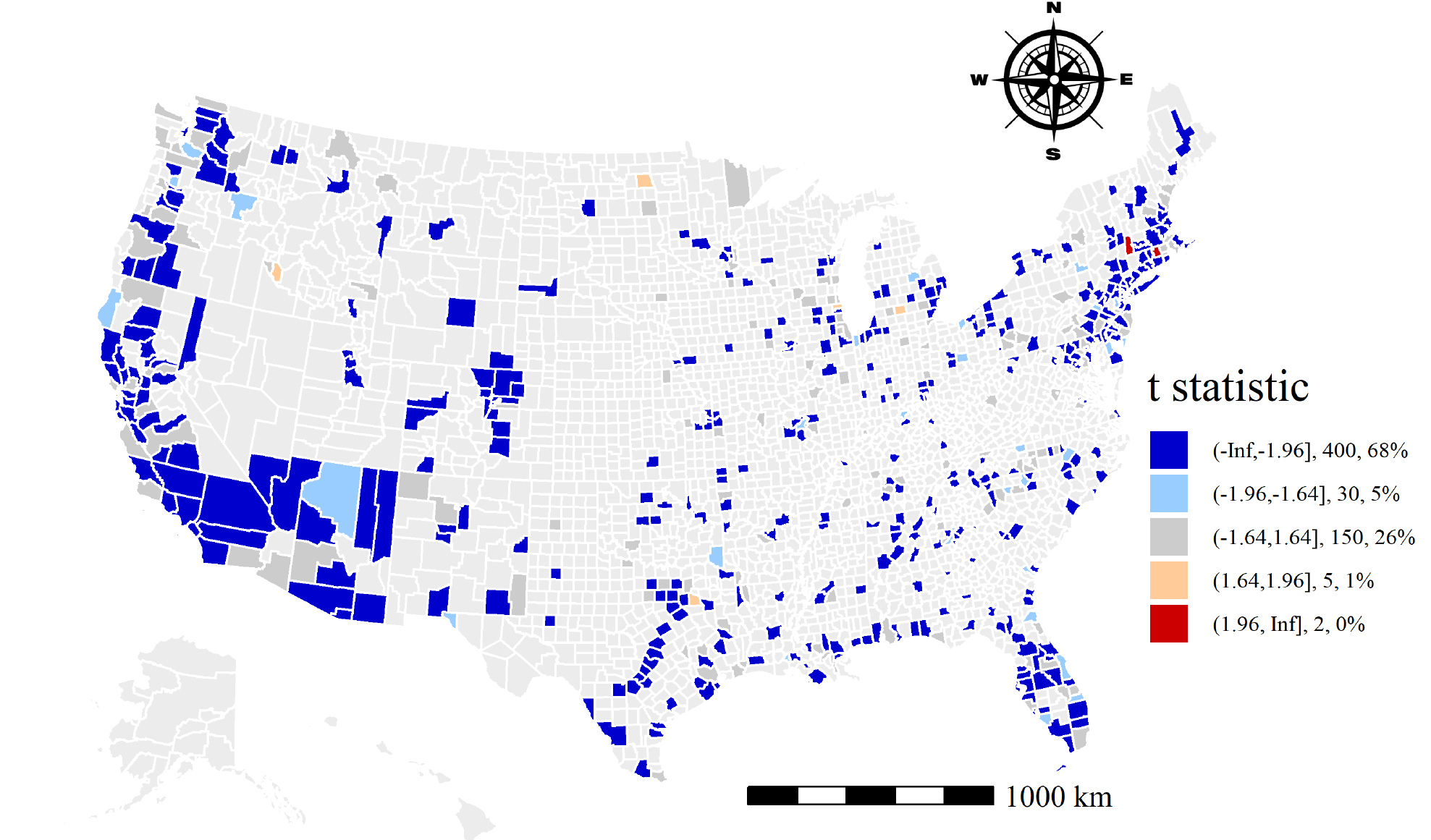}
\label{fig:caseWFH21}}
\subfloat[][$48$th week, WADC model]{
\includegraphics[width=0.5\linewidth, trim={1.5cm 5.5cm 0cm 4.3cm},clip]{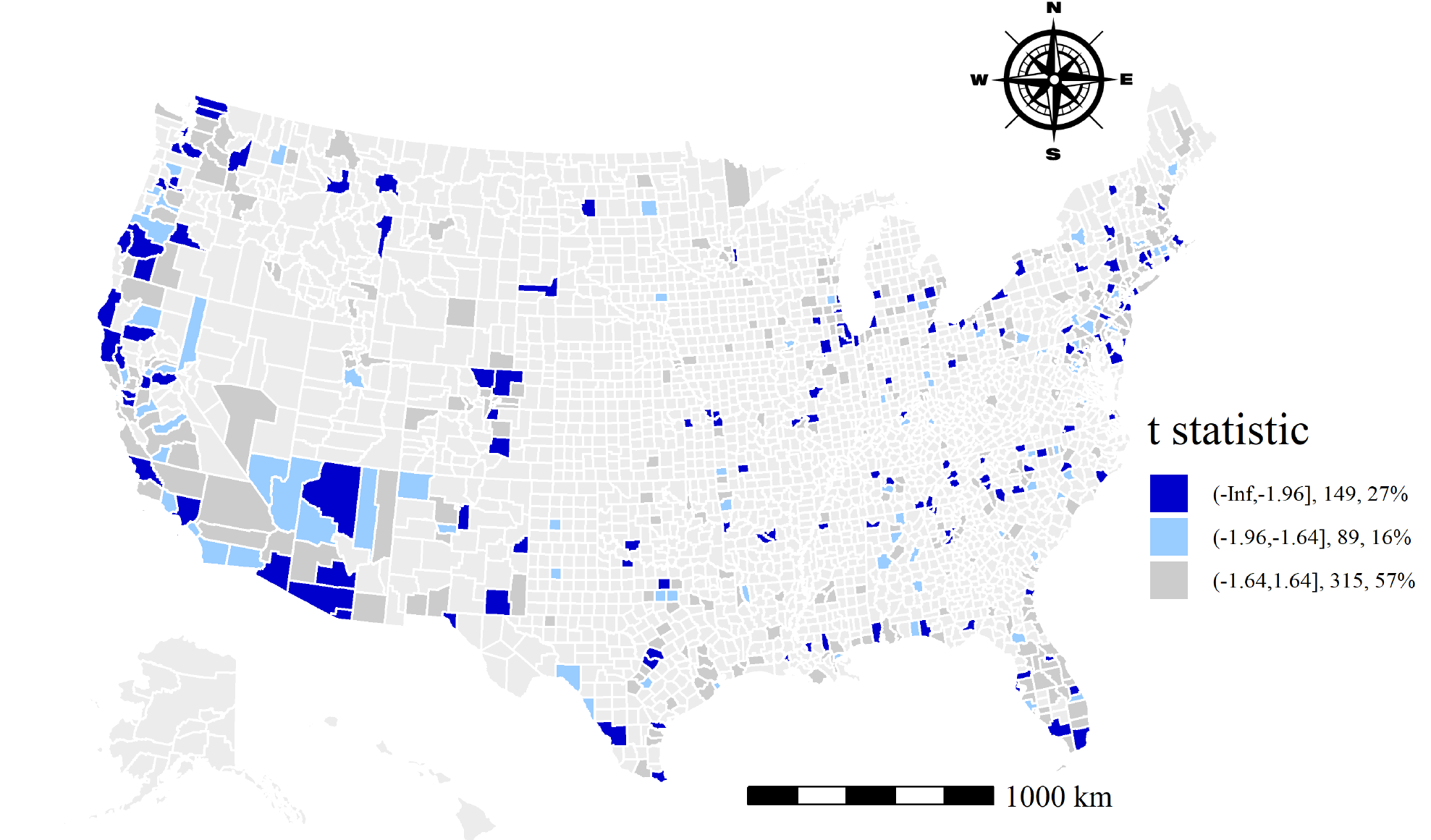}
\label{fig:caseWFH21_tvalue}}
\qquad

\subfloat[][$12$th week, WADD model]{
\includegraphics[width=0.5\linewidth, trim={1.5cm 5.5cm 0cm 4.3cm},clip]{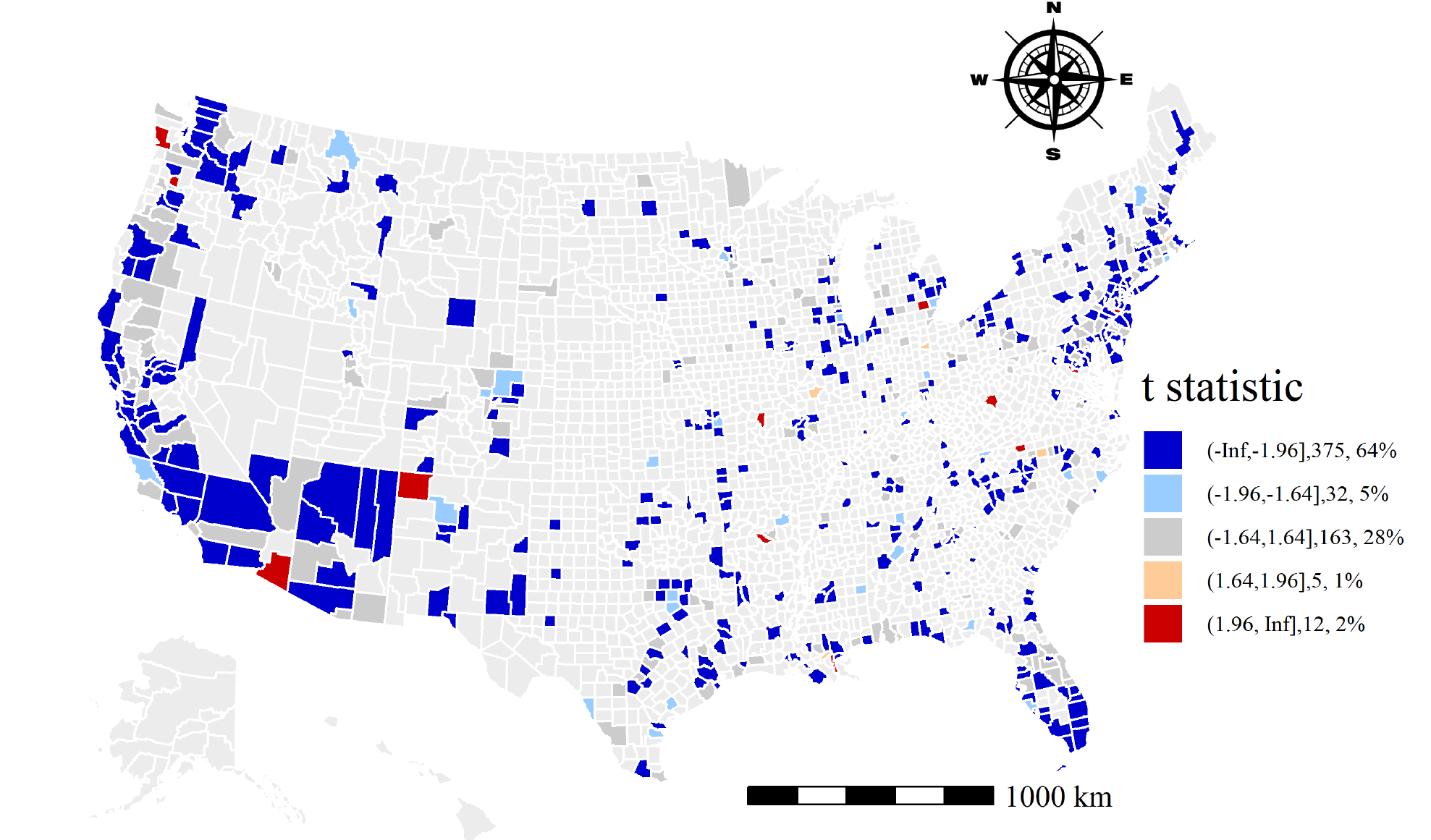}
\label{fig:deathWFH21}}
\subfloat[][$48$th week, WADD model]{
\includegraphics[width=0.5\linewidth, trim={1.5cm 5.5cm 0cm 4.3cm},clip]{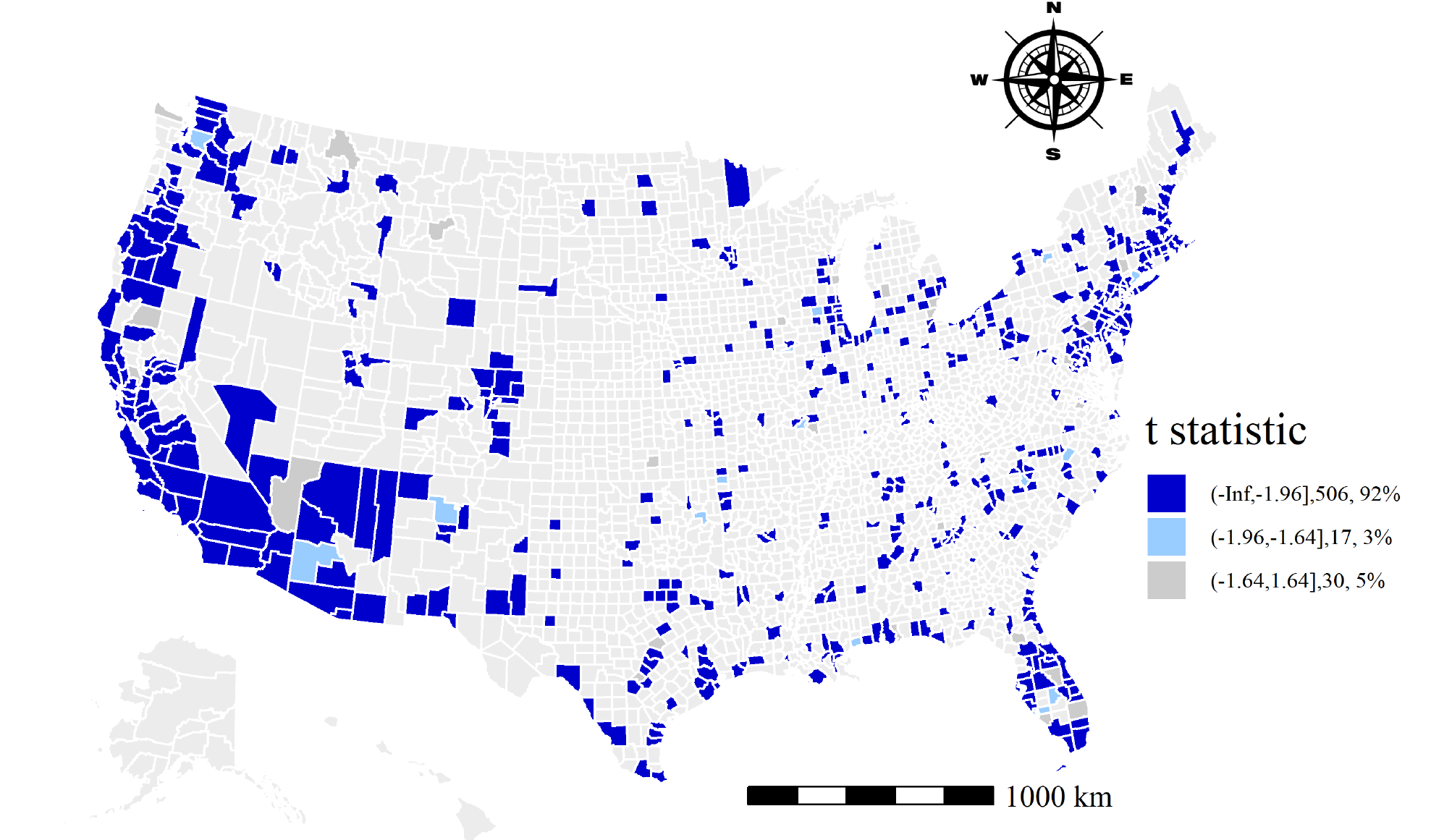}
\label{fig:deathWFH21_tvalue}}

\caption{The t statistic distribution of WFHP in two models}
\label{fig:WFH}
\end{figure}

As for the travel-related effects, the effects of MILE are statistically significant and positive on the WADC and WADD models at all studied counties during the study period. It indicates that the availability of public transportation facility is more likely to increase the number of WADC and WADD. In particular, the median elasticity of the MILE suggests that a 1\% increase in the MILE results in 1.03\% increase in the WADC and 1.04\% increase in the WADD on average. For the travel mode effects, the median elasticity shows that 1\% increase in the MTAT results in 0.22\%(0.95\%) increase in the WADC and the WADD on average. The effects of MTAT on the WADC and the WADD are heterogeneous among counties during the study period. For example, the MTAT for counties in California is positively (t-stat $>$ 1.64) related to WADC and WADD. However, the counties in Arizona have negative effects (t-stat $<$ -1.64) of the MTAT on the WADC and the WADD. The underlying reason might be that the long MTAT in high population density areas intensifies the disease propagation. Whereas, the large scale of urban structure with low population density effectively reduces the contact space among people. \highlighttext{The t statistic estimations for the PTAP is summarized in Figure~\ref{fig:pubtrans}. In particular, about 80\% of counties have positive coefficients in the $12$th week in both models (see Figure~\ref{fig:casepubtran12_tvalue} and Figure~\ref{fig:deathpubtran12_tvalue}).The long contact duration and close proximity among passengers in transit system are likely the causes for this observation~\citep{qian2020modeling}. It also indicates that the public transit closure strategy may effectively curb the propagation of the COVID-19 in the early stage. However, the effects of the PTAP in most counties became less significant or insignificant in the $48$th week. This highlights the temporally varying effects of the modeling determinants and suggests that the PTAP is no longer a determining factor in the later stage of the pandemic. Besides, as shown in Figure~\ref{fig:WFH}, there is a higher number of counties showing that the WFHP is negatively related to WADC and WADD in the $12$th week than in the $48$th week. This finding is consistent with the estimated results of the PTAP and suggests that WFH plays a more significant role in the early stage of the disease. Besides, the evidence of WFH policy applied at Australia~\citep{beck2020slowly} also verifies our insights.}

\begin{figure}[H]
\centering
\subfloat[][$12th$ week, WADC model]{
\includegraphics[width=0.5\linewidth, trim={1.5cm 5.5cm 0cm 4.3cm},clip]{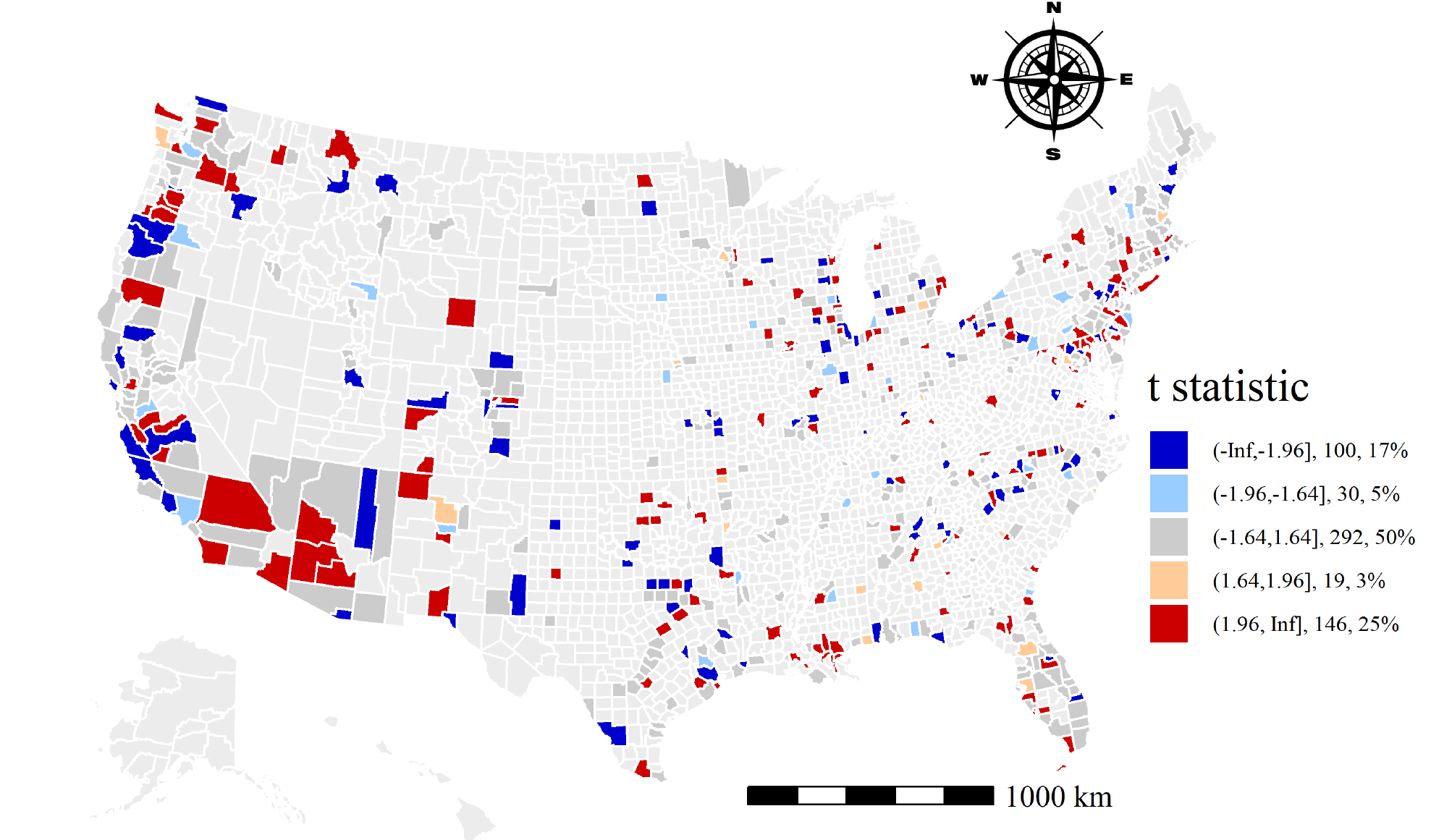}
\label{fig:casegrocery12_tvalue}}
\subfloat[][$48th$ week, WADC model]{
\includegraphics[width=0.5\linewidth, trim={1.5cm 5.5cm 0cm 4.3cm},clip]{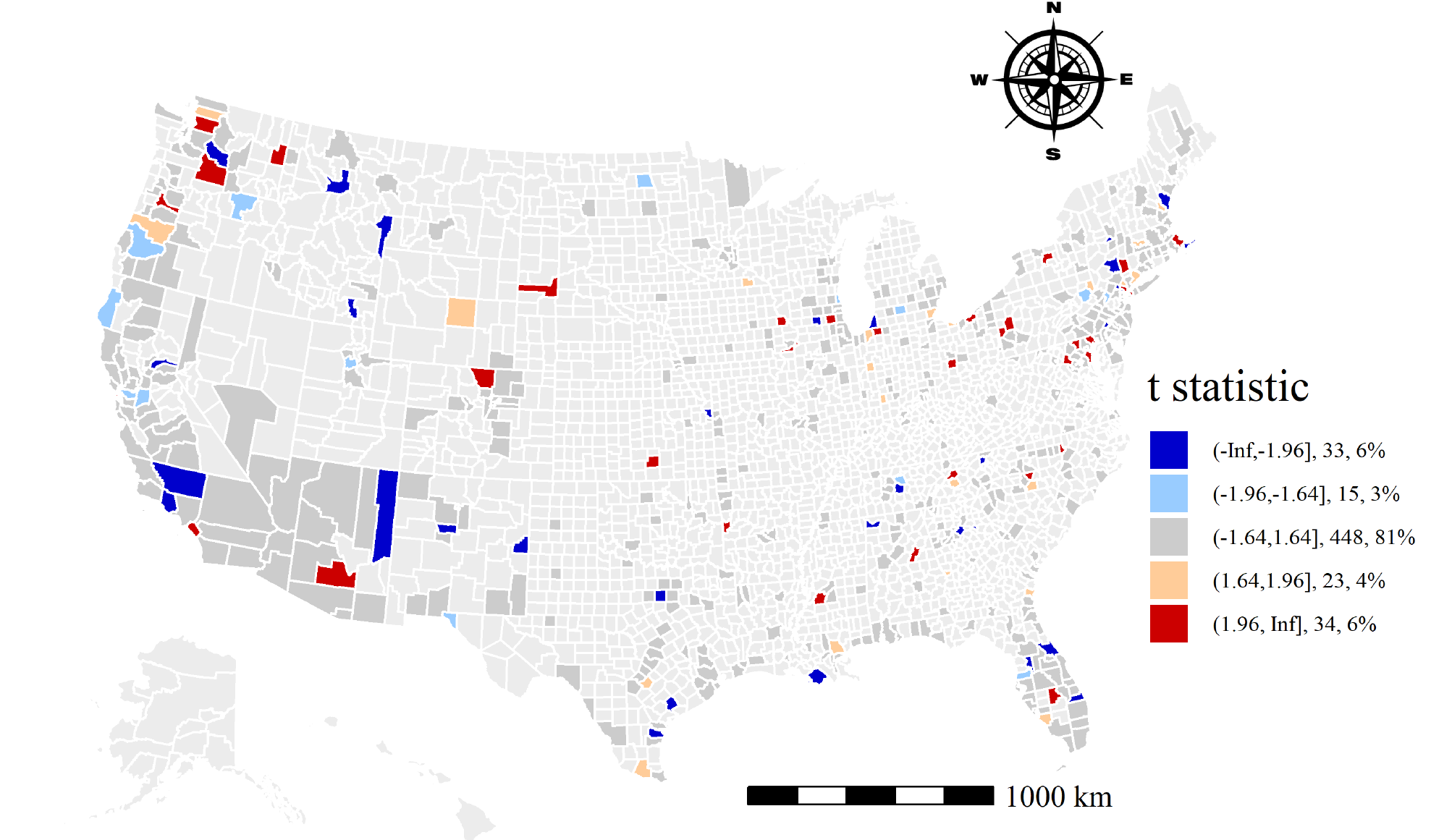}
\label{fig:casegrocery48_tvalue}}
\qquad

\subfloat[][$12th$ week, WADD model]{
\includegraphics[width=0.5\linewidth, trim={1.5cm 5.5cm 0cm 4.3cm},clip]{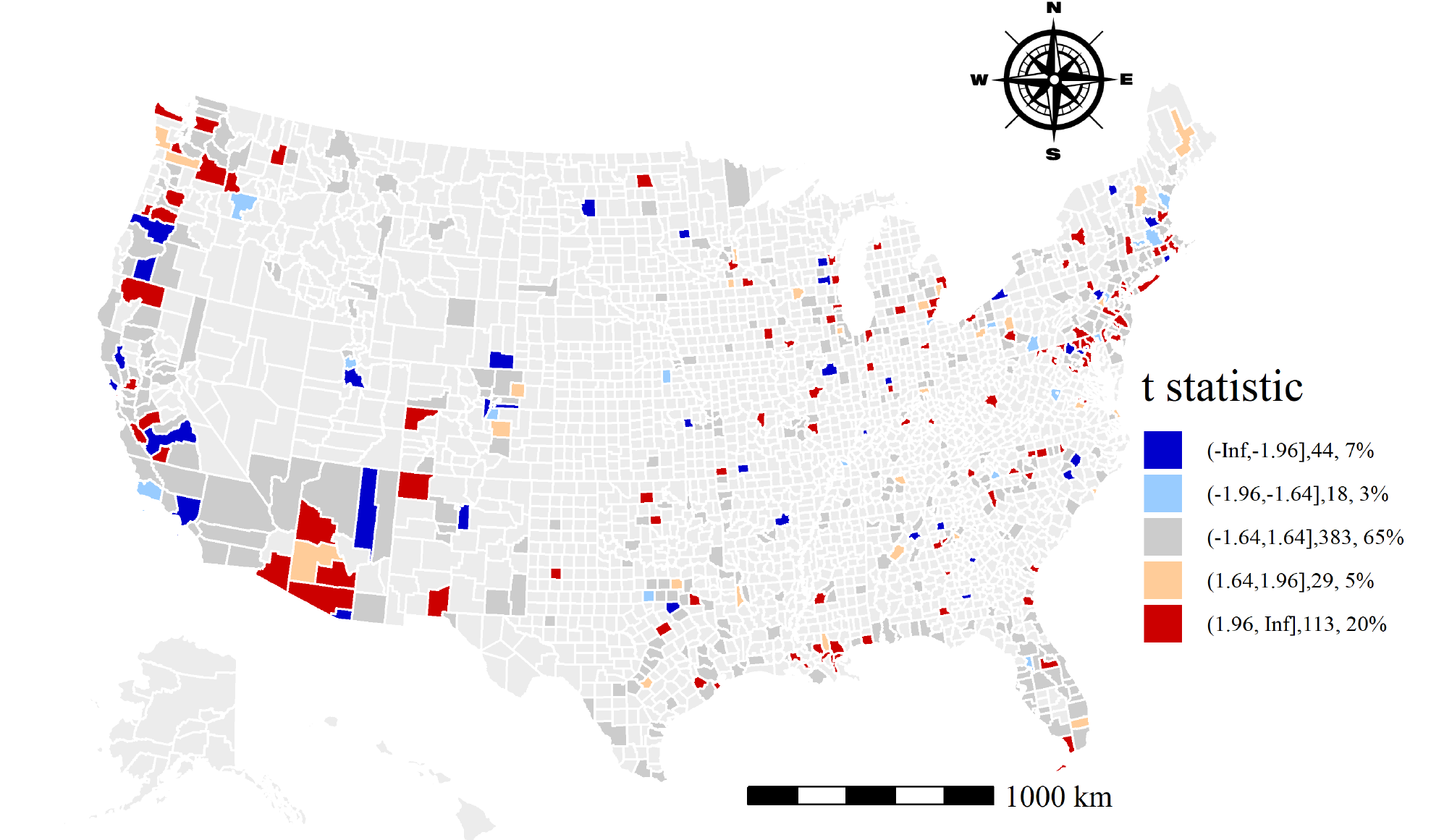}
\label{fig:deathgrocery12_tvalue}}
\subfloat[][$48th$ week, WADD model]{
\includegraphics[width=0.5\linewidth, trim={1.5cm 5.5cm 0cm 4.3cm},clip]{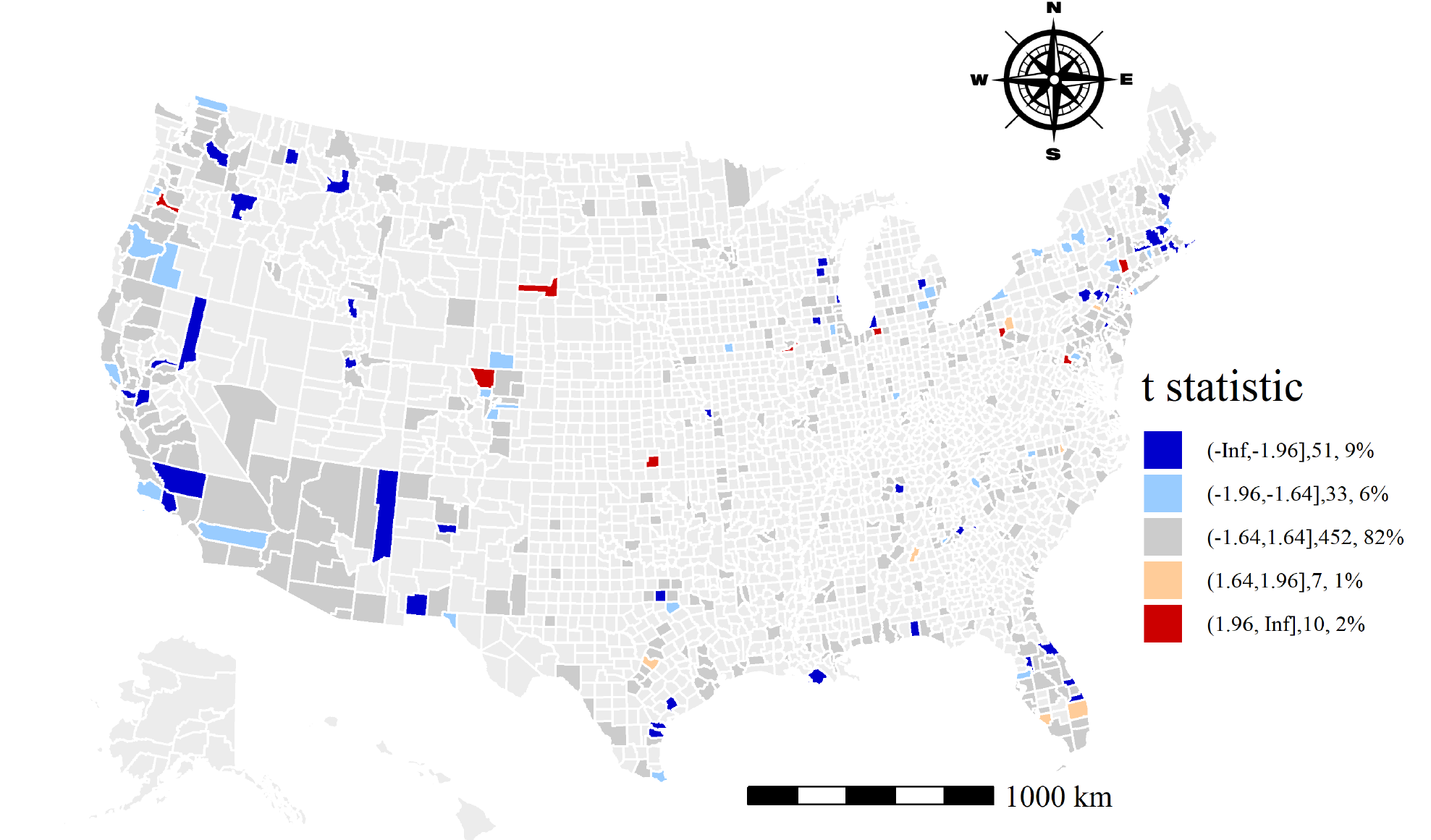}
\label{fig:deathgrocery48_tvalue}}
\qquad
\caption{The t statistic distribution of GPPC in two models}
\label{fig:grocery}
\end{figure}

\begin{figure}[H]
\centering
\subfloat[][$12th$ week, WADC model]{
\includegraphics[width=0.5\linewidth, trim={1.5cm 5.5cm 0cm 4.3cm},clip]{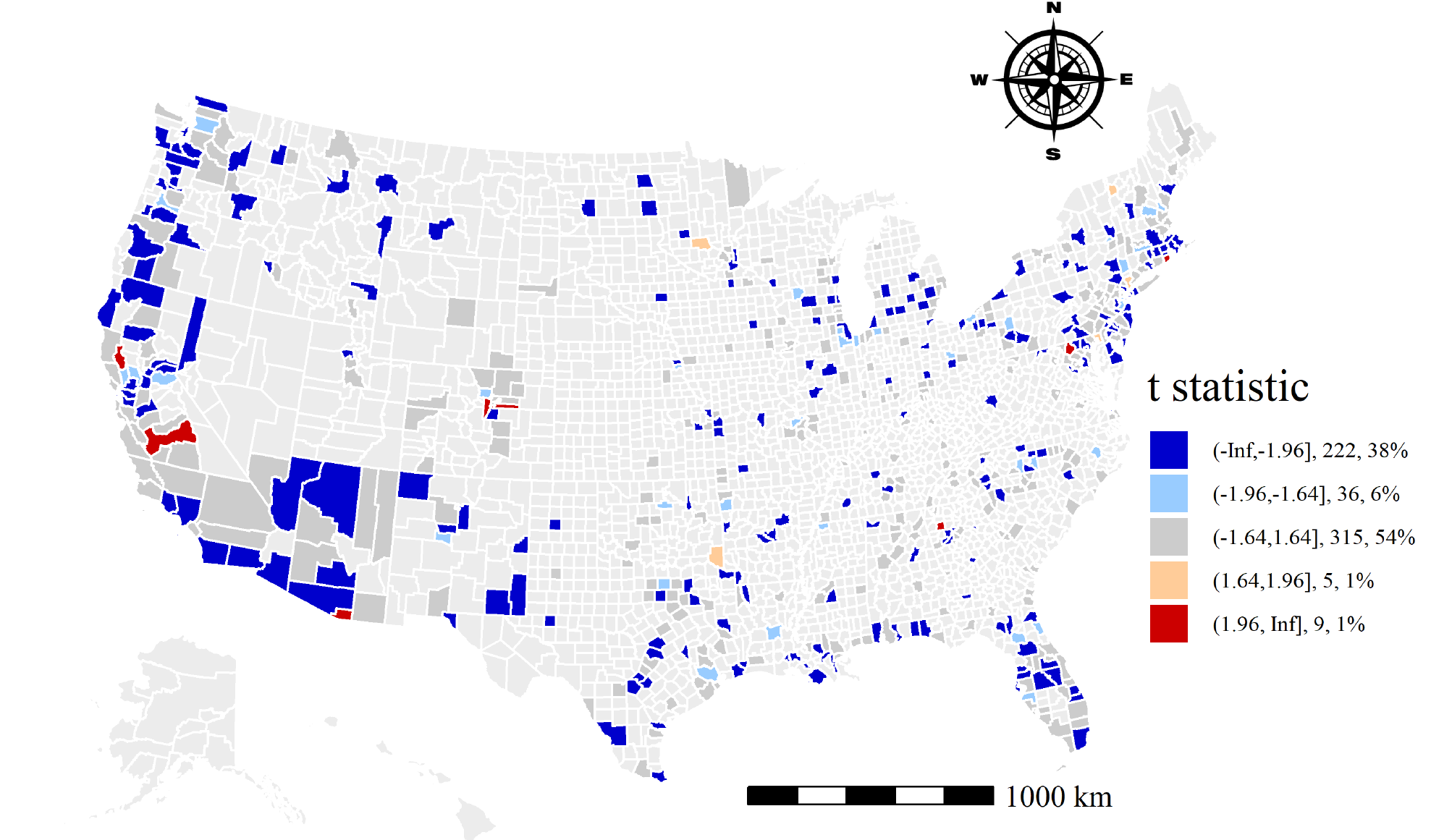}
\label{fig:casetransit12_tvalue}}
\subfloat[][$48th$ week, WADC model]{
\includegraphics[width=0.5\linewidth, trim={1.5cm 5.5cm 0cm 4.3cm},clip]{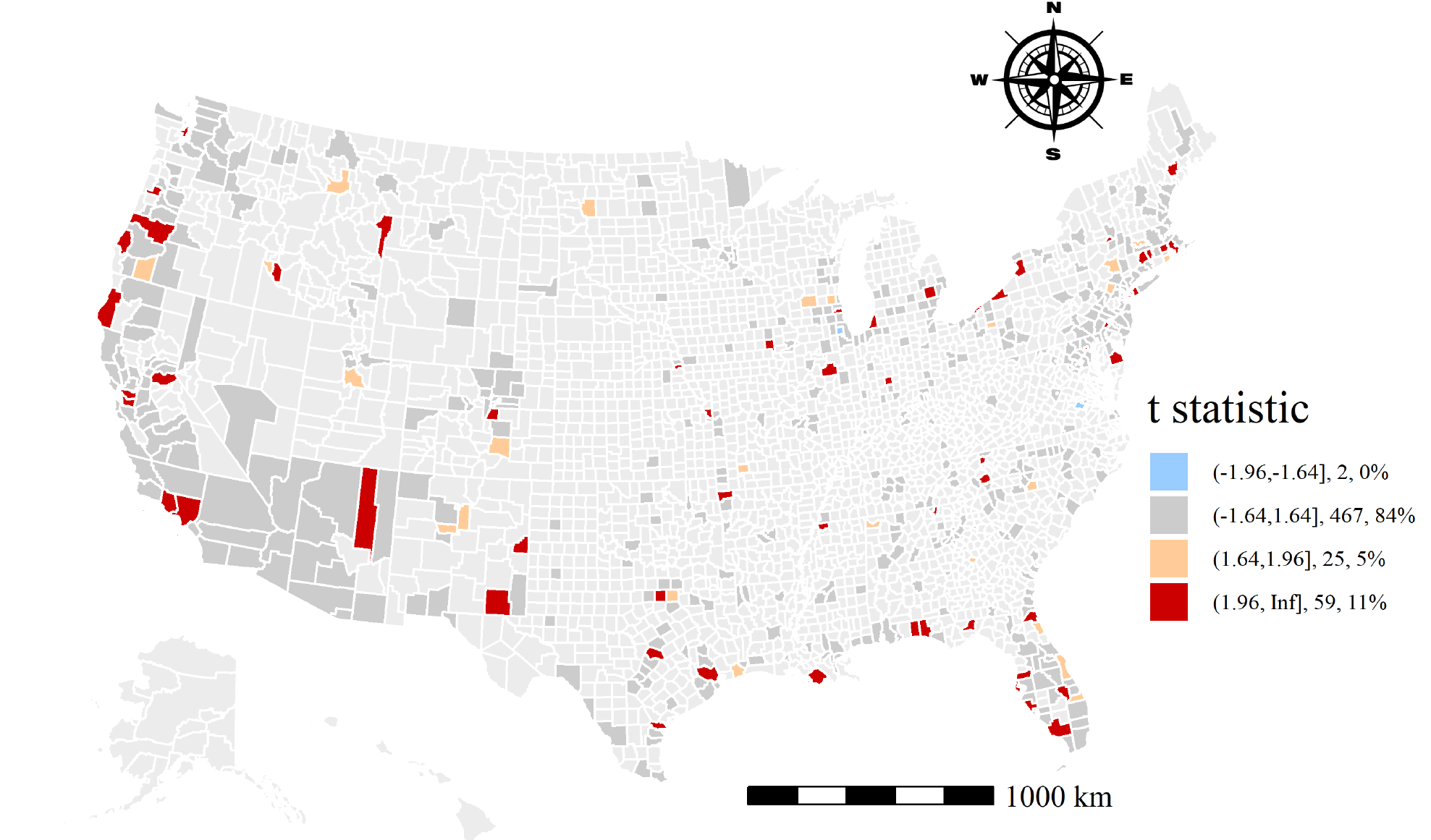}
\label{fig:casetransit48_tvalue}}
\qquad
\subfloat[][$12th$ week, WADD model]{
\includegraphics[width=0.5\linewidth, trim={1.5cm 4cm 0cm 3cm},clip]{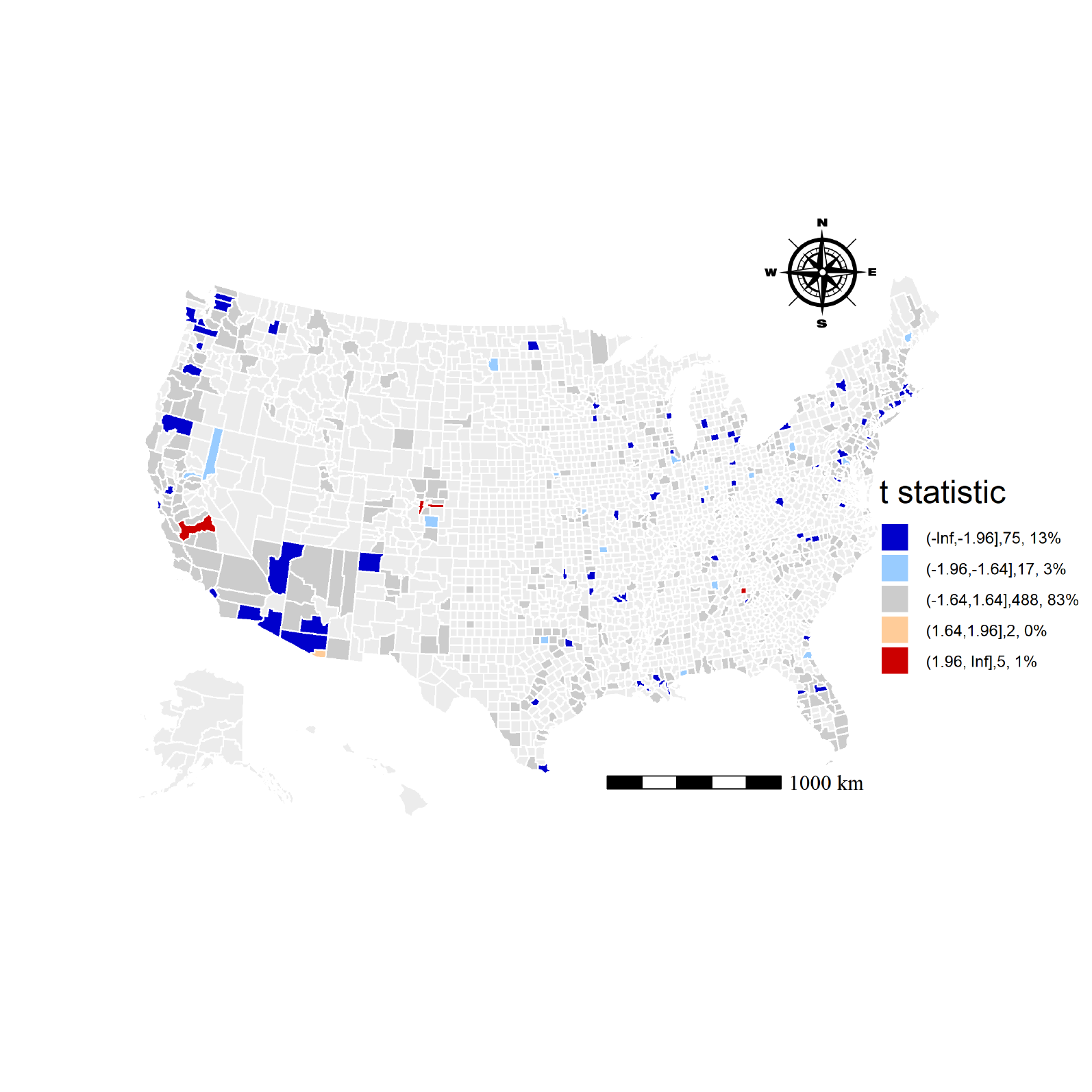}
\label{fig:deathtransit12_tvalue}}
\subfloat[][$48th$ week, WADD model]{
\includegraphics[width=0.5\linewidth, trim={1.5cm 4cm 0cm 3cm},clip]{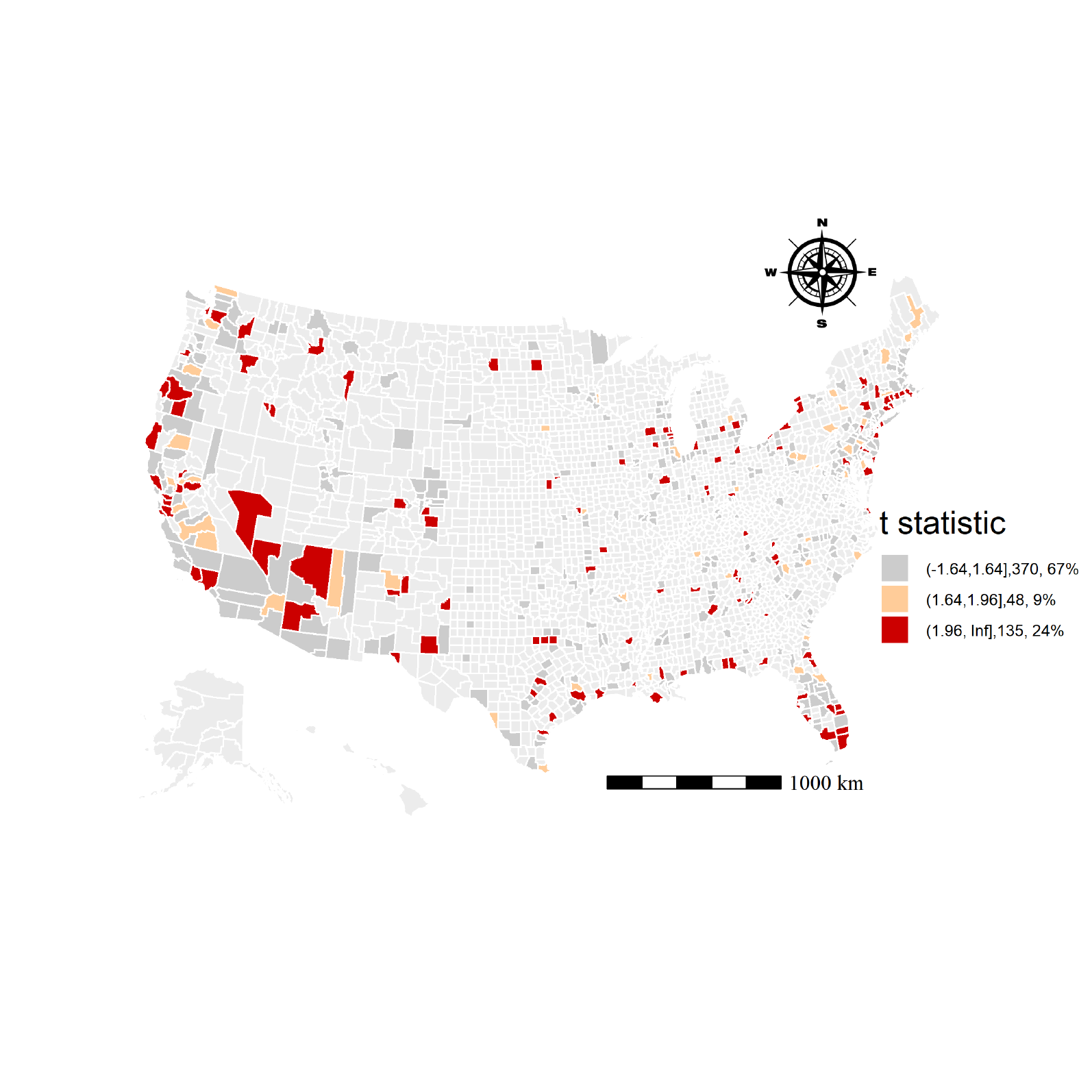}
\label{fig:deathtransit48_tvalue}}
\qquad
\caption{The t statistic distribution of TSPC in two models}
\label{fig:publictransitact}
\end{figure}

The local daily activities directly respond to the WFH policy and travel restriction. Thus, the estimated coefficients of the human activity variables may serve as a measure of the effectiveness of disease prevention and mitigation strategies.  In the M-GTWR model, 
the GPPC is positively (t-stat$>$1.64) related to the WADC and WADD in about 30\% of the counties in the $12th$ week of 2020 as shown in Figure~\ref{fig:casegrocery12_tvalue}, and Figure~\ref{fig:deathgrocery12_tvalue}. 
More importantly, the positive effects are found to be significant in rural counties or low-income areas in California (e.g., Tulare), Arizona (e.g., Yavapai), and New York (e.g., Long Island area). This is on the opposite to the estimated effects of the recreation and parks activities (not present in the model due to the high correlation with the GPPC) in the high population density areas. \highlighttext{Although daily activities increase the disease propagation, people might have better personal prevention (e.g., social distance and mask-wearing) or reduce their daily activities in the high population density areas. Besides, the effects of the GPPC on the WADC and the WADD become less significant in most of the counties in the $48$th week (see Figure~\ref{fig:casegrocery48_tvalue} and ~\ref{fig:deathgrocery48_tvalue}). The WOPC is either negatively (t-stat$<$-1.64) or insignificantly related to WADC and WADD in most counties. In addition, the effects of TSPC are significant in areas with high population density (e.g., Fresno county in California) and low population density areas (e.g., Apache county in Arizona). These might be related to the transit usage policies that apply in different areas. And this discrepancy in terms of TSPC highlight the importance of modeling spatial heterogeneity to more accurately understand the impacts of non-epidemiological factors. The distribution of the t stats of public transit is shown in Figure~\ref{fig:publictransitact}).} In the $12$th week, the number of counties having negative coefficient of TSPC on the WADC and the WADD is more than number of counties having positive coefficient, however, this situation has been changed in the $48$th week. This might be related to the public transit usage restrictions on the early period of pandemic and reopen strategy at the later stage. Therefore, the analysis of all the influencing factors shows that the effect of the influencing factors is spatiotemporal heterogeneous. 

In conclusion, based on our analysis, the coefficients of the POPD and MILE yield persistently positive and statistically significant in studied counties during the studied period. This clarifies the population density and public infrastructure facilities are the primary factors that intensify the number of cases and deaths during the pandemic. On the other hand, the impacts of the several social-demographic variables (e.g., BLAP, MHIC, and PTAP) are observed to become less significant or even insignificant in the later weeks (e.g., the $48$th week). This suggests that static variables may have greater impacts on the disease dynamics in the early stage of the pandemic than in the later stage. Nevertheless, the human daily activity variables (e.g., WOPC, and TSPC) are sensitive to the disease prevention policies and their impacts remain statistically significant during the entire course of the COVID-19. These findings highlight the importance of modeling spatial and temporal heterogeneity to gain more precise understandings of the impacts from non-epidemiological factors. Finally, the insights for disease prevention based on the results of our study are summarized below.

Several implications for the high population density areas (e.g., New York City, counties in California, Washington, Arizona, Virginia, Minnesota, and Florida):
\begin{enumerate}
    \item The intensity of recreation activity is found to be a primary activity factor that facilitates the spread of the COVID-19. Besides, limiting the access to public transit and public office are observed to be effective during the pandemic as suggested in Figure~\ref{fig:publictransitact}. 
    \item Among the counties with a high population density, the percentage of unemployed population(see Figure~\ref{fig:labor}) and population with low education level are the two primary factors that are associated with a higher number of WADC and WADD. 
    \item High population density areas may spend more resources on the older population to reduce their rate of exposure especially in public areas as suggested in the aforementioned analysis of the older population. 
    \item High population density areas that also have a high percentage of black population may consider spending more efforts in alerting the black communities on the risk of the COVID-19 and enforcing the adoption of personal protective equipment (PPE) such as face masks.
\end{enumerate}

Several implications in our study that are important for the low population density counties(e.g., counties in Arizona and counties in Massachusetts):
\begin{enumerate}
    \item The WFH and public transit restriction may be ineffective. Instead, the low population density areas may focus on providing specific strategies to regulate the daily activities of the unemployed populations as suggested in Figure~\ref{fig:caselabor12} and ~\ref{fig:caselabor12_tvalue}. 
    \item The low population density counties should advise the older population to avoid riding public transit and visiting public recreation areas. 
    \item The racial disparities in the infections of the COVID-19 are especially significant in low population density (e.g., counties in New Mexico, Arizona, and Massachusetts). The black community suffers more than other races in most of the low population density counties (see Figure~\ref{fig:caseblack12_tvalue} and ~\ref{fig:caseblack48_tvalue}). Besides, counties in Utah may benefit from improving the COVID-19 prevention among Asian communities (see Figure~\ref{fig:caseasian12_tvalue} and ~\ref{fig:caseasian48_tvalue}).
\end{enumerate}

\section{Conclusion}
In this study, we present the M-GWTR model to account for the spatial and temporal nonstationarity issues in the reported COVID-19 data and investigate the impacts of non-epidemiological factors on the spreading dynamics of the COVID-19 in the U.S. The study serves as a quantitative approach to explore the spatiotemporal effects of the influencing factors on the COVID-19 propagation, and the insights offer effective suggestions for resource allocation policy and disease prevention. 

Specifically, we show that the proposed M-GTWR model is superior to the state-of-the-art benchmarks in capturing the spatiotemporal heterogeneity of disease dynamics during the COVID-19 outbreak. Our results indicate that the concentrated population and the availability of public infrastructures (such as public transit and public road mileage) tend to promote the spread of the disease. Meanwhile, limiting human contacts through reduced human activity levels is found to be largely effective, nevertheless, the effects of which vary significantly over space and time. This indicates that the WFH policy and travel restrictions may not serve as universal mitigation measures, and future strategies should be more tailored to the demographics and socioeconomic of the particular location. Besides, we find that the older, the black, and the Latino are more vulnerable to the COVID-19 than other population groups. The reason may be attributed to either physical weakness or low-risk awareness. The highly educated population is observed more likely to comply with the restrictions during the COVID-19 outbreak. For the commuting time, its median elasticity shows that a 1\% increase in the MTAT to work results in 0.22\% increase in the WADC and 0.95\% increase in the WADD on average. Finally, the change in human activity patterns also presents a mixed impact on disease dynamics. In particular, the scale of the impacts is found to be closely related to the activity intensity and activity types. The grocery and pharmacy activity is found to be significant in low population density areas. And activities associated with public transit usage lead to a positive impact on the WADC and the WADD. This indicates the major role played by the public transit during the COVID-19 and implies the need to restrict public transit usage, especially in high-transit demand areas.  

Based on the findings of the study, we conclude that the general preventative non-pharmaceutical measures, such as work-from-home policy, city lock-down strategy, and travel restrictions,  are unlikely to be universally effective over all subareas of a country. And the effectiveness of which is largely related to the socio-demographic background and travel-related infrastructure in the area and the timing of policy. However, the efficiency of the intervention strategies(e.g., wearing face masks, maintaining social distance, and handwashing) for mitigating the spreading of the COVID-19 is a lack of exploration due to the limited data source. More importantly, since these strategies are at a great cost to the economy, the optimal control strategies to balance the public health and freedom of movement, the economy, and society deserve further investigation. Besides, although we estimated the effects of the several types of activities, we do not differentiate the risk level of detailed activities due to the data limitation (e.g., we estimated the effects of the recreation activities, but the exposure risk of the bar and book store might be different.). And the understanding of the exposure risk of the detailed activities provides directional suggestions for the policy-makers in conducting control strategies for the COVID-19 prevention. Finally, the study indicates the severe disparities respond to the crisis of the disease, which might even intensify the community inequity and racial bias on the more vulnerable groups. Thus, the resource allocation strategies (e.g., health care accessibility, wage guarantee system during the pandemic, and social welfare distribution) should be explored for maintaining a sustainable environment in future crises.

\section{Funding Disclosure}
This research did not receive any specific grant from funding agencies in the public, commercial, or not-for-profit sectors.

\bibliographystyle{elsarticle-harv}

\bibliography{ref}

\end{document}